\documentclass[12pt,preprint]{aastex}

\usepackage{epsf}
\usepackage{dcolumn}
\usepackage{aastexug}

\newcommand{\BE}{\begin{equation}}
\newcommand{\EE}{\end{equation}}
\newcommand{\BA}{\begin{eqnarray}}
\newcommand{\EA}{\end{eqnarray}}

\def\be{\begin{equation}}
\def\ee{\end{equation}}
\def\bea{\begin{eqnarray}}
\def\eea{\end{eqnarray}}

\begin{document}

\renewcommand{\topfraction}{0.8}

\title{\Large\bf
Observation of small scale structure using sextupole lensing}

\shorttitle{Observation of small scale structure using multipole
lensing}

\author{\bf John Irwin and Marina Shmakova}

\affil{ Stanford Linear Accelerator Center, Stanford University,
P.O. Box 4349, CA 94309, USA}

\email{irwin@slac.stanford.edu, shmakova@slac.stanford.edu}

\newpage

\begin{abstract}
Weak gravitational lensing seeks to determine shear by measuring
induced quadrupole (elliptical) shapes in background galaxy
images. Small impact parameter (a few kpc) gravitational lensing
by foreground core masses between $2 \, 10^{9} \mbox{ and } \, 2
\, 10^{12} M_\odot$ will additionally induce a sextupole shape
with the quadrupole and sextupole minima aligned.  This
correlation in relative orientation of the quadrupole and
sextupole provides a sensitive method to identify images which
have been slightly curved by lensing events. A general theoretical
framework for sextupole lensing is developed which includes
several low order coefficients in a general lensing map. Tools to
impute map coefficients from the galaxy images are described and
applied to the north Hubble deep field. Instrumental PSFs, camera
charge diffusion, and image composition methods are modelled in
the coefficient determination process. Estimates of Poisson
counting noise for each galaxy are used to cut galaxies with
signals too small to reliably establish curvature. Curved galaxies
are found to be spatially clumped, as would be expected if the
curving were due to small impact parameter lensing by localized
ensembles of dark matter haloes. Simulations provide an estimate
of the total required lensing mass and the acceptable mass range
of the constituent haloes.  The overdensities and underdensities
of visible galaxies and their locations in the Hubble foreground
is found to be consistent with our observations and their
interpretation as lensing events.
\end{abstract}


\newpage

\section{Introduction} \label{intro}

Weak gravitational lensing methods [for review see
\citep{Bartelmann:1999yn, Mellier:1998pk, Hoekstra:2002nf} and
references therein] allow one to investigate the evolution of
matter clustering and the growth of large-scale structure
\citep{Wittman:2000tc, Mellier:2002vp}, and thus probe the
properties of both dark energy and dark matter. It is a unique way
to investigate both the past and future of the universe. Structure
growth is understood as the growth of the matter density
fluctuations predicted by inflationary cosmology and dark energy
evolution (for review see \citep{Peebles:xt}), hence the structure
of density fluctuations provides information about inflationary
scenarios \citep{Huterer:2000mj, Huterer:2001yu, Linder:2003dr}.
On the other hand, the matter distribution measurements give
bounds on the dark energy equation of state, allowing one to
predict the future of the universe \citep{Linde:2002gj,
Kallosh:2003mt, Kallosh:2003bq}.

The traditional weak gravitational lensing techniques
\citep{Kaiser:2000if}, which locate and quantify the large clumps
of matter such as clusters of galaxies (visible or dark) with
$10^{14} \, M_\odot$, are not sensitive to the substructure of
large clusters or to smaller groups and clumps of matter. We
present here a new method designed to detect the presence of
smaller clumps. We have applied the method to the north Hubble
deep field and have seen a signal which has the features we were
expecting. We call this ``sextupole lensing", since it involves
the measurement of the quadrupole and sextupole lensing strengths,
and may be extended to other sextupole-order and higher terms as
well.

Sextupole-lensing strengths lie between strong and weak lensing
since these lensing events must yield both quadrupole and
sextupole moments the order of the background noise.  In
comparision, the induced quadrupole moment in weak lensing can be
smaller than the background noise and in strong lensing the
magnitude of the lensing shear parameter is close to unity.

In the multipole-lensing view, the familiar strong-lensing arclets
\citep{Bartelmann:1999yn, Mellier:1998pk} are an example of the
general property that the nonlinear 1/r deflections of a light
stream passing a mass concentration will produce, relative to the
stream centroid, a full complement of moments. The curving seen in
the arclet can be understood as the correlated superposition of a
quadrupole and sextupole moment, with the length of the arclet
usually is determined by the local strength of the octupole
moment.

\begin{figure}
\centerline{\leavevmode\epsfysize= 4cm \epsfbox{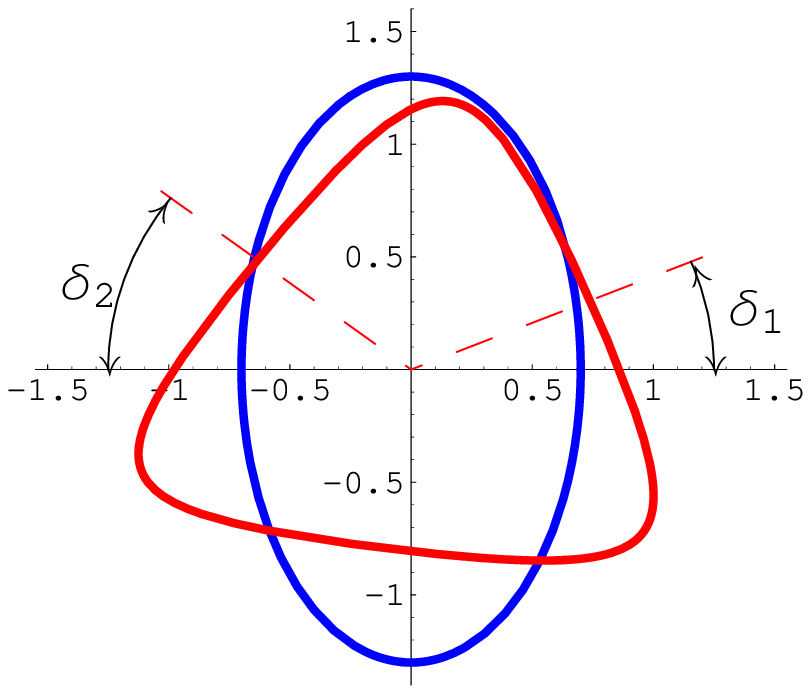}
\includegraphics[width=1.3in,height=4cm]{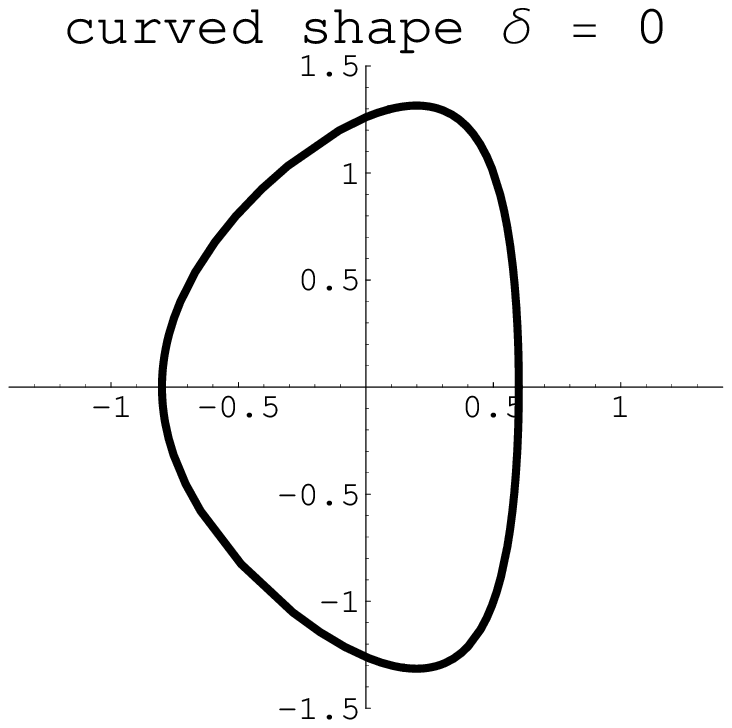}\,\,
\includegraphics[width=1.3in,height=4cm]{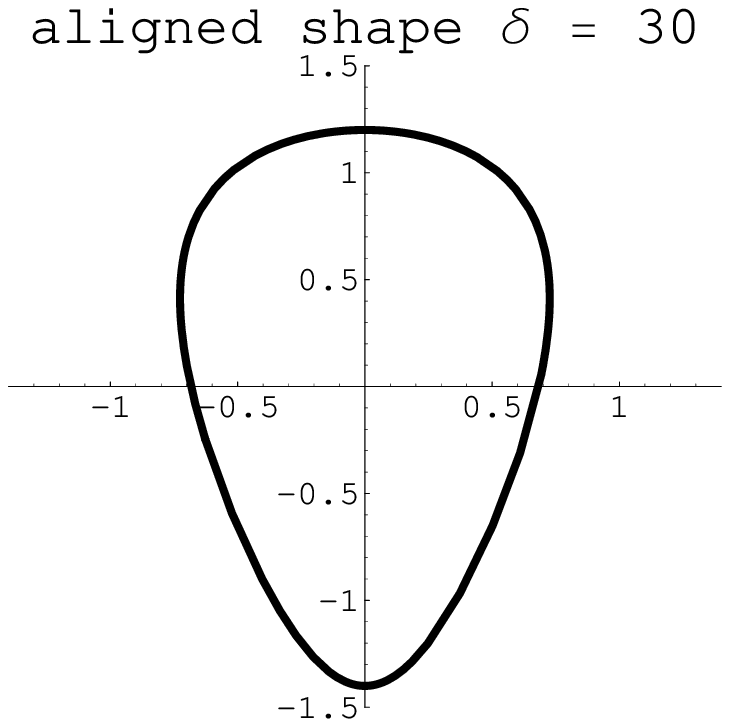}} \caption[qusegdef]
{ "Curved" and "Aligned" galaxies.  $\delta = \delta_{1}$ if
$\vert \delta_1 \vert < 30^\circ$, else $\delta=\delta_2$.
$-30^\circ \le \delta \le 30^\circ,$ where $ \delta \sim 0^\circ $
for "curved" and $ \delta \sim \pm 30^\circ $ for "aligned"
galaxies.} \label{deltaDef}
\end{figure}
The typical cluster identifiable through weak lensing, has a mass
of 10$^{14}$ solar masses and a radius of about 500 kpc. Since the
strength of the quadrupole kick, responsible for the ellipticity,
is proportional to the mass and falls off like 1/r$^{2}$, one
could get the same induced quadrupole moment in a light stream
positioned 5 kpc from an object of 10$^{10}$ solar masses. In the
latter example, since the sextupole moment varies as 1/r$^{3}$,
the sextupole moment becomes 100 times stronger, becoming as large
as the intrinsic sextupole moments of background galaxies. It
occurred to us that if ensembles of dark matter contain an
abundance of lower-mass clumps, the light streams passing through
them might occasionally pass close to a 10$^{9}$ to 10$^{11}$
solar mass object producing an observable sextupole moment in the
image \citep{finalfocus}. Since such ``close-encounter" lensing
has an alignment of the quadrupole and sextupole moments, a signal
for ensembles of dark matter containing a population of lower mass
objects, would be evident as spatial correlations in the
orientation of the sextupole with respect to the quadrupole
moments (``a spatial correlation of an angular correlation").

Additionally, if the light stream penetrates a mass clump at a
location where the mass gradient is non-zero, another 3rd order
moment, like the sextupole moment but with a rotational dependence
of a dipole moment, is created. This moment, which we call the
gradient moment, would be correlated in the same manner.

Let us begin with the simple case of a concentrated point mass.
The $1/r$ deflection of a light stream may be expanded in a power
series. With a dark matter clump at the origin, and the beam
centroid at position $x_0$, the kick $\Delta x^\prime$ is given as

\BA\label{deflection1D} \Delta
x^\prime=-\frac{4MG}{x}=-\frac{4MG}{x_0+ \Delta x}=
-\frac{4MG}{x_0} \left[1 - \frac{\Delta x}{x_0}
+\left(\frac{\Delta x}{x_0}\right)^2 + ... \right].\EA

\noindent $\Delta x$ indicates the offset from the centroid within
the light stream.  Each term in the power series is down from the
previous term by the fraction $\frac {\Delta x}{x_0}$.  In other
words, to see the effects of higher order terms, the impact
parameter can be only a small multiple of the light-stream width.

This power series expansion can be generalized to yield the kick
in both coordinates:\footnote{We will prove this and generalize to
arbitrary mass distributions in the next section.}

\BA\label{deflection2D} \Delta x^\prime + i \Delta y^\prime =
-\frac{4MG}{x_0+ (\Delta x - i \Delta y)}= -\frac{4MG}{x_0}
\left[1 - \frac{\Delta x-i \Delta y}{x_0} +\left(\frac{\Delta x -
i \Delta y}{x_0}\right)^2 + ... \right].\EA

\noindent Concentrating on the linear term, one sees for example
that the horizontal kick is defocussing, while the vertical kick
is focusing.  The image appears larger in the focussed direction
and smaller in the defocussed direction, hence the linear term
changes a circle into an ellipse. It is a general feature that the
image distortions have the opposite sign of the map coefficients.

Equation \ref{deflection2D} may equally well be written in polar
coordinates:

\BA\label{deflection2DPolar} \Delta x^\prime + i \Delta y^\prime =
-\frac{4MG}{x_0} \left[1 - \frac{r}{x_0} e^{-i \theta} +
\left(\frac{r}{x_0}\right)^2 e^{-i 2\theta}+ ... \right].\EA

\noindent The 2nd-order term is the sextupole term.  For $\theta =
0 $ the deflections of the 1st- and 2nd-order terms have the
opposite sign, hence the quadrupole is minimum there and the
sextupole term is maximum.  For $\theta = \pi $ the quadrupole has
its other minimum and the sextupole also has a minimum. To the
right in fig. \ref{deltaDef} are shown the superposition of
quadrupole and sextupole distortions for two distinct orientations
of the sextupole moment with respect to the quadrupole moment.  It
is the curved shape that would arise from a lensing event.

\noindent The plan we will follow is to measure the quadrupole and
sextupole shape of all galaxies, classify each galaxy according to
whether it is ``curved", ``mid-range" or ``aligned", and examine
the distribution of ``curved" galaxies on the sky to determine if
such galaxies are randomly distributed or unusually clumped.

The remaining sections within this paper are :  \textbf{2. Lensing
maps} which introduces the general concept of a nonlinear map, and
derives its coefficients for lensing from a general mass
distribution arranged on a single plane. Examples of a point mass
and a Gaussian radial distribution are provided. \textbf{3.
Extraction of map coefficients from images}, describes three
increasingly sophisticated methods for inferring the nonlinear map
coefficients from images. The first method is based on image
moments, the second uses the best fit of a mapped (initially
symmetric) galaxy shape parameterized in a simple form. The third
method expands upon the second method to include effects of the
point-spread function, charge diffusion in the camera, and
important features of the image composition process.

\textbf{4. Galaxy properties} describes the selection of galaxies
in the Hubble deep fields and their size and intensity. \textbf{5.
Quadrupole coefficient measurements} presents the results of three
methods of map determination to measure the quadrupole
coefficient. An analytical estimate for the noise in the
quadrupole coefficient, based on Poisson noise of the galaxy
image, is presented and noise cuts are implemented. \textbf{6.
Sextupole coefficient measurements} presents the analogous results
for the sextupole. The distribution in the angle between
quadrupole minimum and sextupole minimum, distinguishing
``curved", ``mid-range", and ``aligned" galaxies, is presented.

\textbf{7. Clumping probabilities} employs a ``nearest neighbors"
analysis to quantify the probability of obtaining the observed
spatial distribution of ``curved" galaxies in the field if it were
to arise randomly.  Both ``curved" and ``aligned" and galaxies are
shown to be more clumped than randomly chosen subsets.

\textbf{8. Dark matter lensing?} estimates the mass of the
ensembles required to obtain the observed clumping. The total mass
and constituent mass are discussed. Lensing by overdensities in
dark matter foregrounds is proposed.  \textbf{9. Systematic
effects} addresses possible systematic sources of clumping, such
as correlated curvature originating with the background galaxies
themselves, or correlated curvature from an uncompensated residual
of the point-spread function. Section \textbf{10. Summary}
summarizes the paper.
\smallskip


\section{ Lensing maps }\label{maps}

\begin{figure}[h!]
\centering\leavevmode\epsfysize= 8 cm \epsfbox{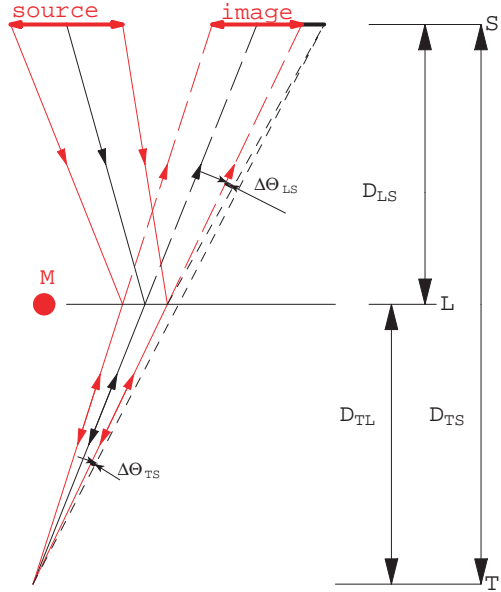}
\caption[geometry] {A diagram showing 3 light rays from a source
image scattered by a single concentrated mass.  The relationship $
D_{TS} \Delta \theta_{TS} = D_{LS} \Delta \theta_{LS}$  expresses
the observed displacement angle in terms of the actual deflection
angle for one ray of the image. Distances are at the epoch of the
lensing deflection.} \label{geometry}
\end{figure}

Fig. \ref{geometry} shows rays from a distant background galaxy
being deflected by a mass concentration. The apparent image, as
seen by the telescope, is defined by an intensity function which
depends only on the angle of each ray as it enters the telescope.
Following a ray backwards, toward the apparent image, it is
deflected by mass distributions, but is known to depart somewhere
from the source galaxy. The ray will have a definite position and
angle at the source galaxy (measured relative to the position and
angle of the centroid ray). This map from telescope variables to
source variables will be symplectic since the light geodesics are
described by a Hamiltonian. From the point of view of the galaxy,
all rays traced backward from the telescope come from the same
point. Hence the two angles (or equivalently, the focal plane
coordinates) at the telescope uniquely describe the trajectory
through space and the initial position $x_S $ and $y_S $ (and
angle ${x}'_S $ and ${y}'_S $)  at the galaxy. Thus the
``backwards'' map can be written as a set of 4 functions $x_S (x_T
,y_T ),\;y_S (x_T ,y_T ),\;{x}'_S (x_T ,y_T ),\;\mbox{and }{y}'_S
(x_T ,y_T )$, where ``T'' designates ``telescope'' and ``S''
designates ``source''. The coordinate system for both the source
and telescope images will be taken to be location on the focal
plane. The position of the centroid trajectory is taken to be
identical for both images, i.e. the dipole kick suffered by the
image as a whole is ignored. Only under special circumstances,
such as strong lensing, will a ray leaving the telescope at
different angles arrive at the same point on the source galaxy. We
will not consider such cases and hence will be able to drop the
two functions ${x}'_S \mbox{ and }{y}'_S $ and concern ourselves
with regions for which the determinant\footnote{We monitor this as
we process each image.}
\begin{equation}
\label{determinant} {\Big\vert} \begin{array}{cc}
 {\frac{\partial x_S }{\partial x_T }} & {\frac{\partial x_S
}{\partial y_T }} \\
 {\frac{\partial y_S }{\partial x_T }} & {\frac{\partial y_S
}{\partial y_T }}\\
\end{array} {\Big\vert} > 0.
\end{equation}
\noindent The two functions $x_S(x_T ,y_T)\mbox{ and } y_S(x_T
,y_T)$ can be combined into one complex function by defining $w_S
=x_S +iy_S $. This complex function can be written in terms of the
variables $w_T =x_T +iy_T $ and $\bar {w}_T =x_T -iy_T $ by
substituting $x_T =\frac{1}{2}\left( {w_T +\bar {w}_T } \right)$
and $y_T =\frac{1}{2i}\left( {w_T -\bar {w}_T } \right)$. The map
equations can then be written as the single function $w_S (w_{T,}
\bar {w}_T ).$  Since the transverse width of the light stream
will be small compared to characteristic dimensions of the
variations of the lensing mass distributions, we may expand this
function in a power series about the stream centroid:
\begin{equation}
\label{map} w_S (w_T ,\bar {w}_T )=w_T +\sum\limits_{n,m=0}^\infty
{a_{nm} w_T^n } \bar {w}_T^m .
\end{equation}
The significance of the variables $w$ and $\bar{w}$ rests on the
fact that products and powers of them are rotation eigenfunctions.
It follows that the terms in the expansion for $w_S$ have a simple
interpretation. The $1+a_{10}$ combination represents a simple
rotation and scaling, the $a_{01} $ term is a quadrupolar
distortion, the $a_{02} $ term is a sextupolar distortion, the
$a_{03} $ term is an octupolar distortion, the $a_{20} $ term is a
cardioid-like distortion, and the $a_{11} $ term is an
$r^2$-dependent translation of circles, and so on. We will be
concerned primarily with the terms  $a_{01} $, $a_{02} $, $a_{03}
$, and $a_{20} $ and refer to them more simply by the letters,
$a,\; b,\; c,\mbox{ and } \bar {d}$, respectively. As we show in
the following paragraph, for a map arising from mass distributed
on a plane, $a_{10}$ is necessarily real and $a_{11} =2\bar
{a}_{20} =2{d}.$  Note that the coefficients $b,\;c, \mbox{ and }
d$ have dimensions.  Since we are using focal-plane position as
the coordinate system, units will usually be given in terms of the
pixel grid on that plane.

In this paper we limit ourselves to maps arising from mass
distributions which can adequately be represented by mass
projected onto a single plane (which we will refer to as the
lensing plane). To determine the Green's function for this case,
consider the effect of a point mass.  A light stream passing at
distance $r$ will be deflected radially by the angle $ \Delta {r}'
= -\frac{4MG}{r}$. The potential for such a kick, the sought
Green's function, is $2\Phi _\delta =4MG\,Ln\left[ r \right]$.
This is just the Green's function for the 2 dimensional Laplace
equation, $\nabla ^2\Phi =4\pi G\rho $, where $\rho$ is the mass
density function.\footnote{$\Phi$ is the potential for the
deflection of a non-relativistic particle.  The factor of 2 is
explicitly retained indicating that that light rays receive twice
this kick.}

In accord with the variables chosen for the map, we expand the
solution to Laplace's equation in the variables $w =x+iy $ and
$\bar {w} =x -iy$.  Variables without subscripts are taken to lie
in the lensing plane.  To enable power series expansions, it is
useful to introduce derivative operators $\partial \equiv
\frac{\partial }{\partial w}\equiv \frac{1}{2}\left[
{\frac{\partial }{\partial x}-i\frac{\partial }{\partial y}}
\right]$  and $ \bar{\partial}\equiv \frac{\partial }{\partial
\bar {w}}\equiv \frac{1}{2}\left[ {\frac{\partial }{\partial
x}+i\frac{\partial }{\partial y}} \right]$, which have the desired
property that $\partial w = \bar{\partial} \bar {w} =1$ and
$\partial \bar {w} = \bar{\partial} w =0$.

\begin{figure}[h!]
\centering \leavevmode\epsfysize= 5.5cm \epsfbox{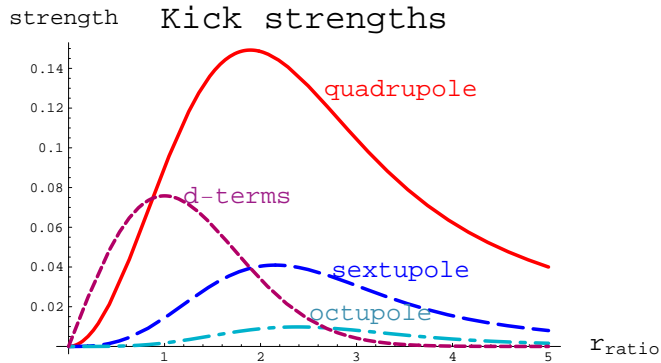}
\caption[gaussianProfile] {\centering Relative strength of the
quadrupole, sextupole, octupole, and cardioid-like kicks within a
Gaussian clump. The horizontal axis is the radius divided by the
rms lens radius. } \label{gaussianProfile}
\end{figure}

Using these variables, the power series expansion for the
potential is $ \Phi =\sum\limits_{n,m=0}^{\infty} \frac { 1}{n!
m!} \Phi_{n m} w^n \bar {w}^m $ where $\Phi_{nm}= {\partial}^n
\bar{\partial}^m \Phi $ evaluated at $w=w_0$.  The reality
condition on $\Phi$ implies relationships among these
coefficients. For example, the $\Phi_{nn}$ are real and $\Phi_{nm}
= \overline {\Phi_{mn}}$.  If both $n$ and $m$ are greater than or
equal to 1, the term will be proportional to the density or a
derivative of the density, since $\Phi_{11}=\frac{1}{4}\nabla^2
\Phi = \pi G \rho$. The two components of the kick, given in the
form $\Delta x^{\prime} + i \Delta y^{\prime}$, are contained in
the single equation,
\begin{equation}
\Delta {w}^{\prime} = -2 \, \bar{\partial} {\left( 2 \Phi
\right)}. \label{kick}
\end{equation}
The geometry of fig. \ref{geometry} implies $w_S=w_T + D_{LS}
\Delta w^{\prime}(w,\bar {w} )$ with $w=\frac
{D_{TL}}{D_{TS}}w_T$. Thus for a general potential the map
coefficients are given by
\BA \label{coeffFromPotential} a_{10}
&=& - 4 D_{LS} \frac{D_{TL}}{D_{TS}} \Phi_{11} = - 4 \pi G D_{LS}
\frac{D_{TL}}{D_{TS}} \rho,\cr
  a &=& - 4 D_{LS} \frac{D_{TL}}{D_{TS}} \Phi_{02},\cr
 b &=& - 4 D_{LS} \left({\frac{D_{TL} }{D_{TS} }} \right)^2
\frac{1}{2} \Phi_{03},\cr c &=& - 4 D_{LS} \left( {\frac{D_{TL}
}{D_{TS} }} \right)^3\frac{1}{3!} \Phi_{04},\, \mbox{ and } \cr d
&=& - 4 D_{LS} \left({\frac {D_{TL} }{D_{TS} }} \right)^2
\Phi_{12} = - 2 \pi G D_{LS} \left({\frac{D_{TL} }{D_{TS} }}
\right)^2 \bar{\partial} \rho  .
 \EA
 The map coefficients for the point source may be found by
evaluating the derivatives of $2\Phi_\delta =4MG\,Ln\left[ r
\right]=2MG\,Ln\left[ w \bar{w} \right]$:
 \BA \label{coeffFromPoint}
  a_\delta & = &D_{LS} \frac{D_{TL}}{D_{TS}
}\frac{4MG}{\bar {w}_0^2 },\cr
 b_\delta & = & -D_{LS} \left(
{\frac{D_{TL} }{D_{TS} }} \right)^2\frac{4MG}{\bar {w}_0^3 },\cr
c_\delta & = & D_{LS} \left( {\frac{D_{TL} }{D_{TS} }}
\right)^3\frac{4MG}{\bar {w}_0^4 }, \,\, \mbox {and} \cr d_\delta
&=& a_{10 \,\delta}= 0. \EA

The map coefficients can be easily found for any symmetric mass
distribution by using Gauss' law ( $\frac {\partial \Phi}{\partial
r}=\frac{2 M(r) G}{r} =\frac{4 \pi G}{r} \int\limits_{0}^{r}
\rho(r')r' dr'$) and the relationship $\bar{\partial} \Phi (r) =
\frac {w}{2r} \frac {\partial \Phi (r)}{\partial r}.$  For a plot
of the coefficient strengths in the case of a Gaussian mass
distribution see fig. \ref{gaussianProfile}.

\section{Extraction of map coefficients from images}\label{determineMap}

 We use three distinct methods to estimate lensing
map coefficients: $1)$ a moment method, $2)$ a radial-fit method,
and $3)$ a model method that takes into account the point-spread
function (PSF), the diffusion of charge between camera pixels, the
dithering of pointing, and drizzle of photon counts onto the final
pixel grid.

The moment method has the advantage of simplicity, but because the
images are necessarily truncated its accuracy is compromised as a
result of edge effects and an inherent ambiguity in including the
effect of the PSF. Furthermore, it does not bring into play the
knowledge that moments derived from the action of lensing will
have a radial strength proportional to the derivative of the
radial profile of the galaxy. The radial-fit method overcomes
these shortcomings. When we take into account the PSF and other
known processes that distort galaxy images, namely the charge
diffusion and image composition, we refer to the method as the
model method. The model method begins with a parameterized radial
profile of the source galaxy and in addition models the PSF,
diffusion and other image composition processes. This latter
method is limited by the imperfect knowledge of the features it
seeks to include (such as PSF and charge diffusion), plus of
course, the noise inherent in background galaxy shapes and photon
counting noise.

We wish to emphasize that the radial-fit and model methods do not
contain an implicit assumption of azimuthal symmetry for the
background galaxies. One can imagine applying either of these
methods to an unlensed galaxy, and finding non-zero map
coefficients, $a_S$ and $b_S$.  Then if that same galaxy is later
lensed one obtains, $a=a_S+a_L$ and $b=b_S+b_L$, where $a_L$ and
$b_L$ are the coefficients of the linear and quadratic terms in
the lensing map. For a proof of this see the appendix.

\subsection{Moment method}\label{momentMethod}
Image moments are defined by the integrals $ M_{nm}   = \int w^n
\bar {w}^m i(x ,y)dx dy$, where $i(x,y)$ is the light-intensity
function normalized to unit integral.  According to the notation
introduced in the previous section, we distinguish the original
source galaxy intensity by the subscript $S$ and the intensity as
observed at the telescope by the subscript $T$. Because of the
symplectic nature of the lensing map, these intensities are
related to one another by $i_S (x_S ,y_S )\,dx_S dy_S = i_T (x_T
,y_T )\,dx_T dy_T $.  Using this relationship one can relate the
moments of the source to the moments of the observed image through
\BA \label{momentEq} M_{nm}^S &=& \int w_S^n  \bar {w}_S^m \, i_S
(x_S ,y_S )dx_S dy_S \cr &=& \int (w_T + \Delta w)^n (\bar {w}_T +
\overline {\Delta w})^m i_T(x_T ,y_T) \, dx_T dy_T , \EA
\noindent
where $\Delta w = a \bar {w}_T + b \bar {w}_T^2 + c \bar {w}_T^3 +
2d{\kern 1pt} w_T \bar {w}_T + \bar{d} {\kern 1pt}w_T ^2$.

Inserting this expression for $\Delta w$ into the expression for
$M_{20}^S$ one can find the following equation for the quadrupole
map coefficient:
\begin{equation}
\label{quadMomentEq}
\begin{array}{c}
 M_{20}^S
 =\int {\left( {w_T +a\bar {w}_T + ...} \right)^2\,} i_T (x_T ,y_T )dx_T dy_T
 \approx M_{20}^T +2aM_{11}^T +a^2M_{02}^T.
 \end{array}
\end{equation}
Since the moments on the right hand side of this equation can be
measured from the telescope image (neglecting problems with
truncation errors), one obtains a quadratic equation for $a$
depending on the value of $M_{20}^S$, which, lacking any
information of its value, is usually set equal to zero. Since $w
\bar{w}=r^2\mbox{ it follows that }M_{11}^T $ is the mean square
radius of the telescope image. Solving for $a$,
\begin{equation}
\label{quadCoeffEq} a\approx -\frac{\Delta M_{20}}{2M_{11}^T }\,
\frac{2}{1+\sqrt{1+\frac{\Delta M_{20} \bar{M}_{20}}{M_{11}^{T
2}}}},
\end{equation}
\noindent where  $ \Delta M_{nm} \equiv M_{nm}^T - M_{nm}^S. $

Likewise an equation for the sextupole moment may be found by
inserting the expression for $\Delta w$ into the expression for
$M_{30}^S$ obtaining
\begin{equation}
 M_{30}^S \approx M_{30}^T +3 b \,M_{22}^T
+3a\,M_{21}^T +\ldots \mbox{ whence } b\,\approx -\frac{\Delta
M_{30}}{3M_{22}^T }-\frac{a\,M_{21}^T }{M_{22}^T }\quad .
\label{sextCoeffEq}
\end{equation}
\noindent Similarly the expression for $c$ is derived from
\begin{equation}
 M_{40}^S \approx M_{40}^T +4 c \,M_{33}^T
+4a\,M_{31}^T + 6 a^2 M_{22}^T + \ldots \mbox{ yielding }
c\,\approx -\frac{\Delta M_{40}}{4M_{33}^T }-\frac{a\,M_{31}^T
}{M_{33}^T } - \frac{3 a^2\,M_{22}^T }{2 M_{33}^T }.
\label{sextCoeffEq}
\end{equation}
\noindent And to get an expression for the d-term, we insert
$\Delta w$ into an expression for $M_{21}^S$  resulting in
\BA
\label{gradientMomentEq} M_{21}^S = M_{21}^T +2 a \bar{M}_{21}^T +
\bar{a} M_{30}^T + 5 d \, M_{22}^T+ \ldots \mbox{ yielding } d
\approx - \left[ \Delta M_{21}^T + 2 a \bar {M}_{21}^T + \bar{a}
M_{30}^T \right] \frac{1}{5M_{22}^T}.
 \EA
If the image were not truncated it would be possible to extend
this method to expressions for the moments in the presence of a
PSF. The resulting corrections to the moment equations provide
insight into the magnitude of PSF corrections. Intensity functions
which have been convolved with the PSF and moments derived from
these functions will be designated by a ``$^\wedge$''.  Let $p$ be
the point-spread function.  Then
\BA  \label{psfMomentEq} \hat
{M}_{nm}^T & = &\int {w_T^n \bar {w}_T^m \,\hat {i}_T (x_T ,y_T
)dx_T dy_T }
  =  \int {\int {w_T^n \bar {w}_T^m \,p(\vec {r}_T -\vec {{r}'}_T )\,i_T ({\vec
{r}}'_T )dx_T dy_T d{x}'_T d{y}'_T } }  \\
 & = & \int { \int { \left( {\Delta w_T +{w}'_T } \right)^n \left( {\Delta \bar
{w}_T +\bar {{w}'}_T } \right)^m  }}  p(\Delta \vec {r}_T )i_T
({\vec {r}}'_T )d\Delta x_T d\Delta y_T d{x}'_T d{y}'_T .\nonumber
 \EA
Here $\Delta x_T \equiv x_T -{x}'_T $ etc.  When the binomial
expressions are expanded, the double integral reduces to the sum
of a product of single integrals. One finds, for example, \BA
\label{psfMomentEx} \hat{M}_{20}^T &=& M_{20}^P + M_{20}^T, \quad
\hat {M}_{30}^T = M_{30}^T + M_{30}^P, \quad \hat {M}_{21}^T =
M_{21}^T + M_{21}^P , \cr \hat {M}_{11}^T &=& M_{11}^P + M_{11}^T,
\quad  \mbox{and} \quad \hat {M}_{22}^T \approx M_{22}^T +
2M_{11}^P M_{11}^T +M_{22}^P. \EA
 In all cases, if the moments of the point-spread
function are known, the moments of the original image can be found
from the smeared image.

Finally, using the approximations $\hat{a} \equiv -\frac
{\hat{M}_{20}^T}{2\hat{M}_{11}^T}$  and $a^P \equiv -\frac
{M_{20}^P}{2M_{11}^P}$ and similar approximations for $\hat{b}$
and $b^P$, and using the relationships of eq.\ref{psfMomentEx},
one can find an approximation for the map coefficients ``before
PSF": \footnote{We remind the reader that these are rather poor
approximations and often inaccurate for truncated or large galaxy
images, and a poor approximation for the PSF because of
contributions to the moments from large radii.}
\BA\label{mapAddition} a = \frac {1}{\lambda_{1}^T}
\left(\hat{a}-\lambda_{1}^P a^P \right) \; \; \; \mbox{and} \; \;
\; b = \frac {1}{\lambda_{2}^T}\left(\hat{b}-\lambda_{2}^P b^P
\right), \EA \noindent where \BA \lambda_{1}^X \equiv \frac
{M_{11}^X}{M_{11}^T+M_{11}^P} \,; \; \; \mbox{ and   } \; \;
\lambda_{2}^X \equiv \frac {M_{22}^X}{M_{22}^T+2M_{11}^T M_{11}^P
+ M_{22}^P} . \EA

\subsection{Radial-fit method}\label{radialFit}

We assume a radial profile for the background galaxy of the form
\footnote{K. Kuijken \citep{Kuijken:1999} introduced the
radial-fit method, taking a sum of Gaussians as the ansatz for the
radial profile.}
\BA\label{radialProfile} F(r_S^2)=(A+B r_S^2+C
r_S^4)_{_+} \, e^{-Dr_S^2}. \EA
\noindent This will be adequate
since this function must depend only on $r_S^2$, and we have a
parameter ($A$) for the strength at the center of the galaxy, a
parameter ($B$) that allows for an arbitrary quadratic behavior at
the origin, there is a cut-off parameter ($D$) that reflects the
image size , and a parameter ($C$) which can modify the behavior
as one approaches the cut-off. The $+$ subscript indicates that if
the polynomial has a value less than zero, it is to be set equal
to zero.  This is necessary to avoid negative intensities, which
would be unphysical.\footnote{The fit method we use requires an
analytic expression for the derivative of the fit function.
Fortunately $(Q_+)'= (Q \theta(Q) )' = Q' \theta(Q) $ since the
value of the polynomial, $Q$, is zero where the step function
$\theta(Q)$ has a non-zero derivative.}  Given a centroid, one may
numerically determine the radial profile of a galaxy (for example,
by dividing each pixel into many smaller pixels), and this may be
compared with the fit results.  In cases we have examined, the two
curves are almost indistinguishable.

For convenience we introduce two additional parameters: each
occurrence of $r_S^2$ is replaced by $(r_S/r_0)^2$, and a constant
$c_0$ multiplies the polynomial.
\BA\label{scaledRadialProfile}
F(r_S^2)=c_0 (A + B s + C s^2)_{_+} \, e^{-Ds} \mbox{ with }
s=(\frac{r_S}{r_0})^2. \EA
\noindent $r_0$ is taken to be near the
rms size of the source image. By changing the size of $r_0$ one
can change the size of the image without affecting its shape. The
factor $c_0$ is introduced so that the shape parameters, which are
now dimensionless, have values $A \approx 1$ and $B \mbox{ and }
C$ can be compared to unity.

The effect of lensing is contained in the parameters of the map of
eq. \ref{map}. One replaces each occurrence of $r_S^2$ by the
expression $w_S \bar{w}_S$.

The parameters of the radial profile and the map are determined by
minimizing the L2 norm: \BA\label{norm} \|{i_{F} - i_{T}
}\|_{\rho}^2 = \int{(i_F - i_T)^2 \, \rho \, dx_T dy_T } \mbox{,
where } i_F = F(r_S^2) \, \vert \frac{\partial w_S}{\partial
w_T}\vert . \EA
\noindent In $i_F$, $r_S^2=w_S \bar{w}_S$ is
understood to be a function of $x_T$ and $y_T$ through $w_{T}$ and
$\bar{w}_T$.  The $\vert...\vert$ is the Jacobian of the
transformation between $S$ and $T$ variables. All map variables in
the Jacobian occur in 2nd order except for the $d$ variable, which
occurs in 1st order. A weight function $\rho$ can be introduced if
desired. However, because this technique ignores truncated pixels
rather than considering them to be zero, edge effects are
inherently smaller compared to the moment method. At each
calculation of $i_F$ the Jacobian is monitored to see that it is
not negative at that or any smaller $r$.

With parameters $a, b, c,$ and $d$ included in the map, then
together with the centroid position, $w_0$, and the shape
parameters $A, B, C,$ and $D$, there are 14 variables to
determine. The fit is done in several steps using a
multi-dimensional Newton's method.  At each step any subset of the
14 variables are allowed to vary.  The curvature matrix for these
parameters is computed, then diagonalized and eigenvectors with
very small eigenvalues are not allowed to contribute to the
function change in that step.  The convergence to a minimum is
controlled by a parameter step size, which finally is required to
be as small as $10^{-15}$.

The map parameters are not strictly orthogonal, but because of
their distinct leading angular behavior, they appear to be stable
and well determined.  We have thoroughly tested the fit program by
fitting many generated matrices, starting from a variety of
initial conditions.  Typically map parameters are reproducible to
parts per thousand at typical moderate parameter strengths.  The
method is less accurate for images with large map parameters.

\subsection{Model method}\label{modelMethod}
The model method begins by constructing an $i_F$ as in the
previous subsection (here on .02" pixels) and convolving it with a
sub-sampled PSF (also on .02" pixels) as provided by the Tiny
Tim\footnote{http://www.stsci.edu/software/tinytim} program. This
convolved image is dropped (25 times) onto a dithered original
pixel grid (0.1" pixels).  A diffusion kernel, \BA\label{kernel} K
=
\left[\matrix{.025&.05&.025\cr.05&.70&.05\cr.025&.05&.025}\right]\EA
\noindent is applied to each resulting image.  The image on each
original pixel is shrunk to half its size in each dimension, and
then ``drizzled" to the final Hubble deep field grid (0.04"
pixels) according to the intersection of the diminished original
pixel area with the pixels in the final grid. We do not attempt to
reproduce the actual offsets of the dithers of the original
camera. The actual process has 9 dithers whose offsets vary across
the field. We are content to capture the main features of the
dither, diffusion, and drizzle process.  Ultimately one would
prefer to avoid the drizzle operation and fit the original
dithered images directly. The model method is well suited for
that.

An additional complication with the model method comes from the
Jacobian condition given the fact that $a$ and $b$ for the fit to
the source are typically larger by a factor of about 2 as compared
to the radial-fit method. In the case where we are fitting only
the $a$ and $b$ map parameters, the Jacobian is given by $1-\vert
a + 2 b \bar{w_T}\vert^2.$ For fixed radius $r$ the condition on
positivity for all angles $\theta$ becomes $\vert a \vert + 2
\vert b \vert r  \leq 1$. We must introduce an additional cut, of
the form $\vert a \vert + 2 \vert b \vert r_{Max} \leq 1$.


\section{Galaxy properties}\label{galaxyProps}

\begin{figure}[h!]
\centering {\leavevmode
\includegraphics[width=8cm,height=5cm]{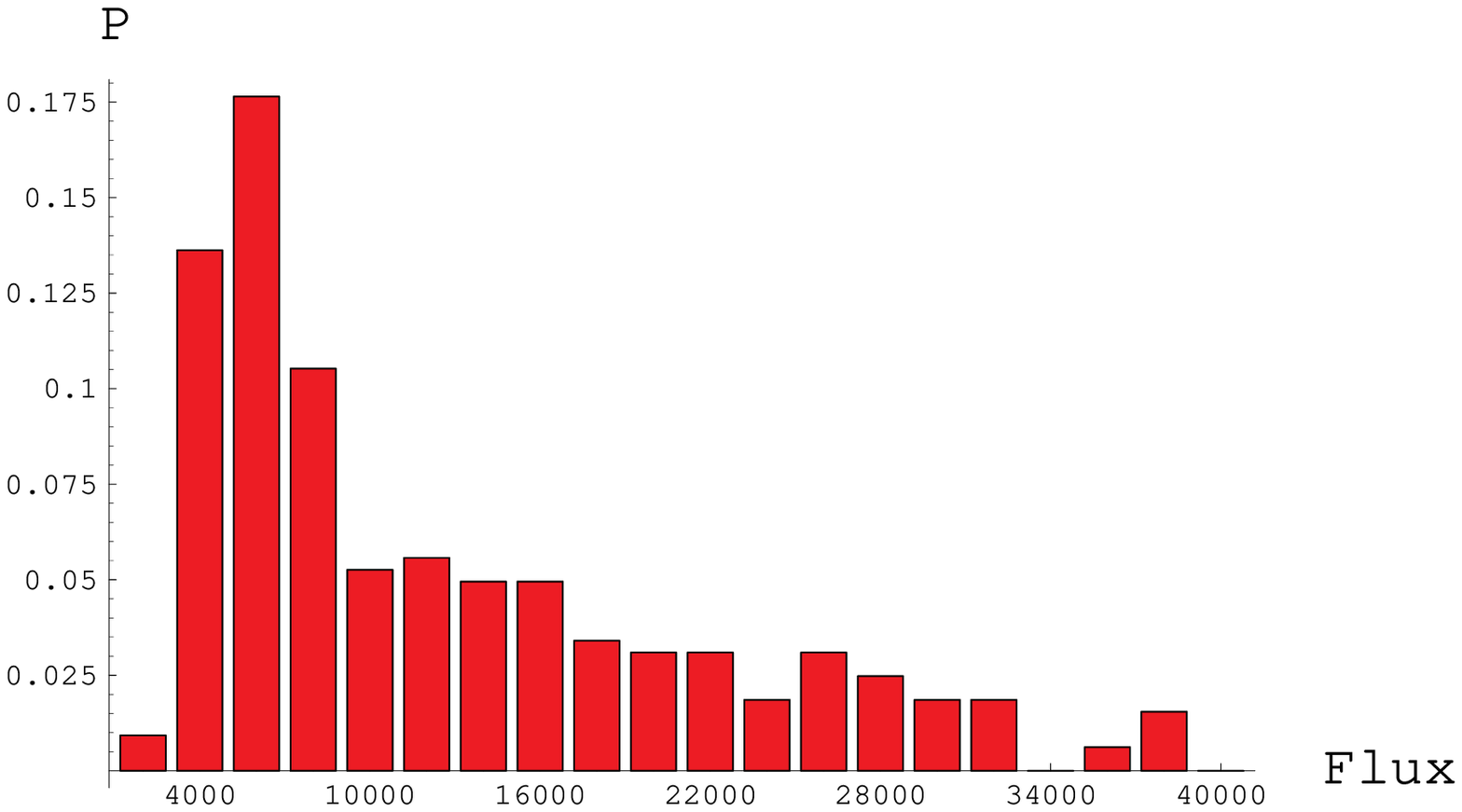}\,\,
\includegraphics[width=8cm,height=5cm]{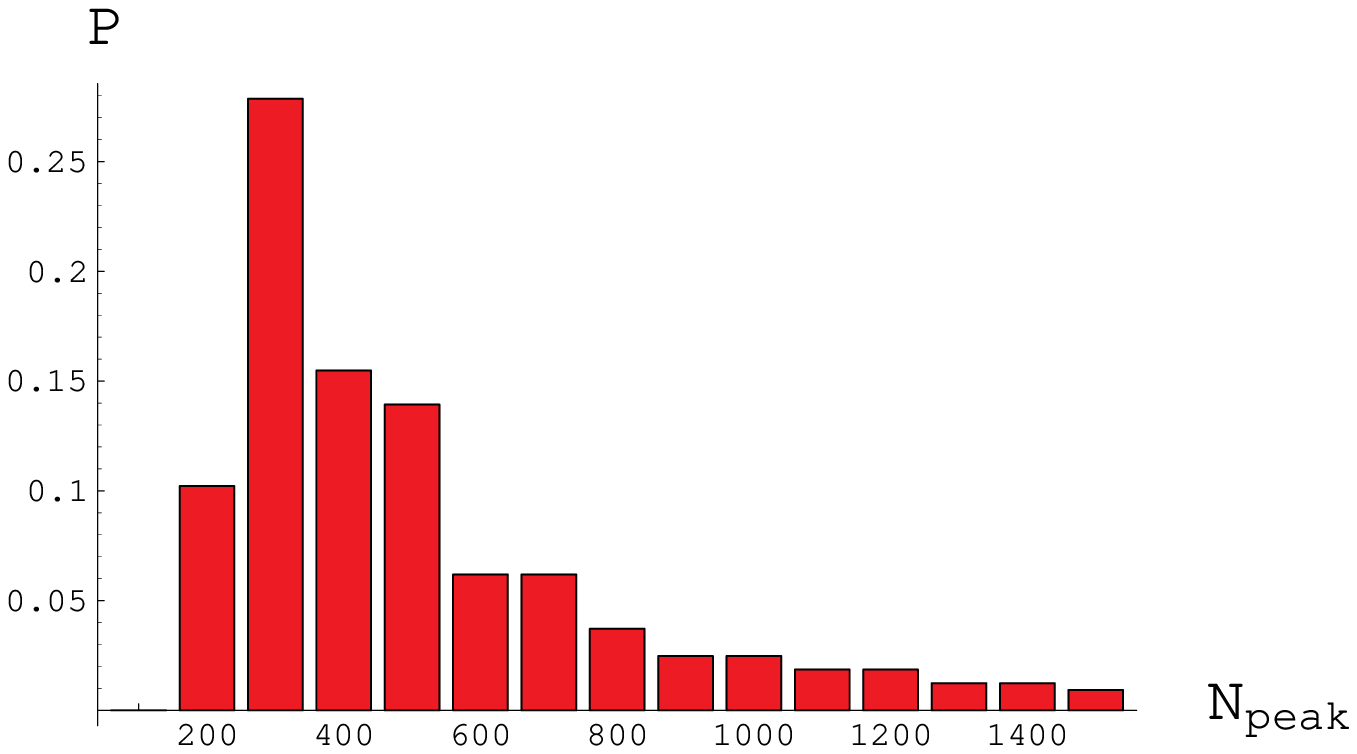}}
\caption[totalFluxpeakHeight] {\centering The distribution of
total photon counts for all galaxies in the north HDF with $z \geq
0.8 $, identified by SExtractor with intensity exceeding $6
\sigma_{NF}$, and having a single major maximum (left plot).  The
photon count of the peak pixel for the same galaxies (right
plot).} \label{totalFluxpeakHeight}
\end{figure}

\begin{figure}[h!]
\centering {\leavevmode
\includegraphics[width=8cm,height=5cm]{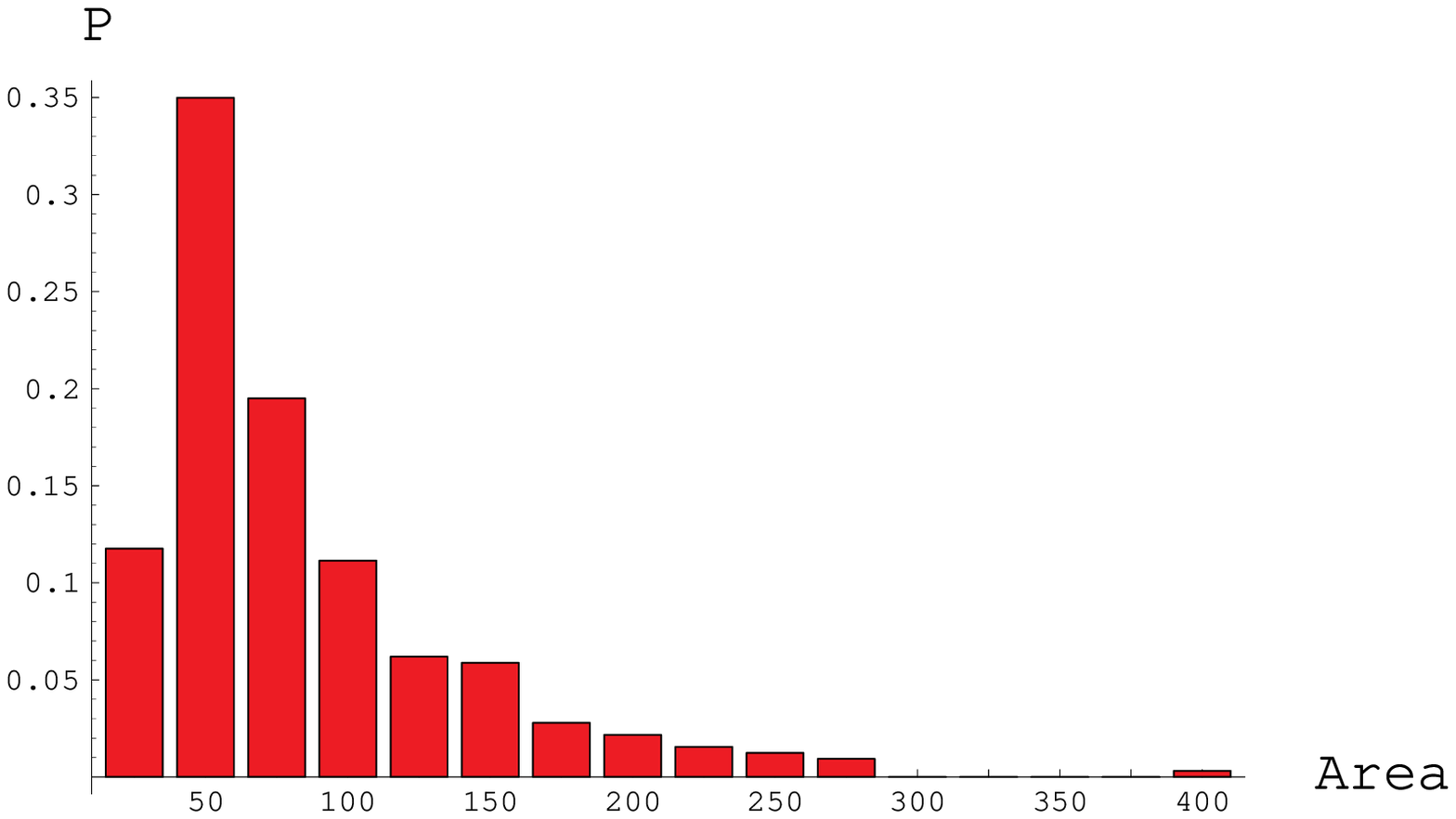}\,\,
\includegraphics[width=8cm,height=5cm]{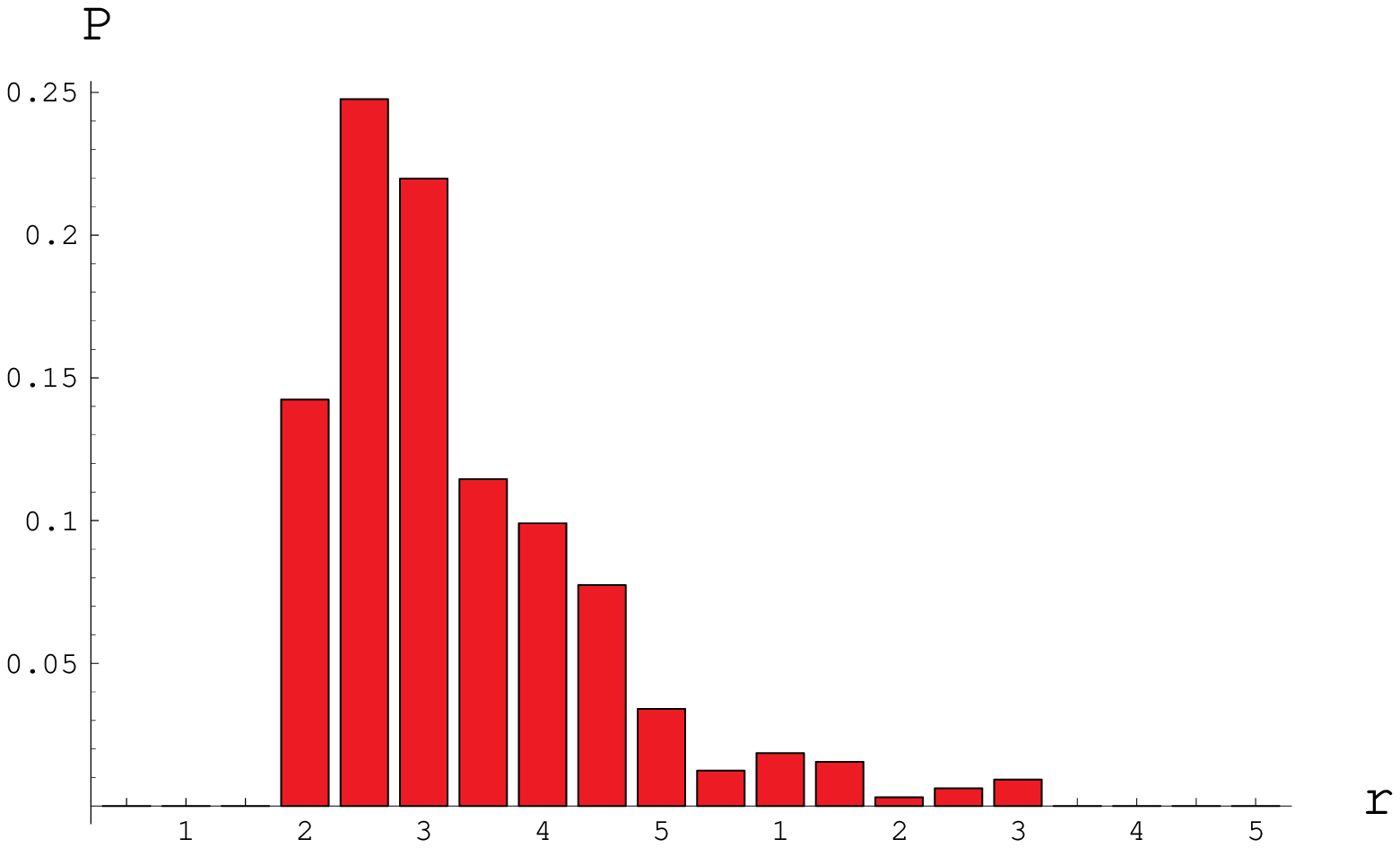}}
\caption[totalArearmsRadius] {\centering Distribution of galaxy
areas (left plot) and distribution of galaxy rms radii (right
plot) for the galaxy set described in fig.
\ref{totalFluxpeakHeight} above.}\label{totalArearmsRadius }
\end{figure}

The software SExtractor \citep{sextractor} was used to select
galaxies from the Hubble deep field and to specify which pixels to
include in the image.  Galaxies were selected which appeared for
both a $4 \sigma_{NF}$ and a $6 \sigma_{NF}$ threshold option  (4
or 6 times the rms noise floor) with the convolution option taken
to be the identity. Only galaxies that had been assigned a
$z$-value with $z>0.8$ were kept.\footnote{We used z-catalogs from
www.ess.sunysb.edu/astro/hdf.html and
bat.phys.unsw.edu.au/~fsoto/hdfcat.html.} There were about 569
galaxies so identified in the north field.  The images used in our
analysis were cut at $6 \sigma_{NF}$ and defined to be the
dominant simply-connected region.

Galaxy images were transferred to the \textit{Mathematica}
programming environment for inspection where galaxies with more
than one maxima were removed. Of the 569 identified galaxies with
$z>0.8$, 427 survived this single-max cut. The total photon count
and peak height of the surviving galaxies are shown in fig.
\ref{totalFluxpeakHeight}, and the total area and rms radius are
shown in fig. \ref{totalArearmsRadius }.

\section{Quadrupole coefficient measurements}\label{quadCoeff}

\subsection{Quadrupole coefficients from moment and radial-fit methods}\label{quadCoeffMFSub}

\begin{figure}[h!]
\centering {\leavevmode
\includegraphics[width=8cm,height=5cm]{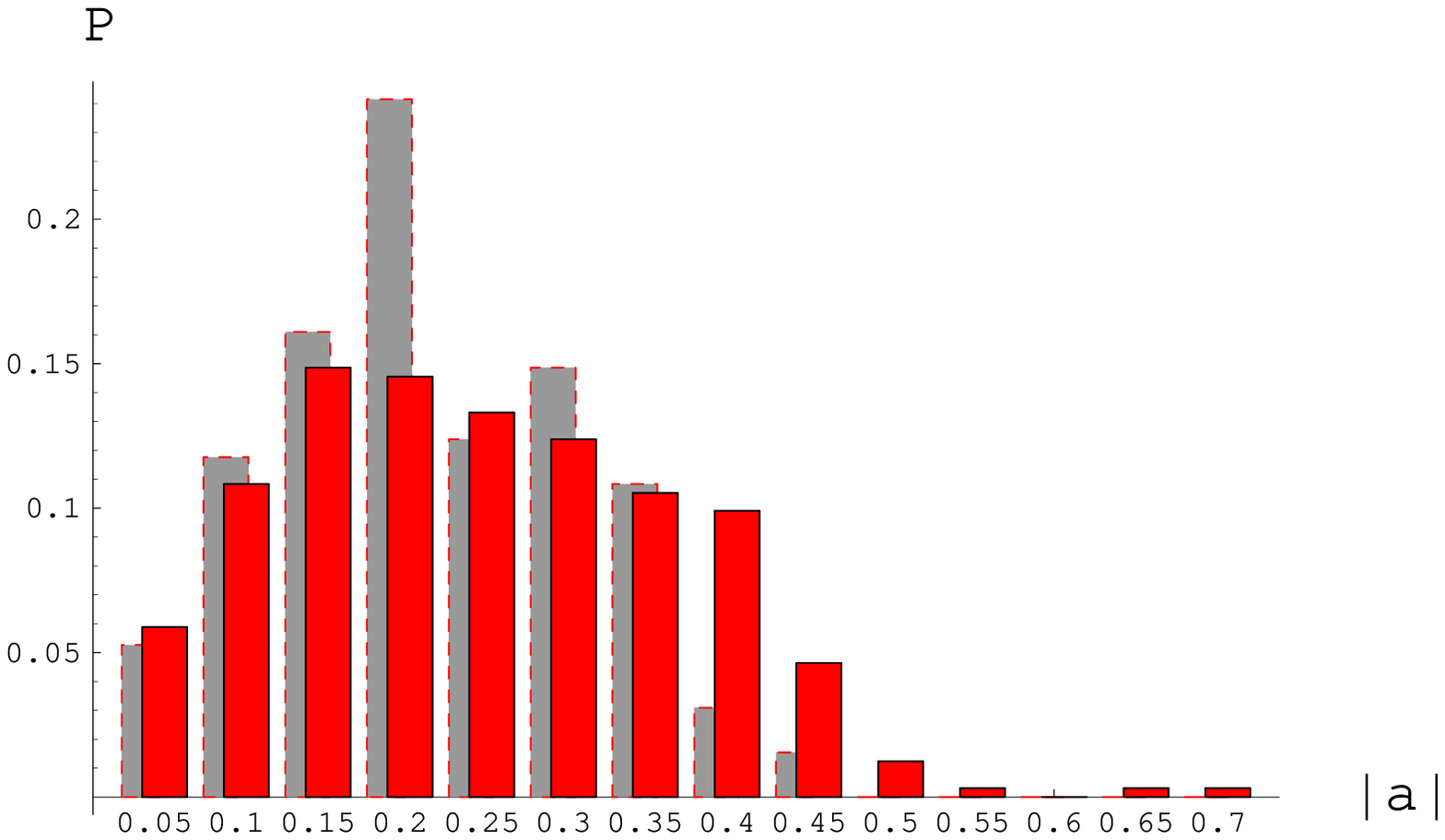}\,\,
\includegraphics[width=8cm,height=5cm]{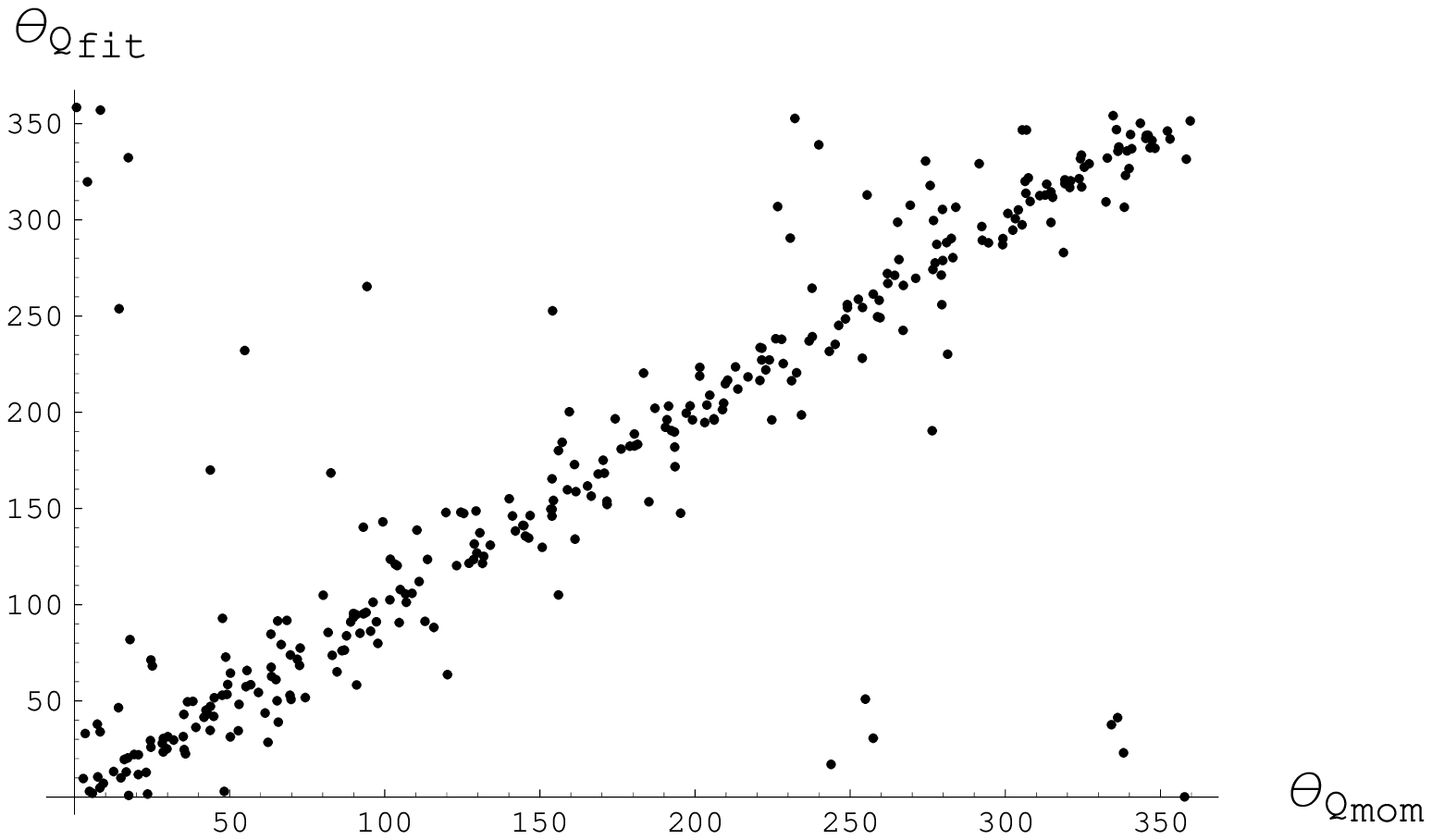}}
\caption[quadMagAngleMF] {\centering Distribution of the
quadrupole map coefficient for all galaxies in the set described
in fig. \ref{totalFluxpeakHeight}, for both the moment method
(background) and the radial fit method (foreground)(left plot) and
comparison of the angular orientation of the quadrupole map
coefficient for the moment method  and the radial fit method
(right plot).}\label{quadMagAngleMF}
\end{figure}

In fig. \ref{quadMagAngleMF} (left) we compare the distributions
of the magnitudes of the quadrupole coefficients using the moment
method and the radial-fit method. The range of magnitudes is
surprisingly similar. This can only arise if the fit to the radial
profile results in an $M_{20}^T$ moment that is significantly less
than the integrated moment.  For if $M_{20}^T$ were arising
dominantly from the mapping of an originally radial galaxy
profile, one can derive a simple analytic expression for the
magnitude of this moment, even when truncated.  To see this,
consider the following expression for one component of $M_{20}^T$.

\BA\label{M20fromRadial} \Re \left[M_{20}^T \right] &=& \int r^2
cos2\theta \, i_S(r_S^2) \,\vert \frac{\partial w_S}{\partial w_T}
\vert \, dA \cr & \approx & \int r^2 cos2\theta \, \{i_S(r^2) + 2
r^2( a_x cos2\theta + a_y sin2\theta) \, \frac {di_S}{dr^2} + ...
\} \, r dr d\theta \cr & \approx & 2\pi a_x \int r^4 \frac
{di_S}{dr^2} dr^2 \approx - 2 a_x \int [i_S(r^2)-i_S(r_T^2)] dA =
- 2 a_x \underline M_{11}^T \EA \noindent The integral here is
only over the footprint of the truncated image, which for the
purposes of this estimate, is assumed to be a disk of radius
$r_T$.  The underline of $M_{11}^T$ indicates that the integrand
is to be defined as the photon-count-per-pixel in excess of the
truncation floor. We will call this an ``above-floor" moment.  The
equation can now be solved for $a_x$ with a result which is
satisfyingly simple and directly comparable with the result found
using the moment method as in eq.\ref{quadMomentEq}: the
denominator $M_{11}^T$ is simply replaced by $\underline
M_{11}^T$. Similar results can be found for the other moments,
which we summarize here for future reference. \BA
\label{allMfromRadial} a & \approx & - \frac{M_{20}^T}{2\underline
M_{11}^T}, \quad \quad \quad \quad b \approx - \frac{M_{30}^T}{3
\underline M_{22}^T}, \cr c & \approx & - \frac{M_{40}^T}{4
\underline M_{33}^T}, \quad \mbox{and} \quad d \approx -
\frac{M_{21}^T}{9 \underline M_{22}^T- 4 M_{22}^T }. \EA
 The irregularity in the expression for $d$ arises from the linear
dependence on $d$ in the Jacobian.

Since the ``above-floor" expressions for the coefficients may have
a significantly smaller denominator than the expressions used in
the straight moment method, these ``above-floor" estimates for map
coefficients could be significantly larger. Based on our results
for the radial-fit method, which does insist on such a
relationship, one can only draw the conclusion that a part of the
moment $M_{20}^T$ is not coming from the distortion of the radial
profile.  This is an interesting result, and indicates that the
radial-fit method is indeed an important tool in making such a
distinction. We would assert that a substantial amount of noise is
projected out with the radial method.  This could be important for
high-resolution weak-lensing surveys such as SNAP.

The orientation of moments is more important to us than their
magnitude. Figure \ref{quadMagAngleMF} (right) compares the map
coefficient orientation of the radial-fit method with the moment
method. We would like to establish the orientation of the
resultant elliptical shape within $10^\circ$. Since the quadrupole
map coefficient advances by $2\phi$ as the ellipse rotates by
$\phi$, the relevant limit for the determination of $\phi$ is
$20^\circ$. One can see in fig. \ref{quadMagAngleMF} that the
angle differences between these two methods for the quadrupole map
coefficient are often larger than $20^\circ$.

\subsection{Quadrupole noise estimates}\label{quadNoise}

The first formula of eq. \ref{allMfromRadial} can be used to
derive a Poisson-noise estimate for the quadrupole coefficient,
$a_{fit}$, derived from the fit method.  An estimate for the
contribution of Poisson noise to $M_{20}$ is required.  For each
component, this can be obtained from the integral
\BA\label{M20Noise} \Re \left[M_{20}^T \right] &=& \frac{\int r^2
cos2\theta \, \zeta \sqrt{n(r,\theta)} dA}{ \int n(r,\theta) dA}
\approx \frac {1}{N} \sqrt { \int r^4 cos^22\theta \, n(r,\theta)
dA }\approx \sqrt { \frac {M_{22}^T}{2 N}} \EA \noindent where
$\zeta$ is a stochastic variable, $n(r,\theta)$ is the number of
counts for each pixel, and $N$ is the total number of counts for
the image.  Similar formulae hold for the other map coefficients.
We summarize the results: \BA\label{allCoeffNoise} \vert a_{x\,N}
\vert &\approx& \sqrt { \frac {M_{22}^T}{2 N}} \frac{1}{2
\underline{M}_{11}^T}, \quad \quad \quad \quad \quad \vert
b_{x\,N}\vert \approx \sqrt { \frac {M_{33}^T}{2 N}} \frac{1}{3
\underline{M}_{22}^T}, \cr \vert c_{x\,N}\vert &\approx& \sqrt {
\frac {M_{44}^T}{2 N}} \frac{1}{4 \underline{M}_{33}^T}, \quad
\mbox{and} \quad \vert d_{x\,N}\vert \approx \sqrt { \frac
{M_{33}^T}{2 N}} \frac{1}{9 \underline{M}_{22}^T - 4 M_{22}^T}.
\EA
\noindent The subscript $x$ indicates that this is an estimate
for one component, which could be in any direction of course. It
is a minimum estimate, in the sense that only the Poisson counting
noise is being included.  For example, contributions to the noise
from edge effects are not included.

\begin{figure}[h!]
\centering \leavevmode\epsfysize= 7 cm \epsfbox{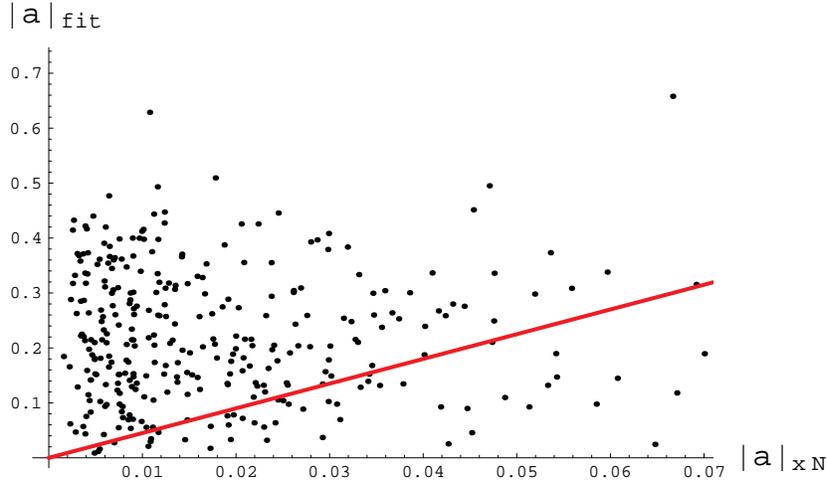}
\caption[quadCut] {\centering  A plot of the quadrupole Poisson
noise estimate (horizontal axis) versus the magnitude of the
quadrupole coefficient.  The galaxies falling below the straight
line with slope 4.5 will be cut as having signal-to-noise ratios
too small.}\label{quadCut}
\end{figure}

The actual $\vert a \vert$ should be large enough so that if the
perpendicular component to $a$ was changed by an amount $\vert
a_{x\,N}\vert$ the angle would change by less than the required
resolution on angles.  The magnitude of the relative orientation
of the quadrupole and sextupole shapes ($\delta$ in fig.
\ref{deltaDef}), runs from $0^\circ$ to $30^\circ$ . Since we
choose to designate ``curved" galaxies as those for which $\vert
\delta \vert < 10^\circ$, ``aligned" galaxies as those for which
$\vert \delta \vert > 20^\circ$, and ``mid-range" galaxies as the
remaining ones, and considering positive as well as negative signs
for $\delta$, the regions describing curved and aligned are
actually $20^\circ$ wide.  Hence we may take $10^\circ$ (=0.17
rad) as an estimate for the required resolution on $\delta$. Since
$\delta$ is the difference between the orientation of the
quadrupole and sextupole shapes, the resolution on $\delta$ could
be achieved by requiring the sum of the resolution squared of each
shape individually to be less than $(0.17)^2$ .  As the shapes
rotates by $\phi$ the quadrupole map coefficient angle changes by
$2\phi$ and the sextupole map coefficient angle changes by
$3\phi$.  Thus this resolution condition on $a_{fit}\mbox{ and }
b_{fit}$ could be written \BA\label{cutCondition}
\left[\frac{1}{2} \tan^{-1} \vert \frac{a_{x\,N}}{a_{fit}}\vert
\right]^2 + \left[\frac{1}{3}
\tan^{-1}\big\vert\frac{b_{x\,N}}{b_{fit}}\big\vert\right]^2 <
(0.17)^2 \EA
We could choose to satisfy this condition by
requiring that the quadrupole contribution be less than
$6.3^\circ$ and the sextupole contribution be less than
$7.8^\circ$. The quadrupole cut would then be equivalent to
requiring that $\vert a \vert
> 4.5 \vert a_{x \, N}\vert$ and the sextupole cut would be equivalvent
to $\vert b \vert > 2.3 \vert b_{x \, N} \vert$. The results of
the quadrupole cut are shown in fig. \ref{quadCut}, where the
vertical axis is the quadrupole strength and the horizontal axis
is the estimated Poisson noise from eq. \ref{allCoeffNoise} for
that galaxy. The straight line is the cut condition and galaxies
below this line would be rejected as having a signal-to-noise
ratio which is too small.

\begin{figure}[h!]
\centering {\leavevmode 
\includegraphics[width=8cm,height=5cm]{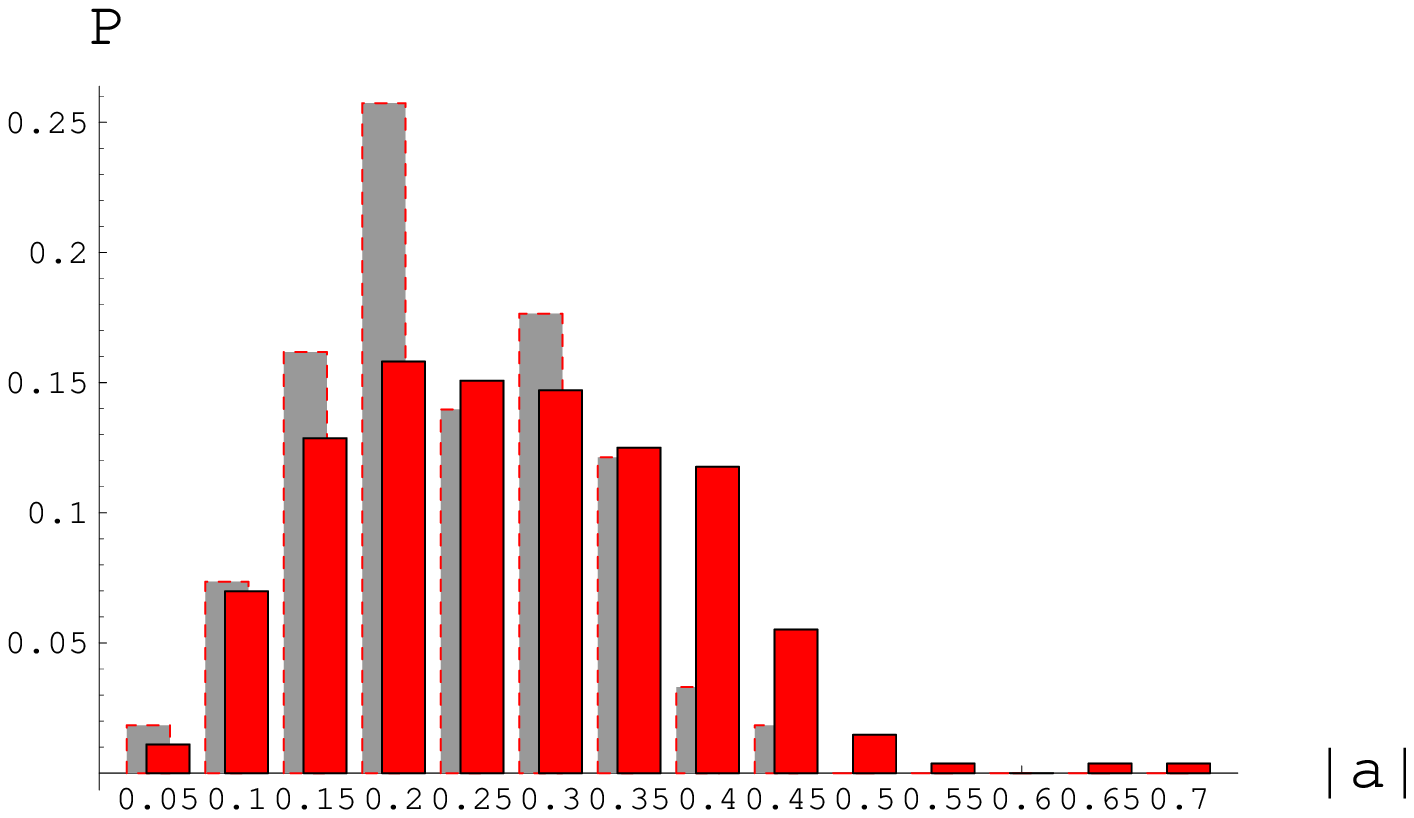}\,\,\,\,
\includegraphics[width=8cm,height=5cm]{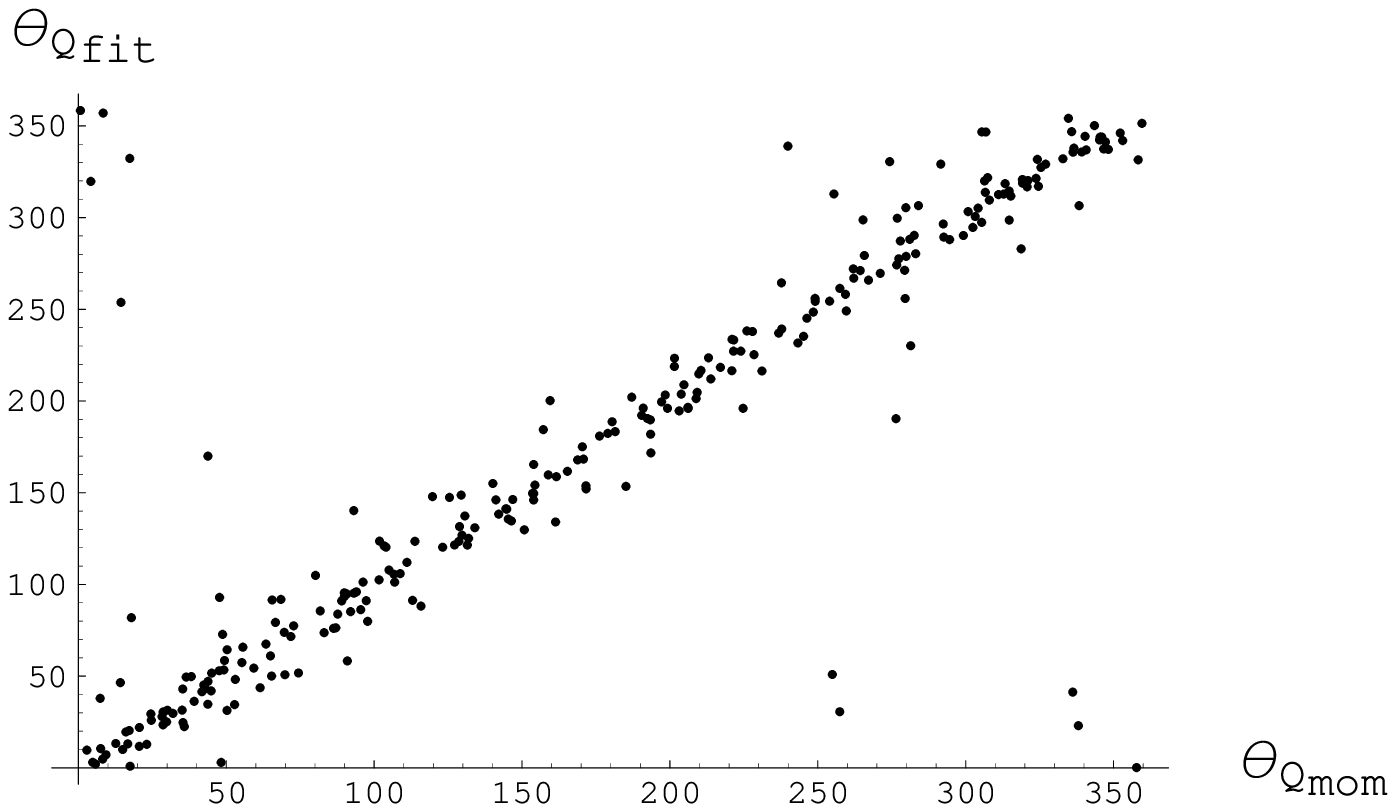}}
\caption[quadMagAngleMFAfterCut] {\centering Comparison, after
signal-to-noise cut, of the magnitude of the quadrupole map
coefficient for both the moment method (background) and the radial
fit method (foreground) (left plot) and of the angular orientation
of the quadrupole map coefficient for the moment and the radial
fit methods (right plot).  Compare with fig.
\ref{quadMagAngleMF}}\label{quadMagAngleMFAfterCut}
\end{figure}

In fig. \ref{quadMagAngleMFAfterCut} we show the comparison
between the moment and radial-fit methods after this
signal-to-noise cut for magnitude distributions and angle
distributions.  The orientations using the two methods are now in
better agreement.

\subsection{Quadrupole coefficients from the model method}\label{modelMethod}

\begin{figure}[h!]
\centering {\leavevmode 
\includegraphics[width=8cm,height=5cm]{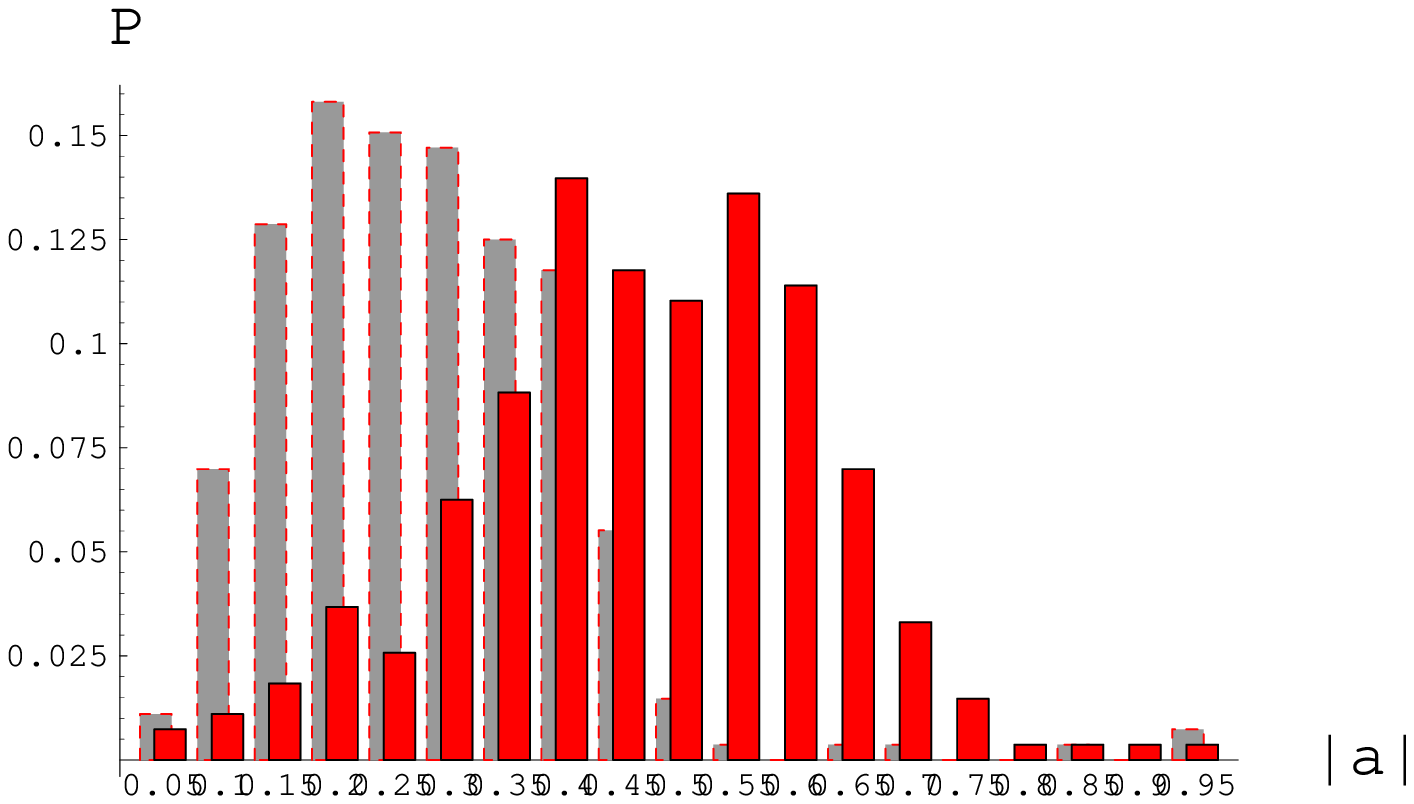}\,\,\,\,
\includegraphics[width=8cm,height=5cm]{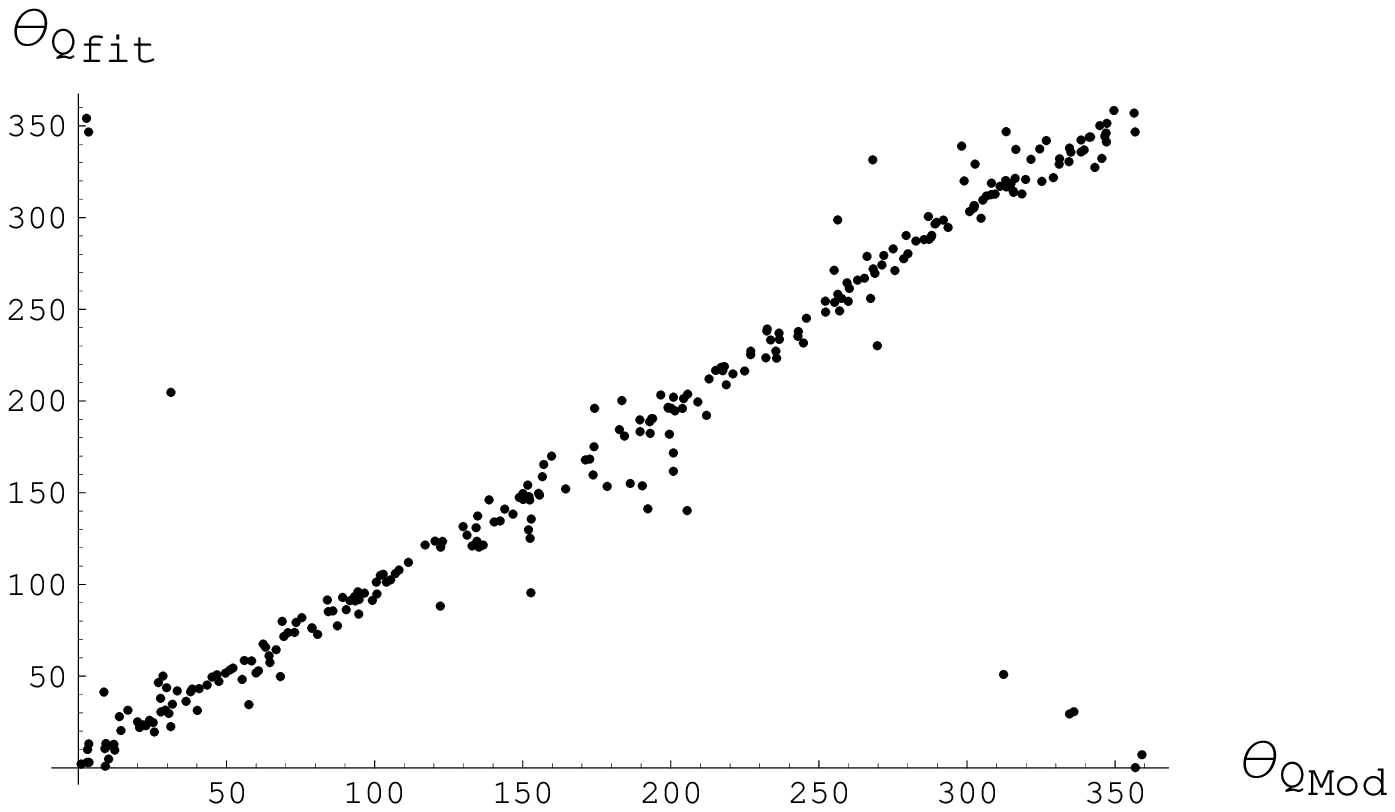}}
\caption[quadMagAngleFitModel] {\centering Distribution  of the
magnitude of the quadrupole map coefficients using the model
method (foreground) compared with the radial-fit method
(background) (left plot) and of the angular orientation of the
quadrupole map coefficient for the model and the radial fit
methods (right plot).}\label{quadMagAngleFitModel}
\end{figure}

In fig. \ref{quadMagAngleFitModel} we show the magnitude of the
quadrupole coefficient (left panel) using the model method. Of the
427 selected galaxies having one apparent maximum, 47 resulted in
an L2 norm squared ($\Vert i_F-i_T \Vert^2$) greater than $0.05$.
These galaxies were cut from the sample, for the reason that their
shape does not sufficiently well conform to the model we are using
to estimate map-coefficient strengths.
The magnitudes of the quadrupole coefficients are clearly larger
in the model case than the radial-fit case because the rms size of
the source image is smaller.  This increase is represented by the
factor $1/\lambda_T$ in eq. \ref{mapAddition}.

Figure \ref{quadMagAngleFitModel} (right plot) compares the
orientations of the quadrupole map coefficient of the radial-fit
method with the model method.  The differences are now much
smaller.  The diffusion, being symmetric, will not change the
orientation, but the PSF of the Hubble is known to have a
substantial quadrupolar component which varies dramatically across
the field, frustrating attempts to do weak lensing with the
Hubble.

\begin{figure}[h!]
\centering {\leavevmode
\includegraphics[width=6cm,height=5cm]{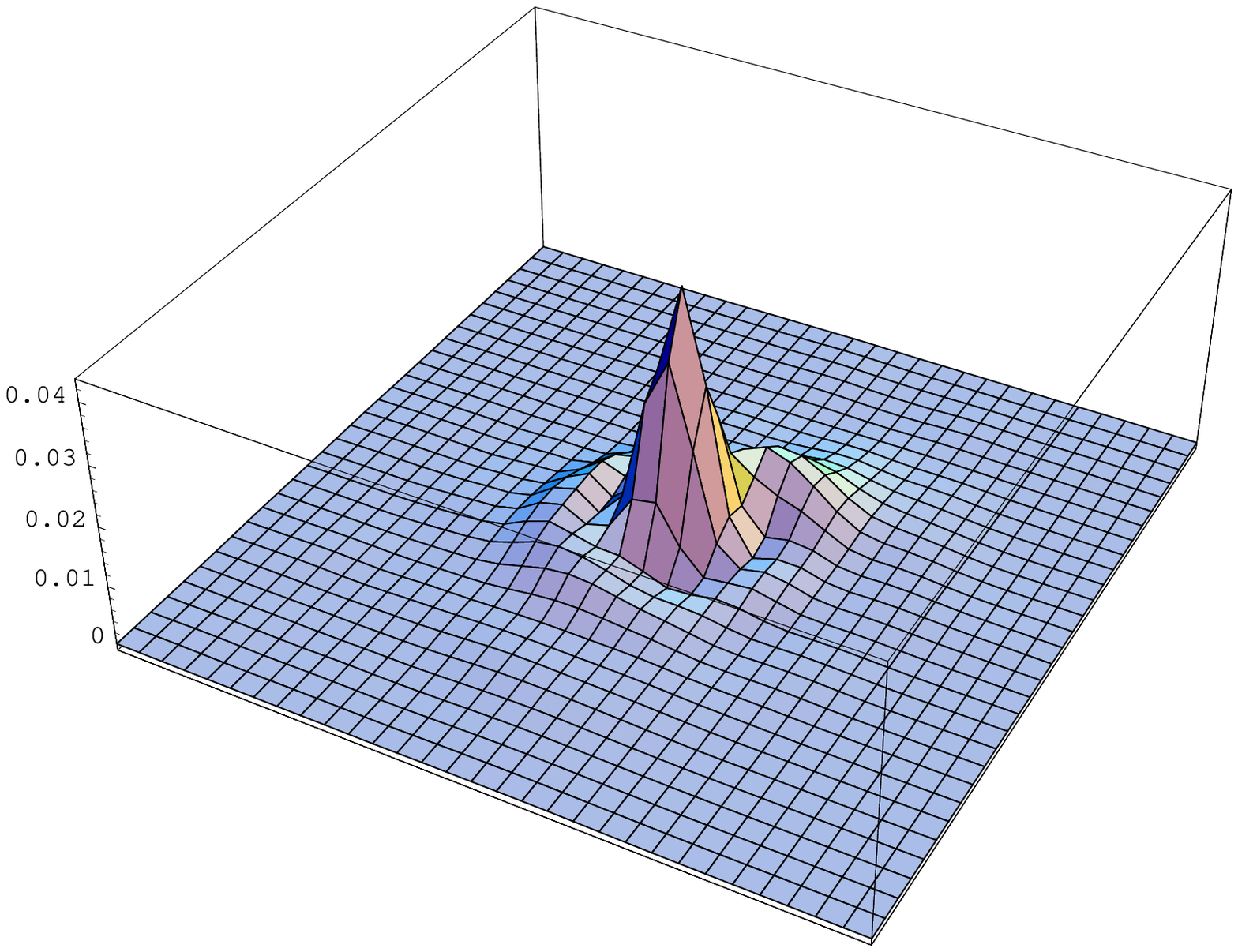}\,\,\,\,
\includegraphics[width=6cm,height=5cm]{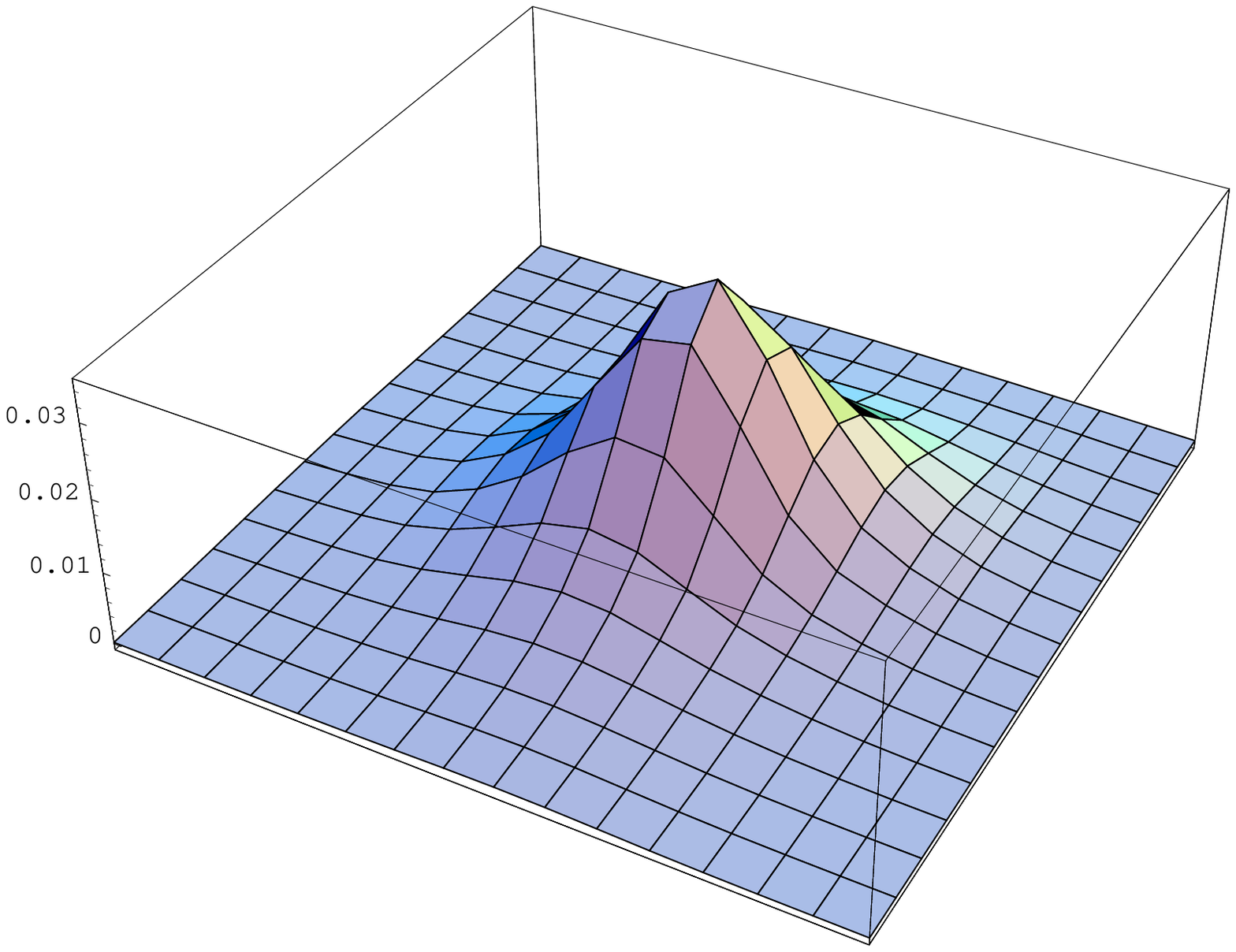}}
\caption[tinyTimPSF] {\centering A picture of a Hubble WPFC2 PSF
as produced by the Tiny Tim program on the left. The pixel size is
5x smaller than the Hubble camera pixels, in other words 2x
smaller than the drizzled HDF pixels (left plot).   On the right,
the same PSF diffused and drizzled .}\label{tinyTimPSF}
\end{figure}

An example of a 5x sub-sampled PSF, as obtained from Tiny Tim, is
shown in Fig. \ref{tinyTimPSF} (left plot).  We have specified the
the F606 filter at the location of each of the 427 galaxies in our
sample. There is typically a diffraction ring at a radius of 2 HDF
pixels which contains about 30\% of the counts. The radius of the
central peak is too small to give significant moments, so the
moments of concern arise from the ring.

\begin{figure}[h!]
\centering{\includegraphics[width=3in,height=3in]{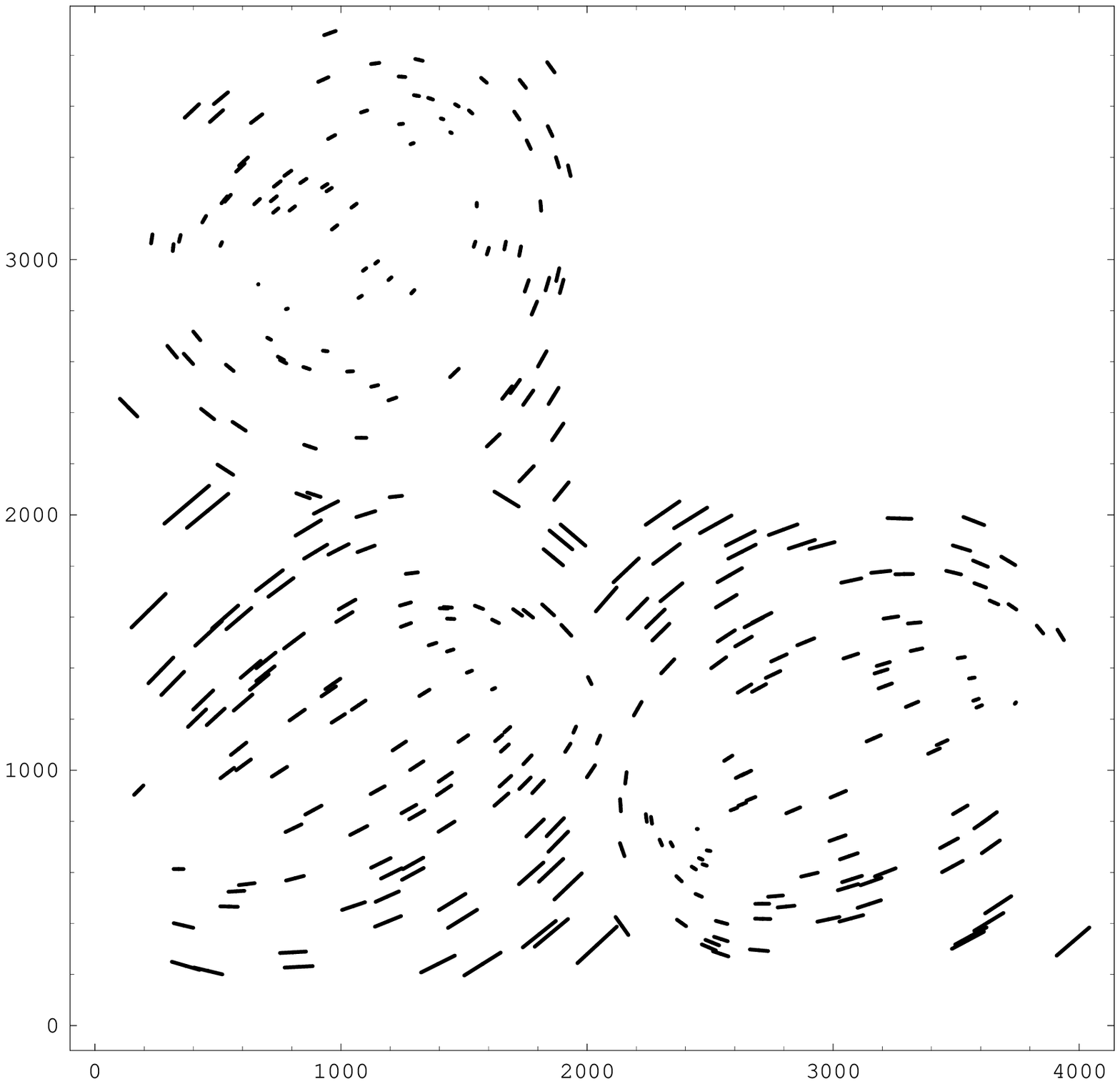}\,\,\,\,
\includegraphics[width=9cm,height=6cm]{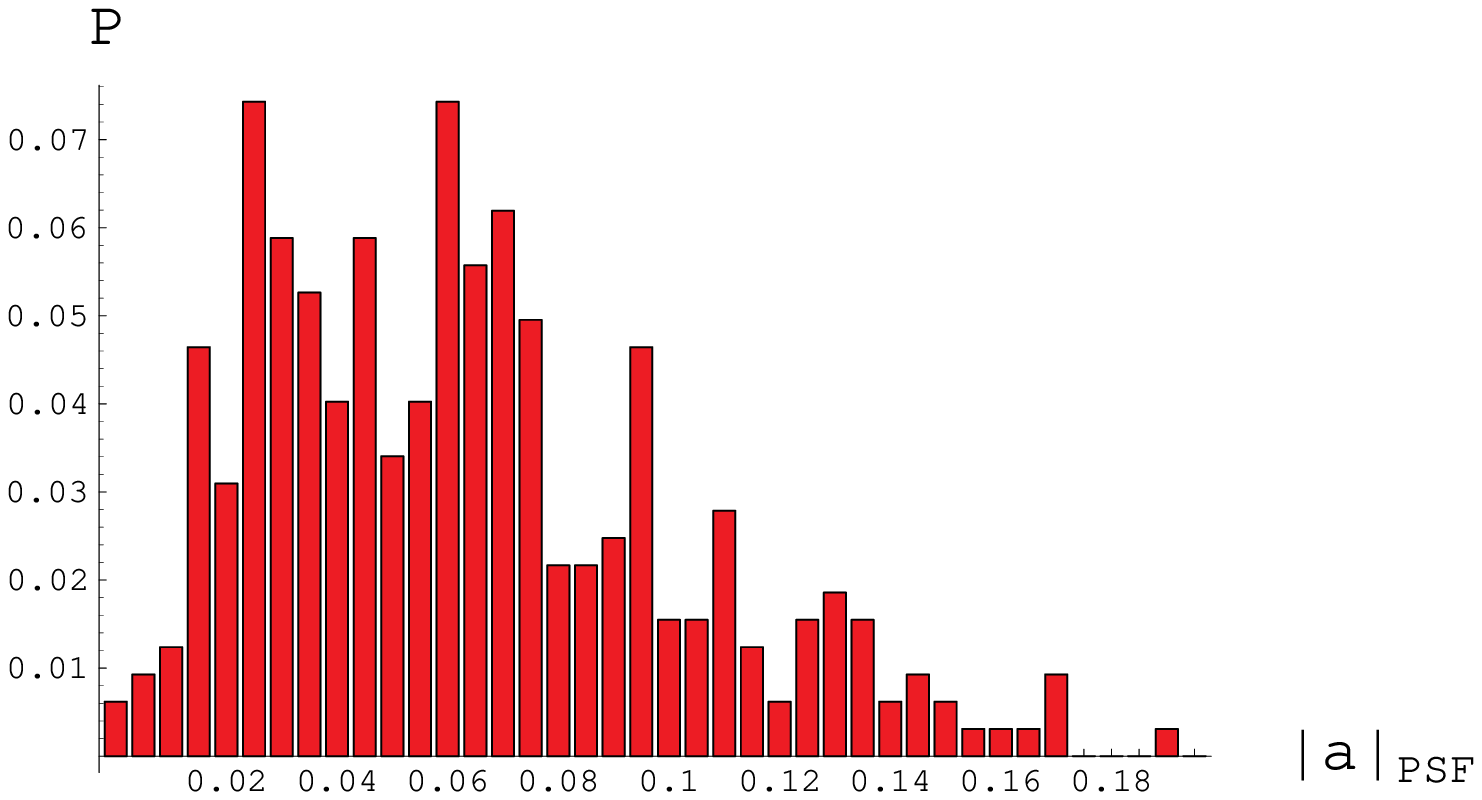}}
\caption[quadPSFcoeff] {\centering  Spatial ellipticity
distribution (left) and the quadrupole coefficient magnitude
distribution (right) for the 5x subsampled and drizzled Tiny Tim
PSFs for the WPFC2 camera on the Hubble. } \label{quadPSFcoeff}
\end{figure}

Fig. \ref{tinyTimPSF} (right plot) shows this PSF after it has
been dithered, diffused and dropped onto the final Hubble 0.04"
grid.  We have taken these PSFs which have been dithered, diffused
and drizzled, and fit them using the radial fit method.
Fig.\ref{quadPSFcoeff} shows the resulting distribution of the
magnitude of the quadrupole coefficients (right) as  well as their
spatial distribution and orientation (left).

\section{Sextupole coefficient measurement}\label{sextCoeff}

\subsection{Sextupole coefficients from moment and radial-fit methods}\label{sextMethods}

\begin{figure}[h!]
\centering {\leavevmode 
\includegraphics[width=8cm,height=5cm]{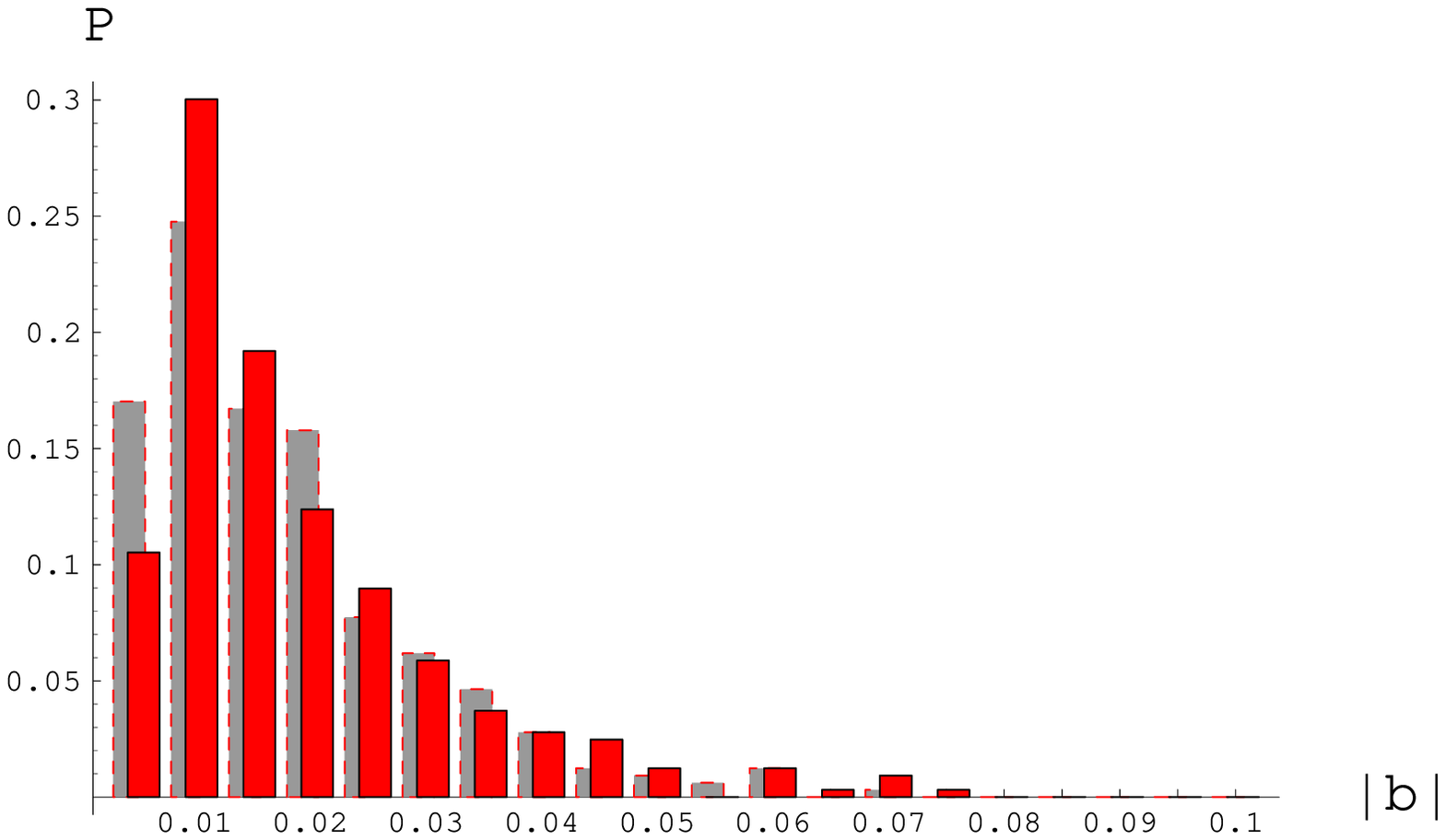}\,\,\,\,
\includegraphics[width=8cm,height=5cm]{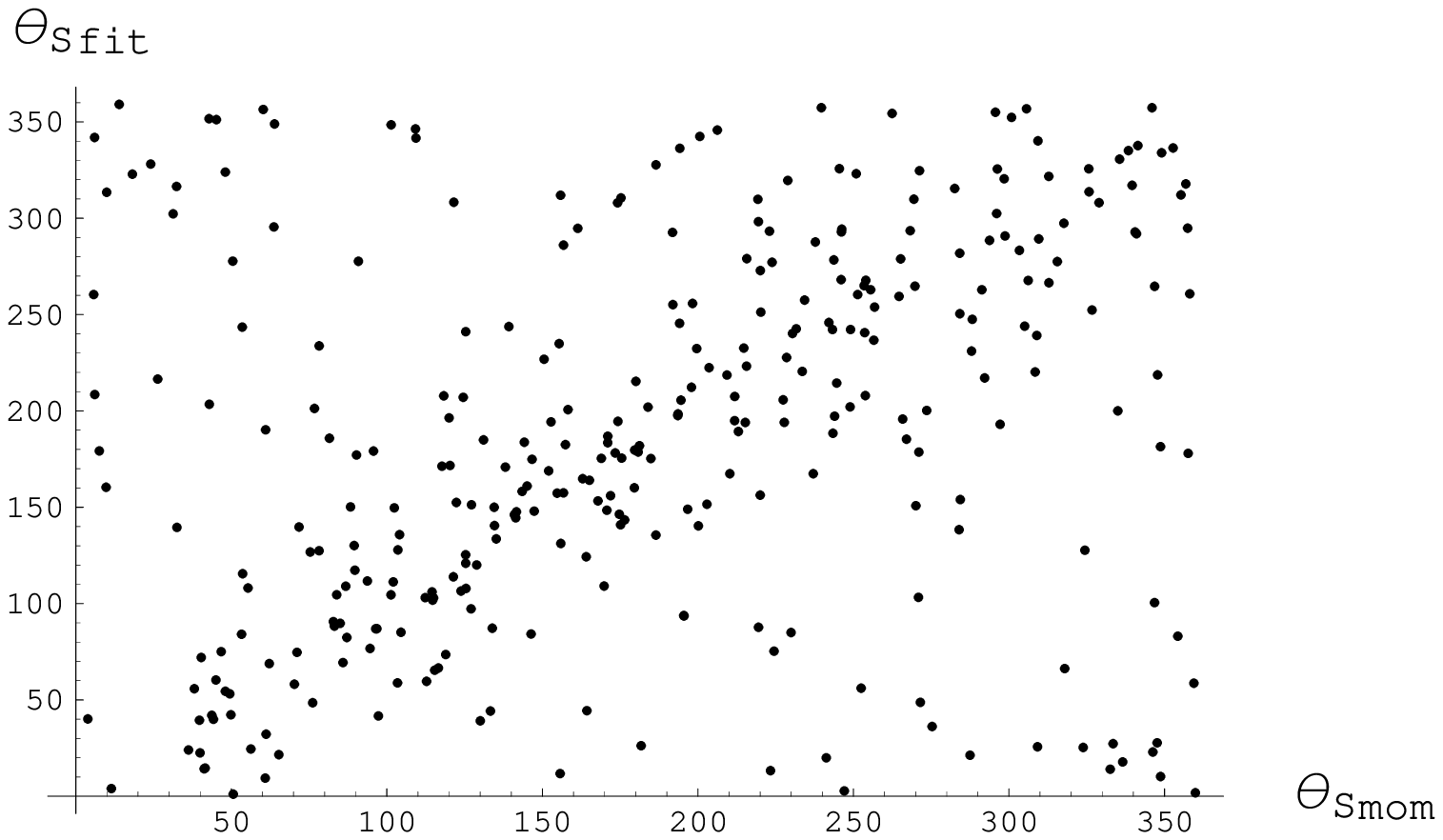}}
\caption[sextMagAngleMF] {\centering The left plot is a
distribution of the magnitude of the sextupole map coefficient,
for the radial fit method (foreground) and the moment method
(background).  Units are``per HDF pixel". The right plot provides
a comparison of the orientation of the sextupole map coefficients
for the radial-fit method (vertical axis) and the moment method
(horizontal axis).}\label{sextMagAngleMF}
\end{figure}

We now follow the sequence of the previous section for the
sextupole coefficient measurements.  Figure \ref{sextMagAngleMF}
(left) shows a comparison of the sextupole coefficient magnitudes
for the moment method and the radial-fit method.  Again the
radial-fit method would have been larger if the entire $M_{30}^T$
arose from a distortion of the radial profile, since the
$M_{22}^T$ will typically be much larger than $\underline
M_{22}^T$.  As before, we conclude that the radial-fit method is
projecting out a substantial portion of the background-galaxy
sextupole-moment noise.  Figure \ref{sextMagAngleMF} (right) shows
a comparison of the sextupole angle measurement for the moment and
the radial-fit method.  Since for this moment a rotation of the
image by $\phi$ results in a change of the coefficient angle by
$3\phi$, an angular difference in the coefficient of $30^{\circ}$
is a measure of significance.

\subsection{Sextupole noise estimates}\label{sextNoise}

\begin{figure}[h!]
\centering \leavevmode\epsfysize= 4.8 cm \epsfbox{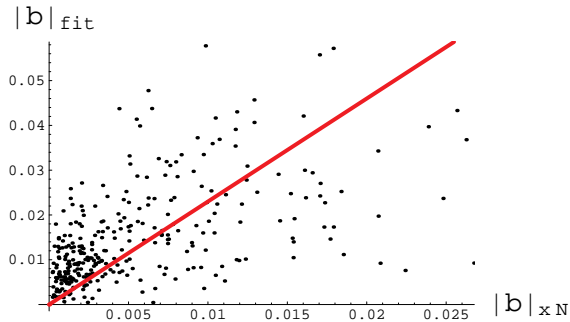}
\caption[sextCut] {\centering  A plot of the sextupole Poisson
noise estimate (horizontal axis) versus the magnitude of the
sextupole coefficient.  The galaxies falling below the straight
line with slope 2.3 will be cut as having signal-to-noise ratios
too small.} \label{sextCut}
\end{figure}

\begin{figure}[h!]
\centering {\leavevmode
\includegraphics[width=7cm,height=4.5cm]{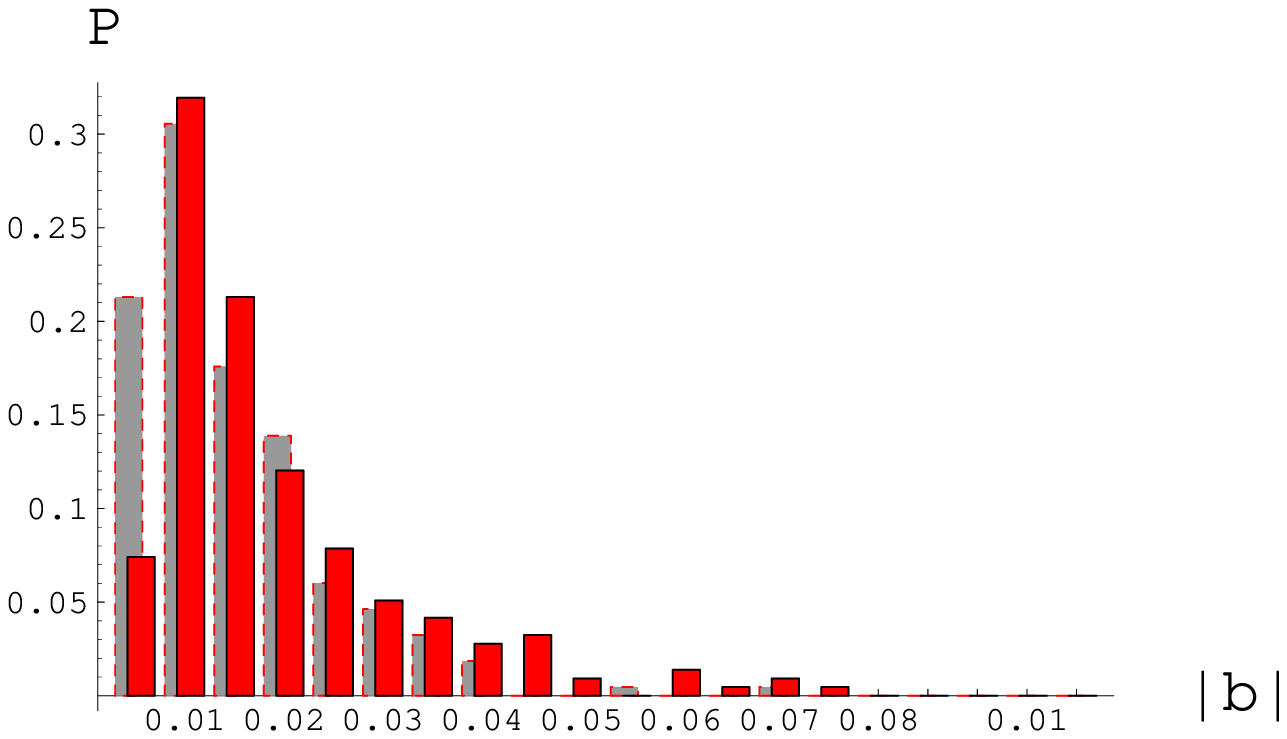}\,\,\,\,
\includegraphics[width=7cm,height=4.5cm]{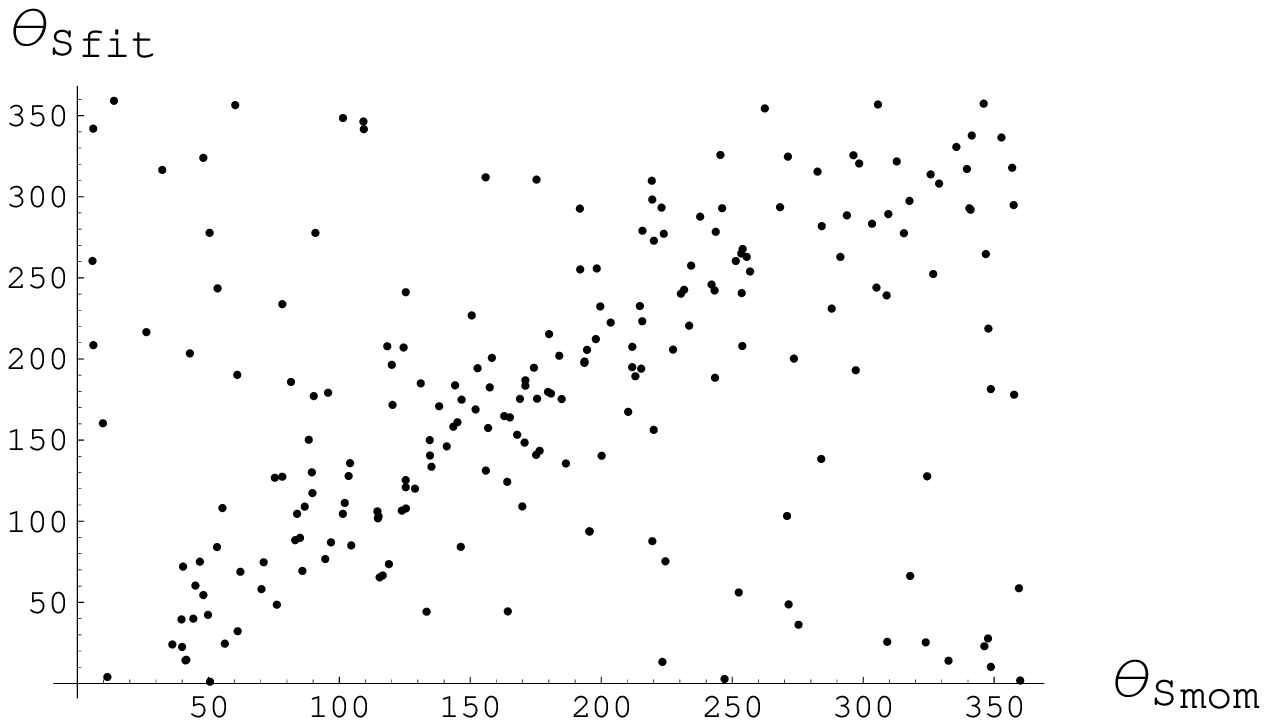}}
\caption[sextMagMFAfterCut] {\centering Comparison, after
signal-to-noise cut, of the magnitude of the sextupole map
coefficients for the moment method (background) and the radial fit
method (foreground) (left plot) and of their angular orientation
(right plot).}\label{sextMagMFAfterCut}
\end{figure}

Following the discussion of subsection \ref{quadNoise}, where we
chose to define the sextupole signal-to-noise by requiring the
angle to change by less than $3 \times 7.8^\circ = 23.4^\circ $
when the Poisson noise estimate was added perpendicular to the
vector representing the sextupole signal.  This implies that $
\vert b \vert > 2.3 \vert b_{x \, N} \vert$.  In fig.
\ref{sextCut} we plot the sextupole strength on the vertical axis
versus the sextupole Poisson-noise estimate on the horizontal
axis.  The straight line has the slope 2.3, showing the cut
criteria for sufficient sextupole signal-to-noise.  In fig.
\ref{sextMagMFAfterCut} we show the distribution comparison
between the moment and fit methods after this signal-to-noise cut
and  the angle comparison after this cut.  Clearly scatter is
reduced as compared to fig.\ref{sextMagAngleMF}, indicating much
of the scatter in the angular measurement could be attributed to
noise.

\begin{figure}[h!]
\centering \leavevmode\epsfysize= 6 cm \epsfbox{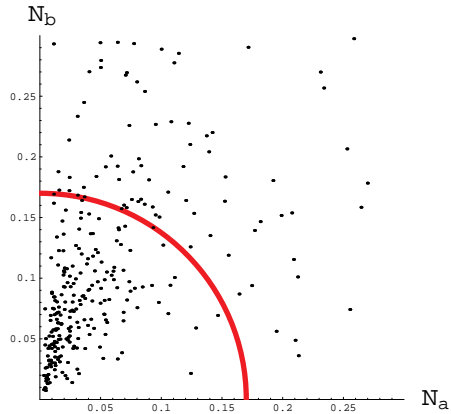}
\caption[combinedCut] {\centering  A plot of the ratio
$N_a=\frac{1}{2}
\left[\tan^{-1}\vert\frac{a_{x\,N}}{a_{fit}}\vert\right]$  versus
the ratio $N_b=\frac{1}{3}
\left[\tan^{-1}\big\vert\frac{b_{x\,N}}{b_{fit}}\big\vert\right]$.
The circle shows the ``cut" condition of eq. \ref{cutCondition}.
The galaxies falling outside the circle will be cut as having the
quadratically-combined signal-to-noise ratios too small.}
\label{combinedCut}
\end{figure}

Finally, fig. \ref{combinedCut} addresses the combination of
resolutions that can be expected for $a$ and $b$, showing the
condition formulated above in eq. \ref{cutCondition}.  It is this
cut that we will impose when we consider the relative orientation
of the quadrupole and sextupole map coefficients. 217 galaxies
survive this cut.

The following list summarizes the results on the several galaxy
cuts. If a line is indented with respect to a line above it, then
the cut is for the combination of these conditions.

\noindent 569 \quad \quad \quad \quad  in the z-catalog with
$z>0.8$ and found by SExtractor for thresholds 4$\sigma_{NF}$ and
6$\sigma_{NF}$ \hfil

\noindent \quad 427 \quad \quad \quad \quad having only one
prominent maximum \hfil

\noindent \quad \quad 370 \quad \quad \quad \quad larger than 5
pixels in both x and y \hfil

\noindent \quad \quad \quad 323 \quad \quad \quad \quad with
radial-fit having its L2 squared norm less than 0.05 \hfil

\noindent \quad \quad \quad \quad 276 \quad \quad \quad \quad with
quadrupole signal/noise greater than 4.5 \hfil

\noindent \quad \quad \quad \quad 216 \quad \quad \quad \quad with
sextupole signal/noise greater than 2.3 \hfil

\noindent \quad \quad \quad \quad 217 \quad \quad \quad \quad
satisfying the joint-variable signal-to-noise cut condition of eq.
\ref{cutCondition} \hfil

\noindent Only a very few galaxies had $\vert a \vert + 2 \vert b
\vert r_{Max} > 1$, so this cut was not enforced. Further
signal-to-noise cuts, arising for the $c$ and $d$ terms, must be
combined with the above cuts to establish their orientation with
respect to the imputed direction to the scattering center.  Since
these cause the surviving statistical sample to become
uncomfortably small, we will postpone a discussion of the
relevance of the $c$ and $d$ terms, if any, to a subsequent paper
with larger statistical samples.

\subsection{Sextupole coefficients from model method}\label{sextMethods}

\begin{figure}[h!]
\centering {\leavevmode
\includegraphics[width=7cm,height=4.5cm]{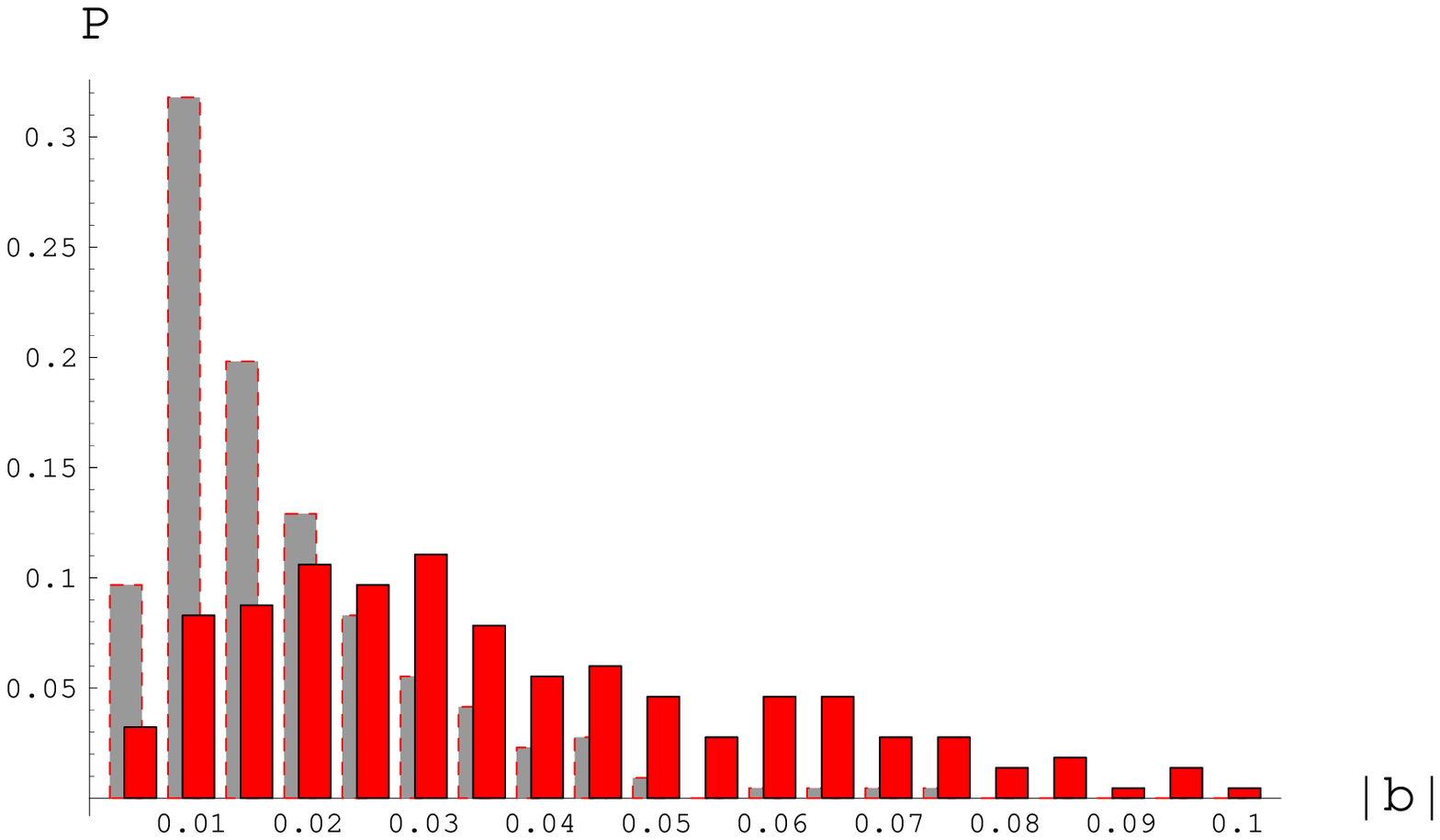}\,\,\,\,
\includegraphics[width=7cm,height=4.5cm]{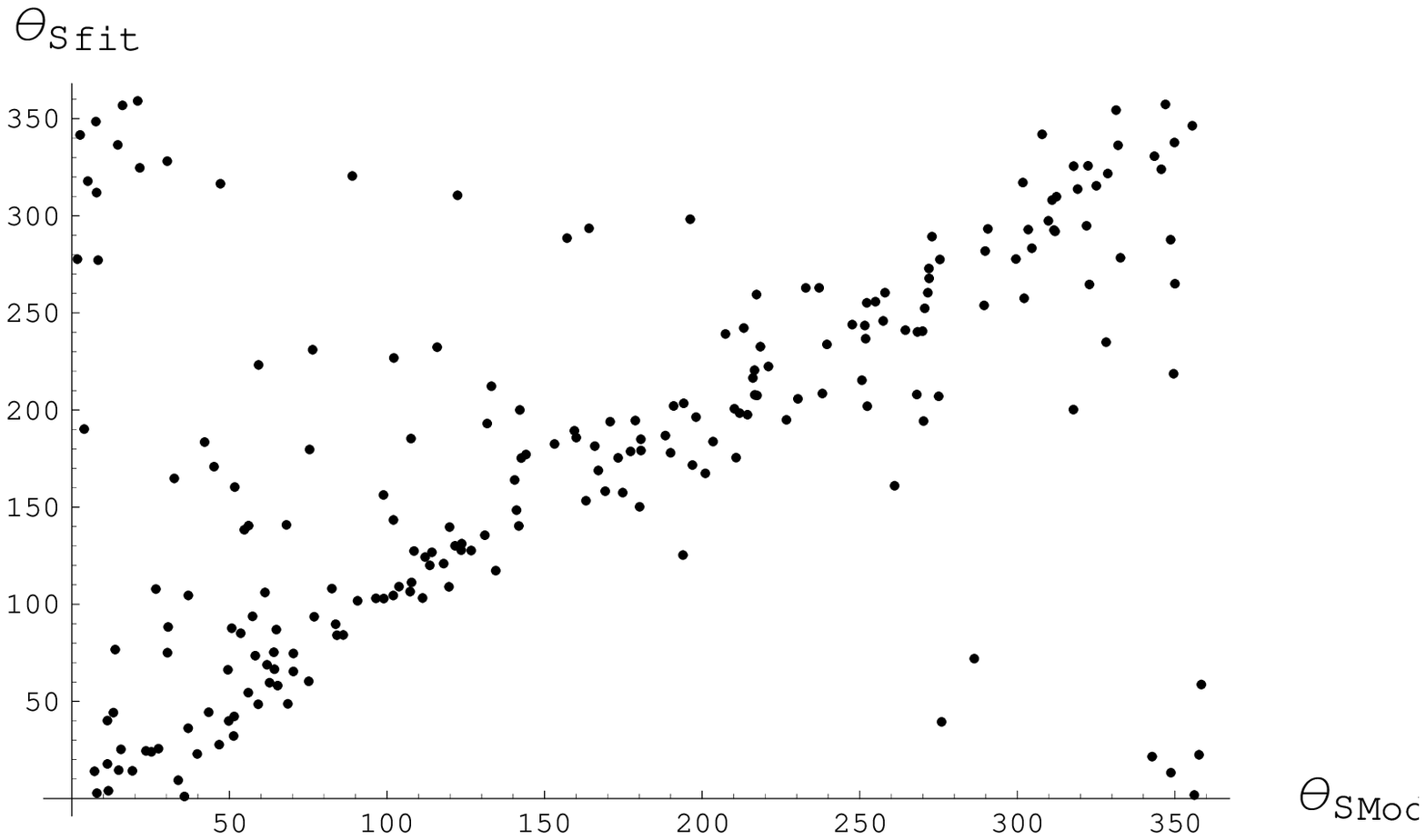}}
\caption[sextMagMFAfterCut] {\centering The left panel shows the
distribution of the magnitude of the sextupole map coefficient,
for the model method (foreground) as compared with the radial-fit
method (background). The units are "per HDF pixel". The right
panel compares the angular orientation of the sextupole map
coefficient for the model method and the radial fit method.}
\label{sextMagAngleModel}
\end{figure}

The distribution of the sextupole-coefficient magnitudes for the
model method is shown in fig. \ref{sextMagAngleModel} (left plot),
and the comparison of the orientation of the sextupole coefficient
orientation for the model and radial-fit method are shown in fig.
\ref{sextMagAngleModel} (right plot).  The relevant difference in
sextupole angle, as noted above, is $30^\circ$.  Many galaxies
have changed by that amount, indicating the importance of the PSF
correction.

\begin{figure}[h!]
\centering{\includegraphics[width=3in,height=3in]{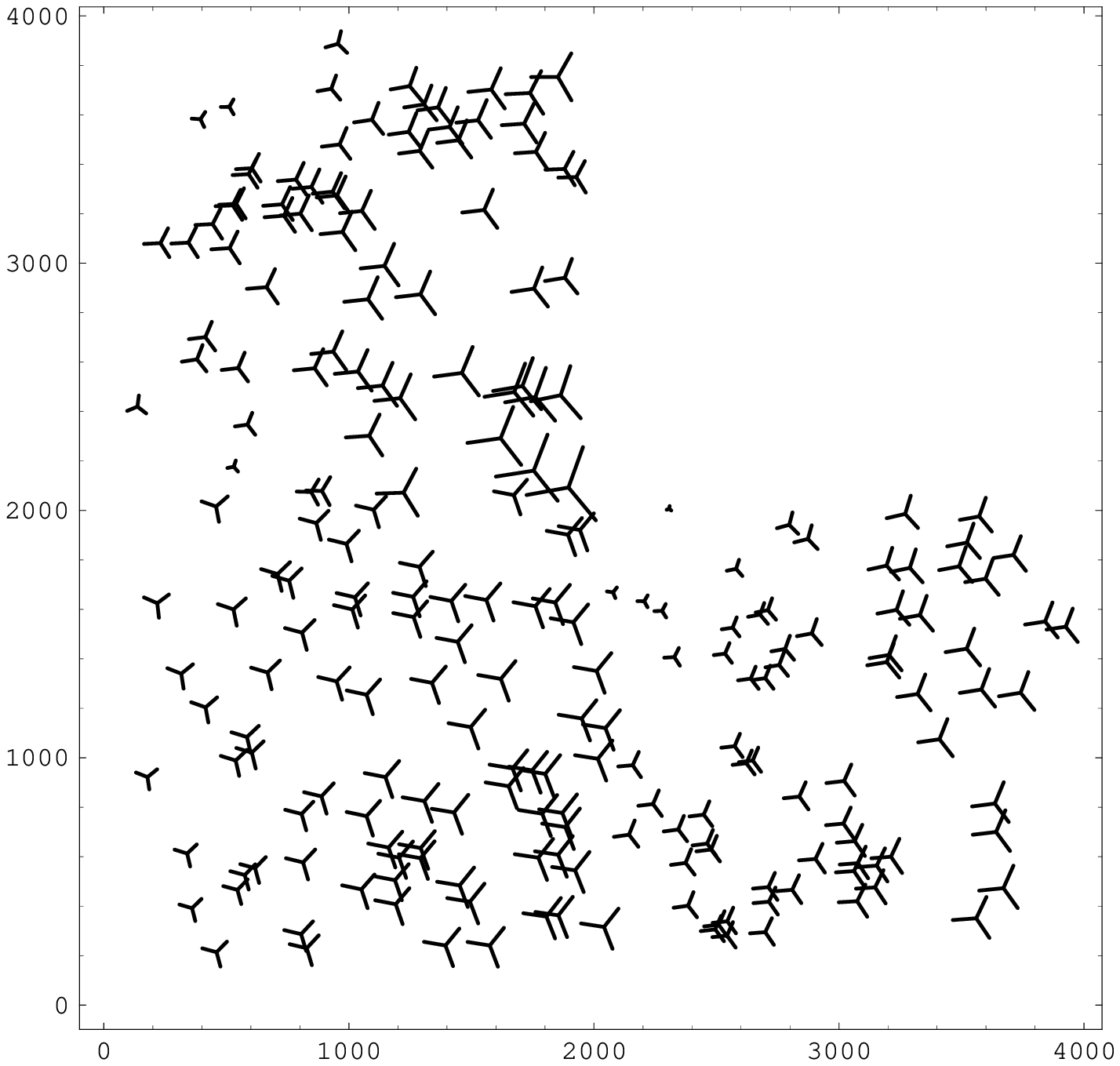}\,\,\,\,
\includegraphics[width=9cm,height=6cm]{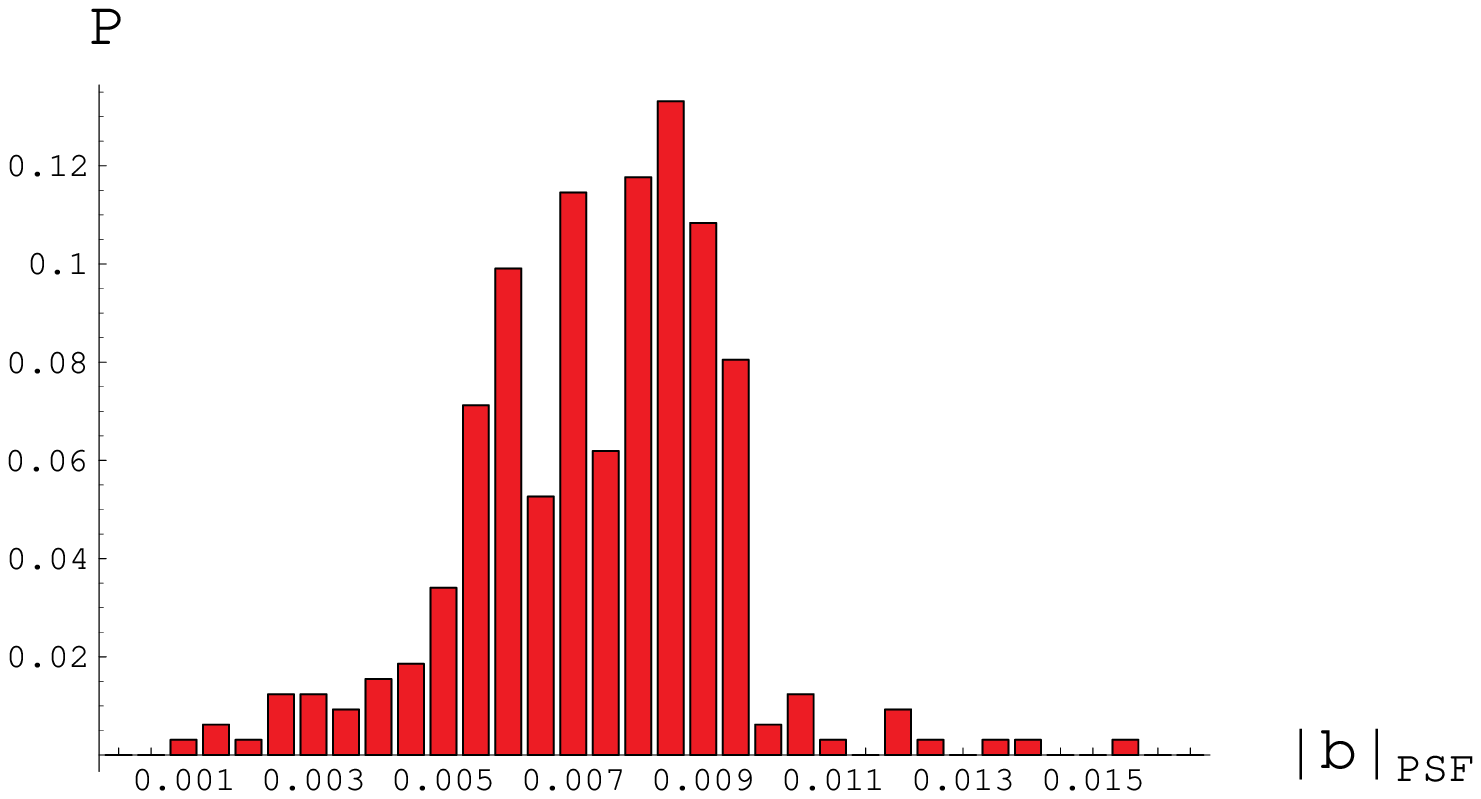}}
\caption[sextPSFcoeff] {\centering  Spatial distribution of
sextupole orientations (left) and the sextupole coefficient
distribution (right) for the 5x subsampled and drizzled Tiny Tim
PSFs for the WPFC2 camera on the Hubble. }  \label{sextPSFcoeff}
\end{figure}

Fig. \ref{sextPSFcoeff} (left) shows the distribution of sextupole
coefficients that were found by computing moments of the PSF after
it had been dithered, diffused and drizzled. The right panel shows
the orientation of the sextupole moment, with the lines of each
symbol pointing toward the three sextupole shape maximum.
Remarkably, the sextupole-moment orientation is uniform across
each chip.  This may be helpful in following time variation of the
sextupole coefficient, if any is present.

\subsection{Relative orientation of the quadrupole and sextupole
map coefficients }\label{orientation}

\begin{figure}[h!]
\centering{\leavevmode
\includegraphics[width=6cm,height=4.5cm]{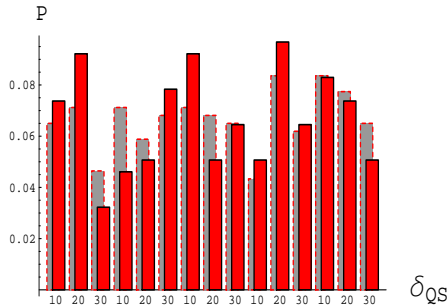}}
 \caption[deltaBarAll] {\centering The
distribuiton of the magnitude of $\delta$, the smallest angle
between a quadrupole minimum and a sextupole minimum, using the
model method, for galaxies in the north HDF surviving the cut
condition of eq. \ref{cutCondition} (foreground) and all galaxies
surviving the L2 cut (background). } \label{deltaBarAll}
\end{figure}

The orientation of the sextupole map coefficient with respect to
the quadrupole coefficient is of primary interest to us. A plot of
the (smallest) angle between sextupole and quadrupole minima using
the model method is shown in fig. \ref{deltaBarAll} for the
galaxies surviving both the L2 norm $<$ 0.05 cut and the
signal-to-noise cut of eq. \ref{cutCondition}. This angle, which
we refer to as $\delta$, runs from $0^\circ$ to $30^\circ$.  For
$\delta = 30^\circ$ the shapes will have aligned maxima, and we
refer to such galaxies as ``aligned". They are pear shaped
galaxies, as compared to the ``curved" galaxies which resemble
bananas.  See fig. \ref{deltaDef}.
\begin{figure}[h!]
\centering{\leavevmode
\includegraphics[width=8cm,height=6cm]{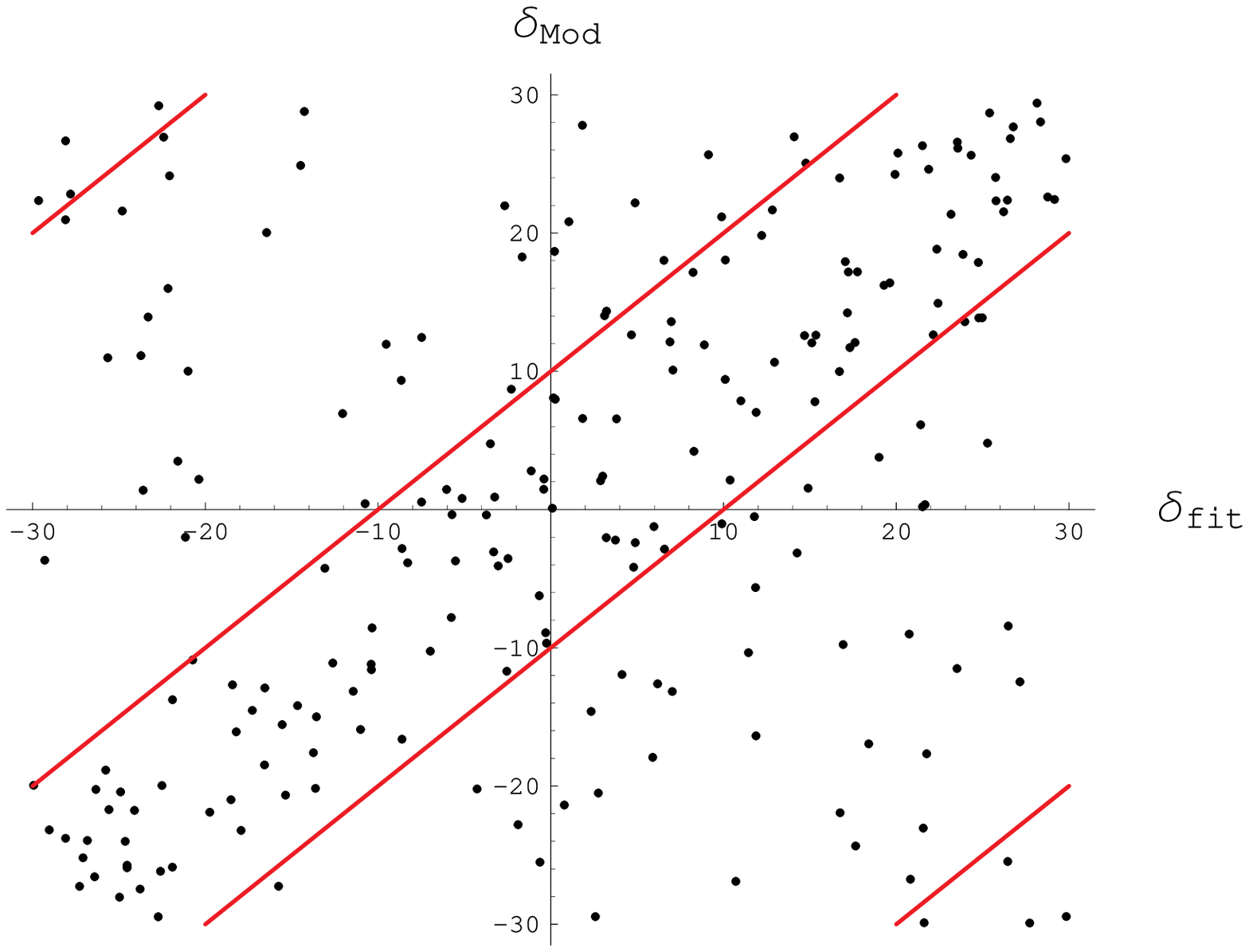}
\,\,\,\,
\includegraphics[width=7cm,height=5cm]{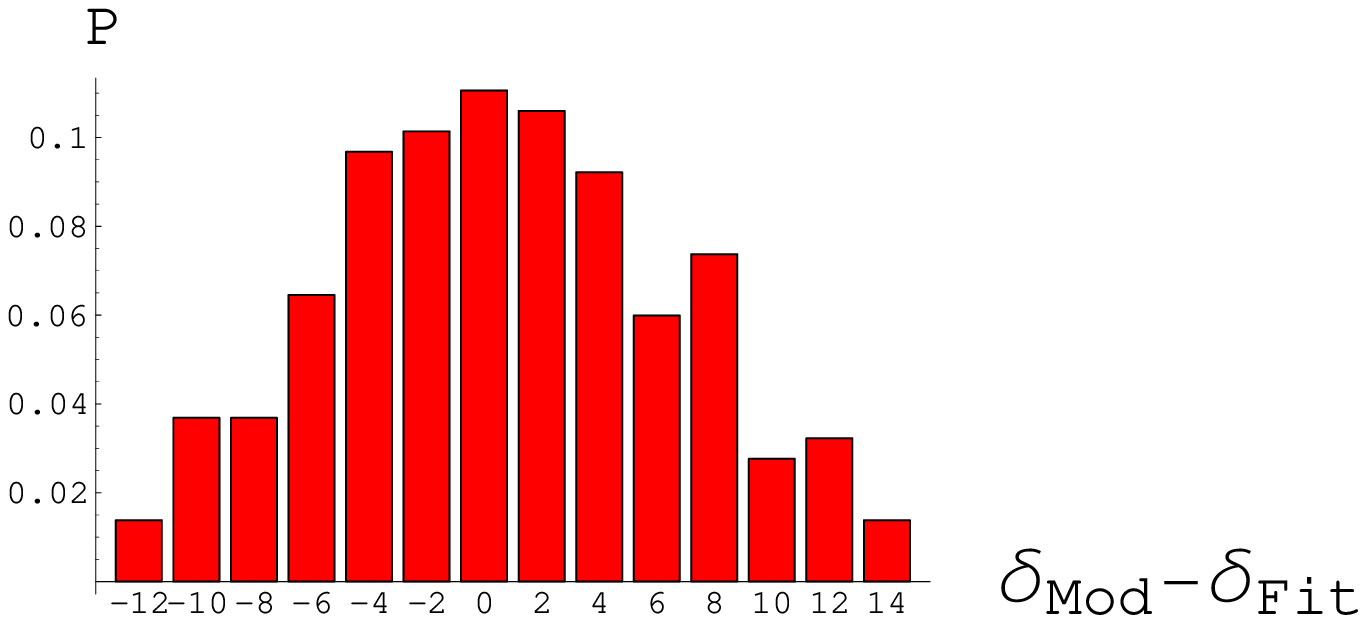}}
\caption[deltaBarAllplus] {\centering The left panel compares the
angle $\delta $ as measured by the radial-fit method and the model
method.  The points within the diagonal band (or in the corners)
have a change in $\delta < 10^\circ $ . The distribution for
$\delta_{Mod}-  \delta_{Fit}$  is shown
 on the right panel.}\label{deltaBarAllplus}
\end{figure}

Figure \ref{deltaBarAllplus} compares $\delta$ found with the
radial-fit and model methods.  About one-third of all galaxies
show a change in $\delta$ or greater than $10^\circ$.  This is a
concern because it means that the point-spread function does
indeed play an important role in distinguishing ``curved" from
``non-curved" galaxies.  This would not be as great a concern if
the PSF was well known, but unfortunately the PSF is known to vary
with time.  We will return to a discussion of the role of the PSF
in subsection \ref{PSFclumps}.

We now proceed to a study of the spatial distributions of the
``curved'' and ``aligned" galaxies as determined in this section.

\section{ Clumping probabilities}\label{clumping}

To quantify clumping for a particular galaxy subset with ${\it N}$
(e.g. the number of curved) members, we draw a circle of fixed
radius $R$ about each member of the subset and count the members
of that subset which lie within the circle. We then compare the
distribution of the number of galaxies having the number of
neighbors $N \,= \,0,\, 1,\, 2,\, 3,$ etc. within the circle, with
a large number of such distributions derived from randomly chosen
subsets having the same number (${\it N}$) of galaxies in two
distinct ways:

\begin{enumerate}
\item for an informative but qualitative comparison, we compare
the distribution of the particular subset with the average
distribution of the random subsets, and

\item for a quantitative comparison, for the particular subset
being studied and for each of 500 randomly selected subsets having
the same number of galaxies as the original subset, we sum the
galaxies having $N \ge N_{Min}$ or more neighbors (usually $3 \le
N_{Min} \le 7$). Each random subset is thereby associated with a
single number, $n_G$, the number of galaxies of that subset having
$N_{Min}$ or more neighbors. We then create a bar graph, showing
for each value for $n_G$ the number of random subsets which had
$n_G$ galaxies with $N_{Min}$ or more neighbors. This distribution
is thus a property of random subsets with ${\it N}$ members, with
specified neighbor distance $R$, for the specified neighbor range
($N \ge N_{Min}$). Since the initial subset will have a certain
number, $n_{G0}$, of galaxies having $N_{Min}$ or more neighbors,
we can ask the question ``what fraction of randomly chosen galaxy
subsets have $n_{G0}$ or more galaxies with $N_{Min}$ or more
neighbors?" We thereby determine a probability that this
configuration could occur by chance.
\end{enumerate}

\noindent In a variation of this, the randomly chosen subsets, may
be constrained in some way. For example, we may wish to look only
at randomly chosen subsets whose members are selected to have the
same z-distribution as the galaxies in the original set.

In this section we will show examples of: a) bar graphs showing
the galaxy fraction ($P$) versus number of neighbors, for both the
observed subset and for the average of 500 random subsets, whether
the observed subset be ``aligned", ``mid-range", or ``curved"; b)
bar graphs showing the number of galaxies having $N$ or more
neighbors for 500 random subsets of the background galaxies (after
cuts); and c)field plots showing the spatial location in the field
of the ``curved" galaxies and the ``aligned" galaxies. On these
field plots may also be shown: i) all background galaxies after
cuts, ii) the direction of the curvature of ``curved" galaxies,
iii) the orientation of the ``aligned" galaxies, iv) the circular
areas defining the neighbors of each ``curved" or ``aligned"
galaxies, v) the combined circular areas of distinct groups, vi)
the z-values of the` `curved" or ``aligned" galaxies, or viii)
some combination of these.

\subsection{ ``Curved" galaxies}\label{curvedGal}

\begin{figure}[h!]
\centerline{\leavevmode
\includegraphics[width=2in,height=3.5cm]{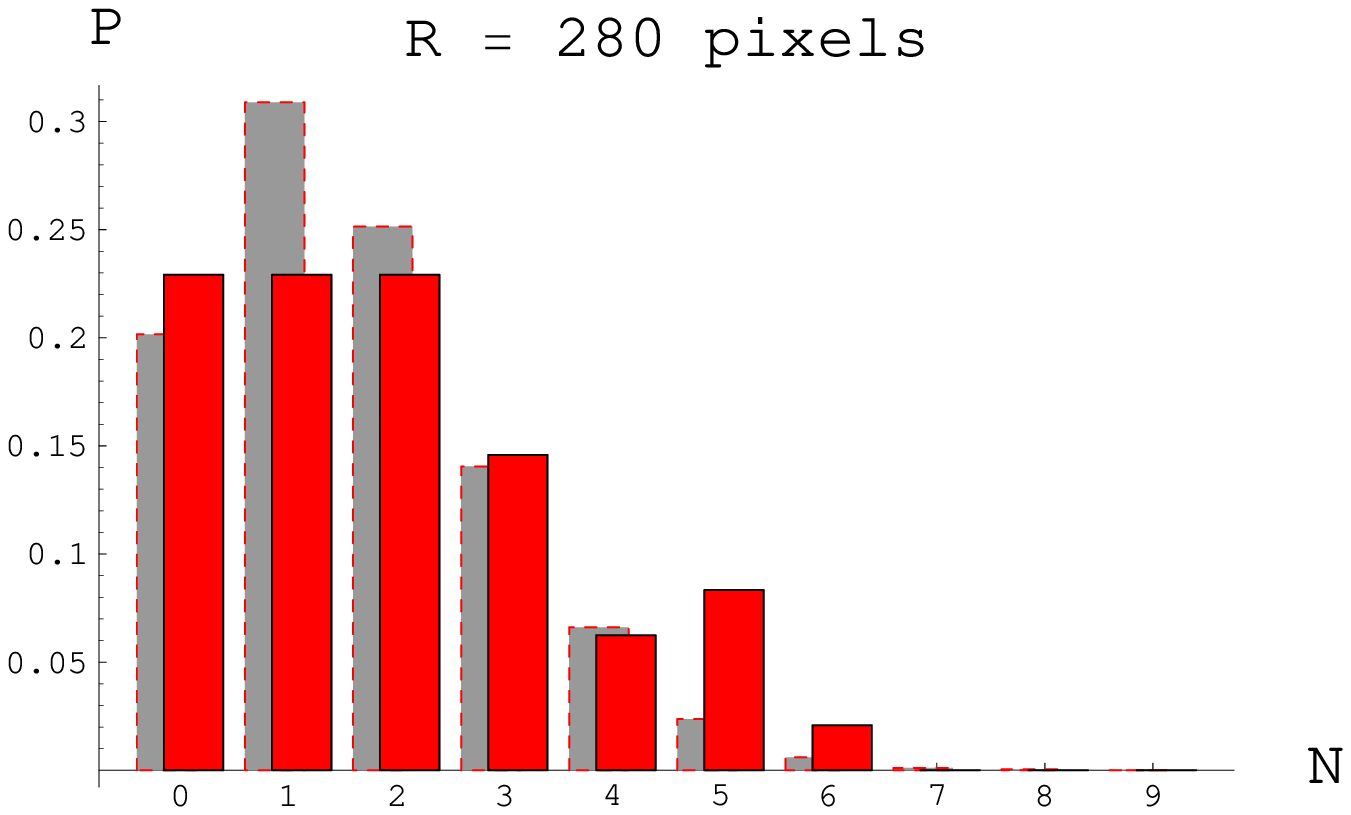}\,\,\,\,
\includegraphics[width=2in,height=3.5cm]{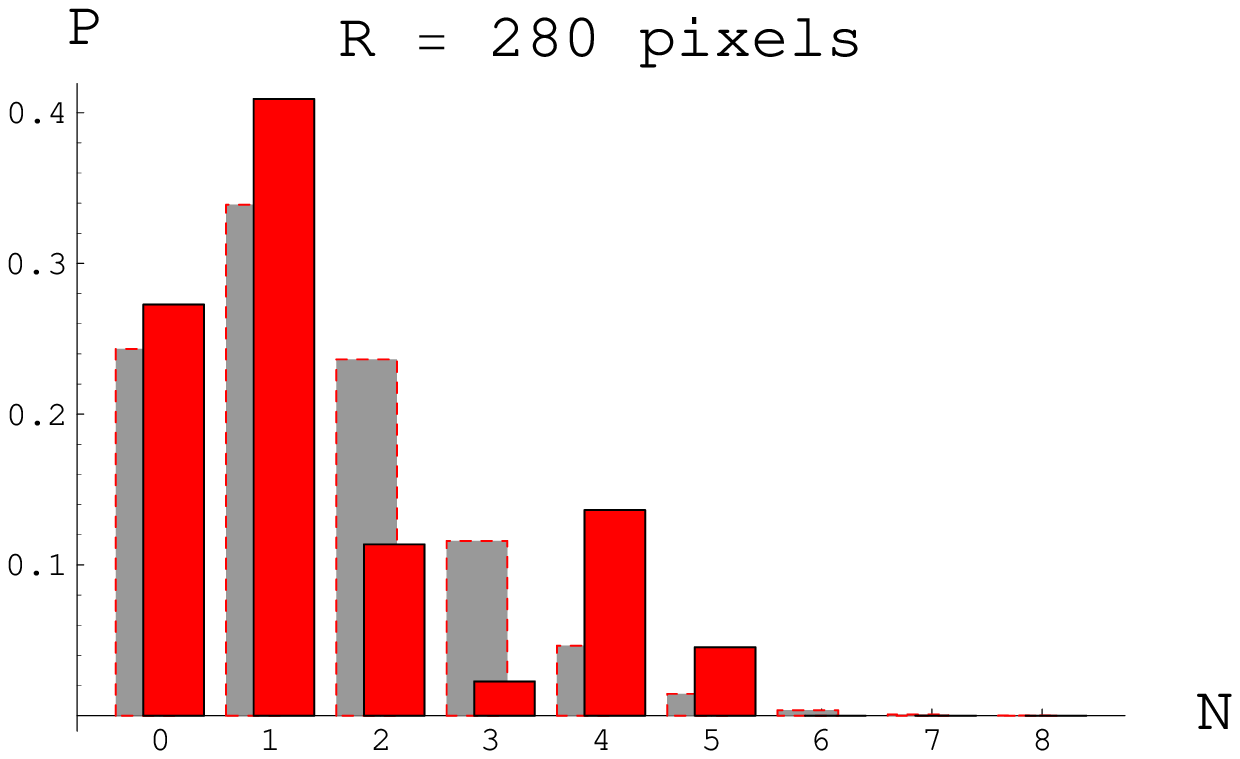}}
\centerline{\leavevmode
$$ \,\,(a)\,\, {\rm ``Radial-fit"} \hskip4cm  (b)\,\, {\rm ``Model"} $$}
\caption[curvedNeighBarMomFitMod] { For two methods, a histogram
showing the fraction of curved galaxies having a specific number,
N, of curved neighbors within a circle of radius 280 pixels,
compared with the average of such a distribution for 500 randomly
chosen subsets having the same number of galaxies.}
\label{curvedNeighBarMomFitMod}
\end{figure}

For ``curved" galaxies we begin with bar graphs of type $1$. In
fig. \ref{curvedNeighBarMomFitMod} (a), we show the distribution
of the number of neighbors in a circle of radius $R=280$ for the
``curved" galaxies using the radial-fit method (which we now take
to be $\delta < 9^\circ$.) In fig.
\ref{curvedNeighBarMomFitMod}(b) we show the same distribution
using the model method.

\begin{figure}[h!]
\centerline{\leavevmode
\includegraphics[width=2in,height=3.5cm]{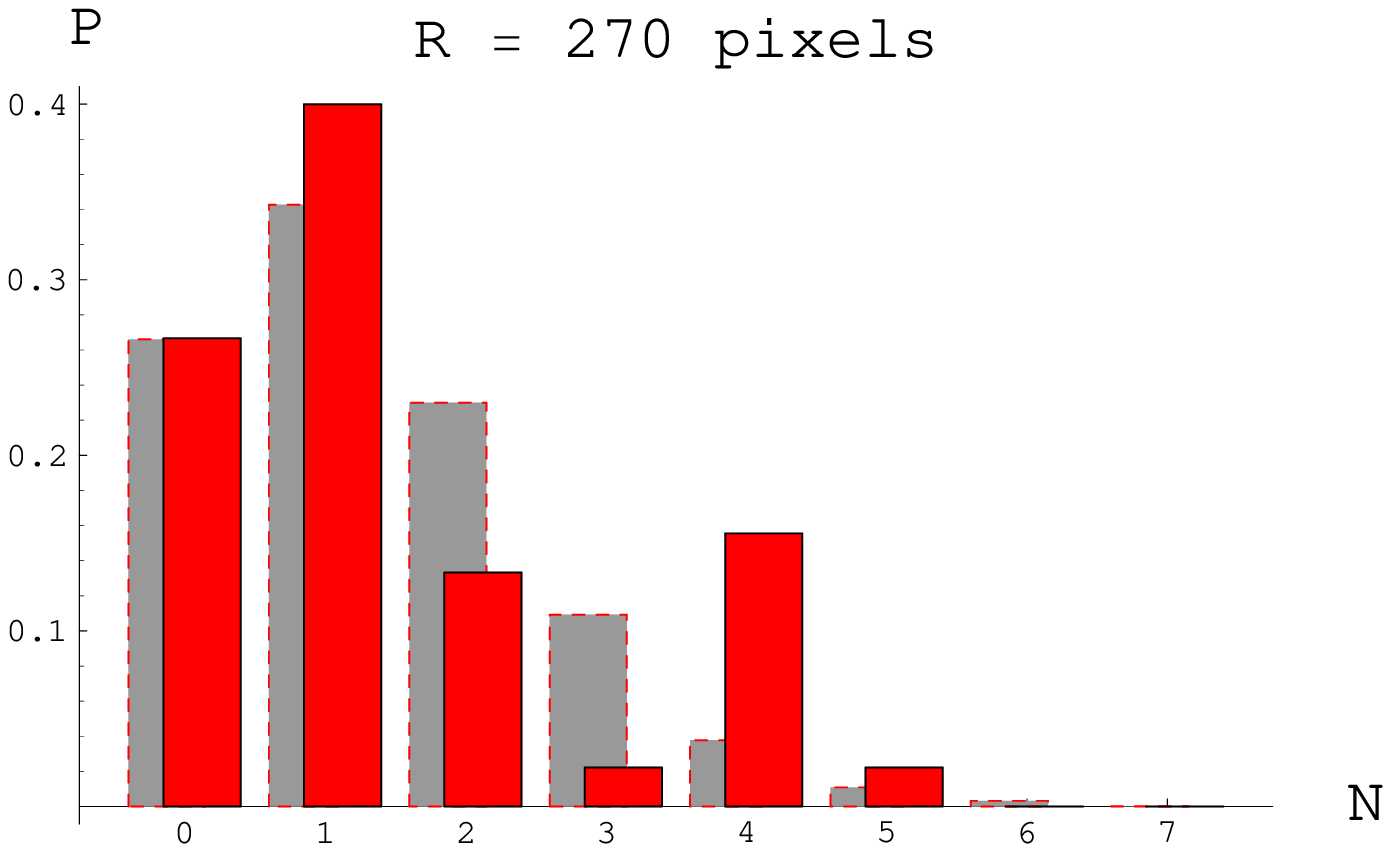}\,\,
\includegraphics[width=2in,height=3.5cm]{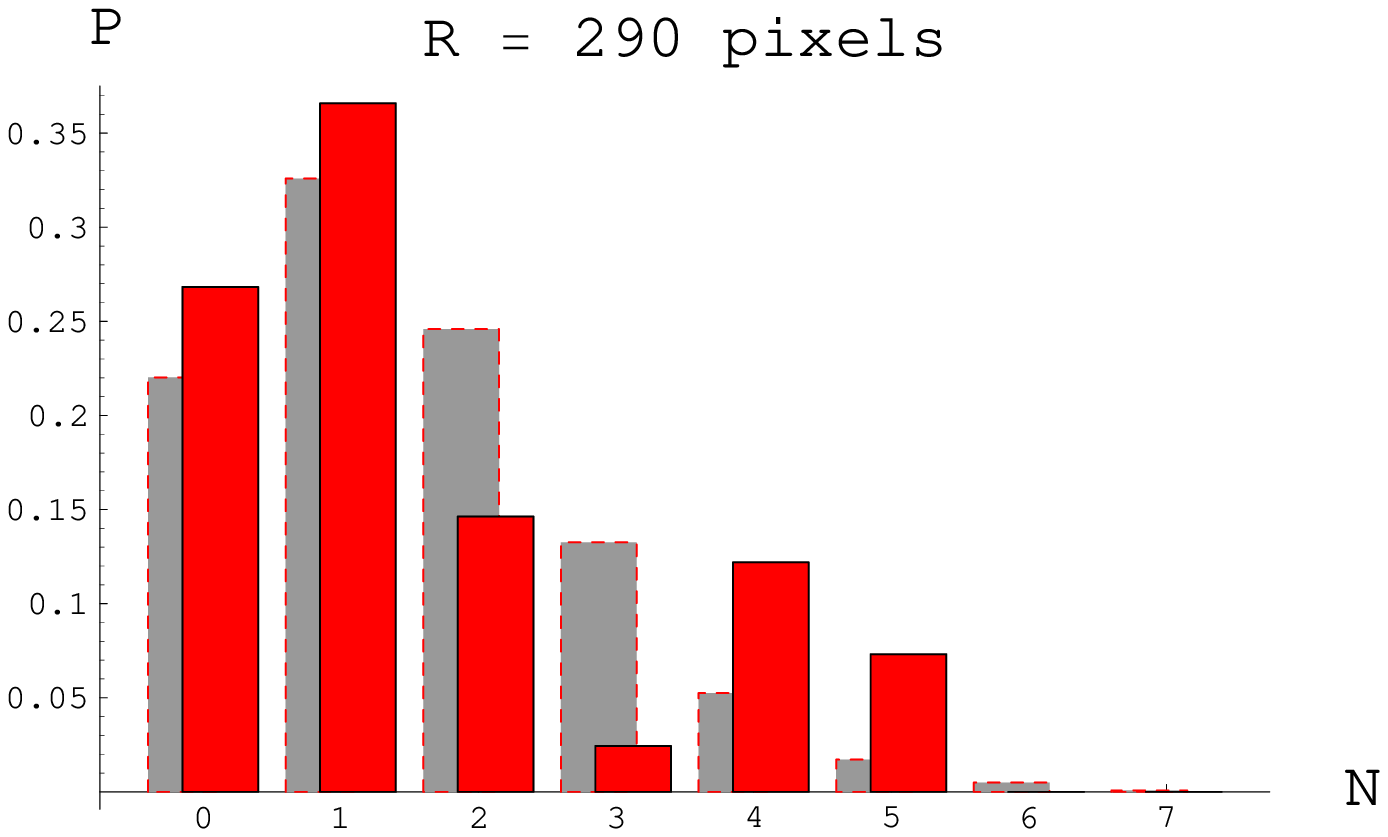}\,\,
\includegraphics[width=2in,height=3.5cm]{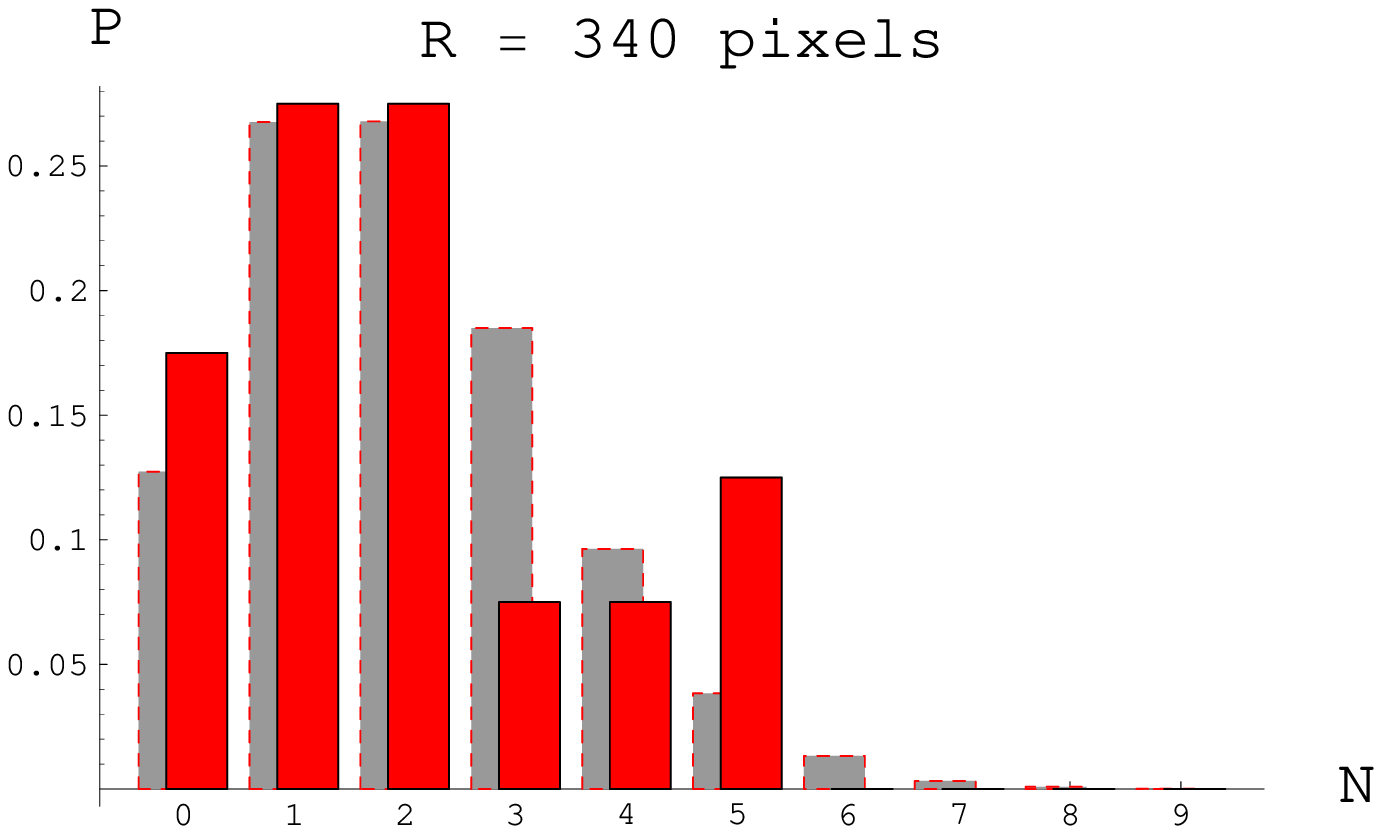}}
\caption[curvedFourRadii]{The same ``neighbors'' histogram for
curved galaxies in the north HDF in the circle with  R = 270, 290
and 340 pixels, using the model method.} \label{curvedFourRadii}
\end{figure}

\noindent In fig. \ref{curvedFourRadii} we show for the model
method how this distribution changes as the radius of the circle
is varied.\footnote{ The Hubble deep field images have a drizzled
pixel size of 0.04 arc sec. At $z$ =0.6 for current cosmological
parameters (dark matter 23{\%}, baryons 4{\%}, dark energy 73{\%})
the distance scale would be 6.67 kpc per arc sec. 280 pixels
corresponds to 75 kpc. }

\begin{figure}[h!]
\centerline{\leavevmode
\includegraphics[width=2.4in,height=3.5cm]{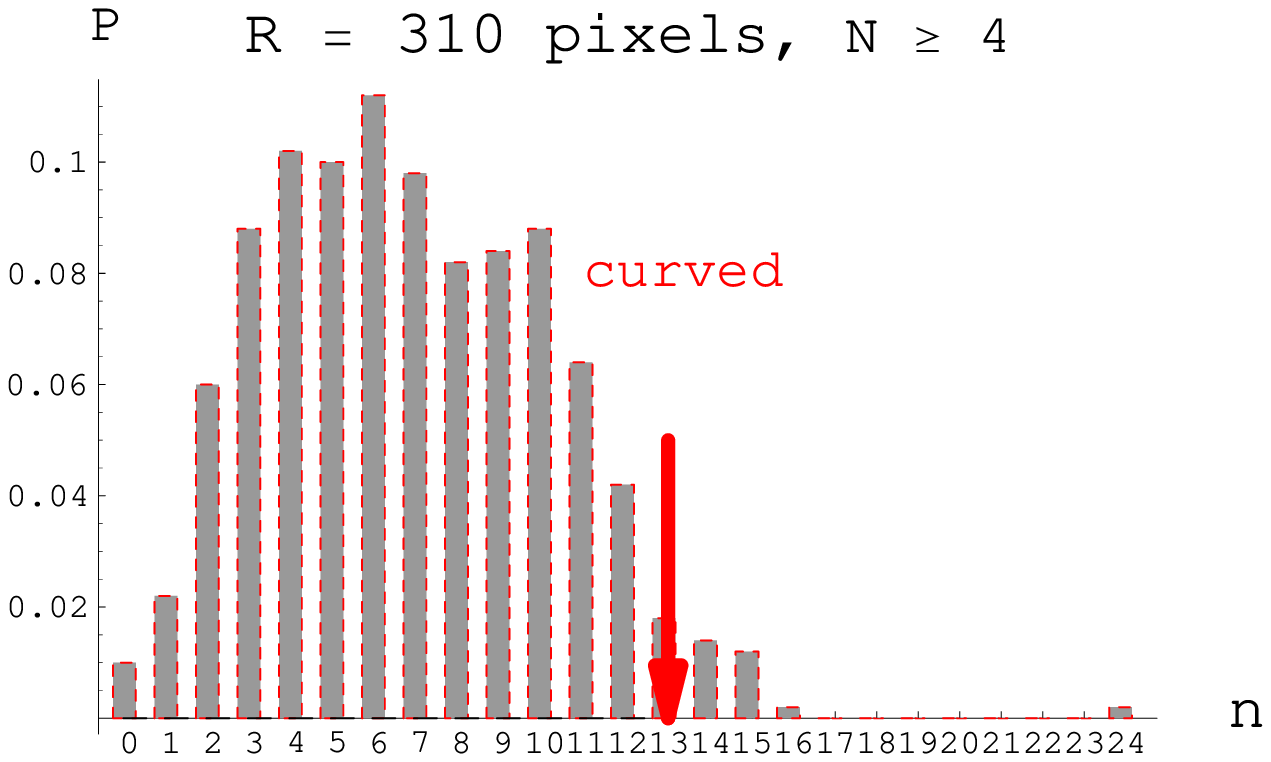}\,\,
\includegraphics[width=2.4in,height=3.5cm]{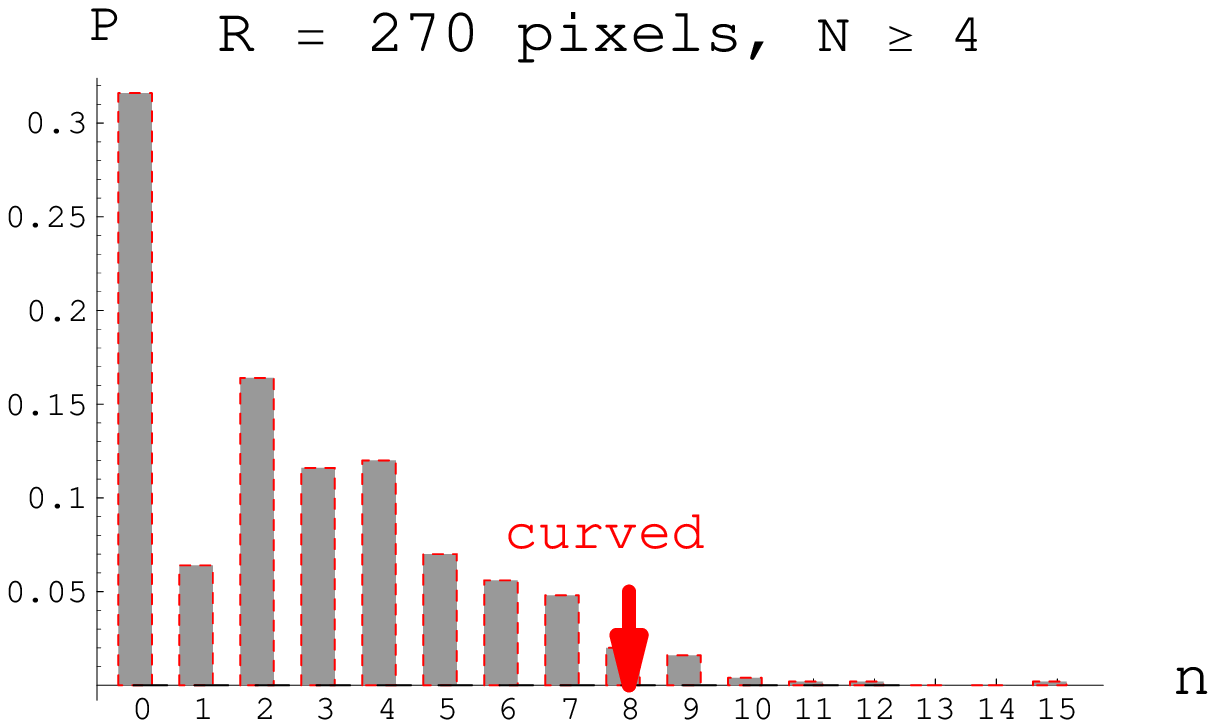}}
\centerline{\leavevmode
$$ \,\, (a)\,\, {\rm ``Radial-fit"} \hskip3cm  (b)\,\, {\rm ``Model"} $$}
\caption[curvedThreeOrMoreMomFitMod] { Histogram of numbers of
galaxies in 500 randomly chosen sets having  4  or more neighbors
in the circles  of 310 and 270 pixels. The red arrow indicates the
number of ``curved" galaxies with N or more neighbors for the
``curved'' set.} \label{curvedThreeOrMoreMomFitMod}
\end{figure}

Fig. \ref{curvedThreeOrMoreMomFitMod}(a) displays an analysis of
type $2$ for the radial-fit method.  (The number
$n_{G0}$corresponding to the ``curved'' set is indicated by an
arrow in this bar graph.) For the optimum radius, of 310 pixels,
there will be 21 out of 500 sets that have as many galaxy-circles
with counts equal to or greater than the original curved set. In
other words, the probability of achieving the curved set by chance
is equal to or smaller than 4{\%}.  Fig.
\ref{curvedThreeOrMoreMomFitMod}(b) displays an analysis of the
type $2$ for the model method.  The optimum radius is now 270
pixels, and the probability to achieve the curved set by chance is
equal to or smaller than $4\%$.
\begin{figure}[h!]
\centerline{\includegraphics[width=3in,height=3in]{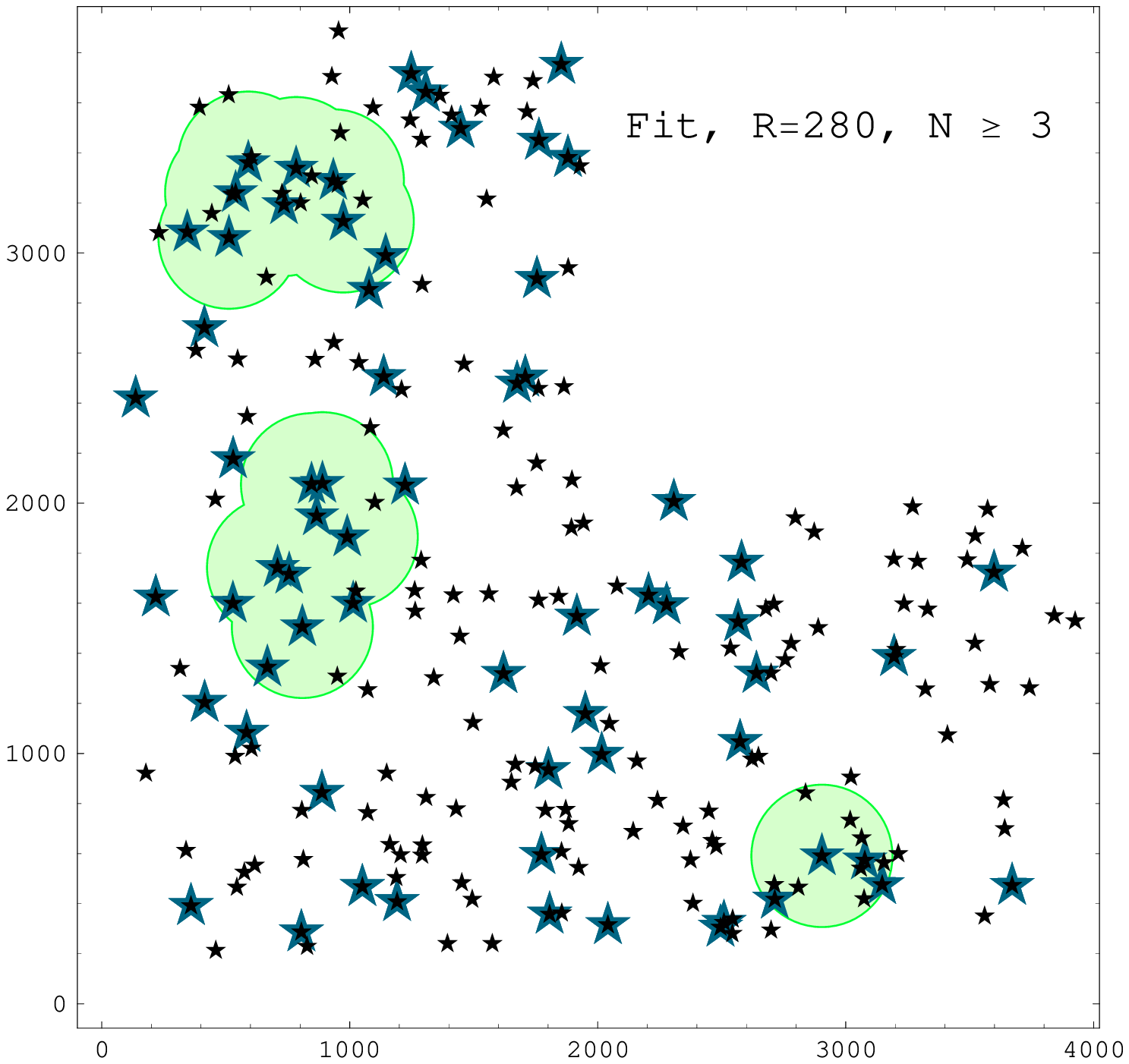}\,\,
\includegraphics[width=3in,height=3in]{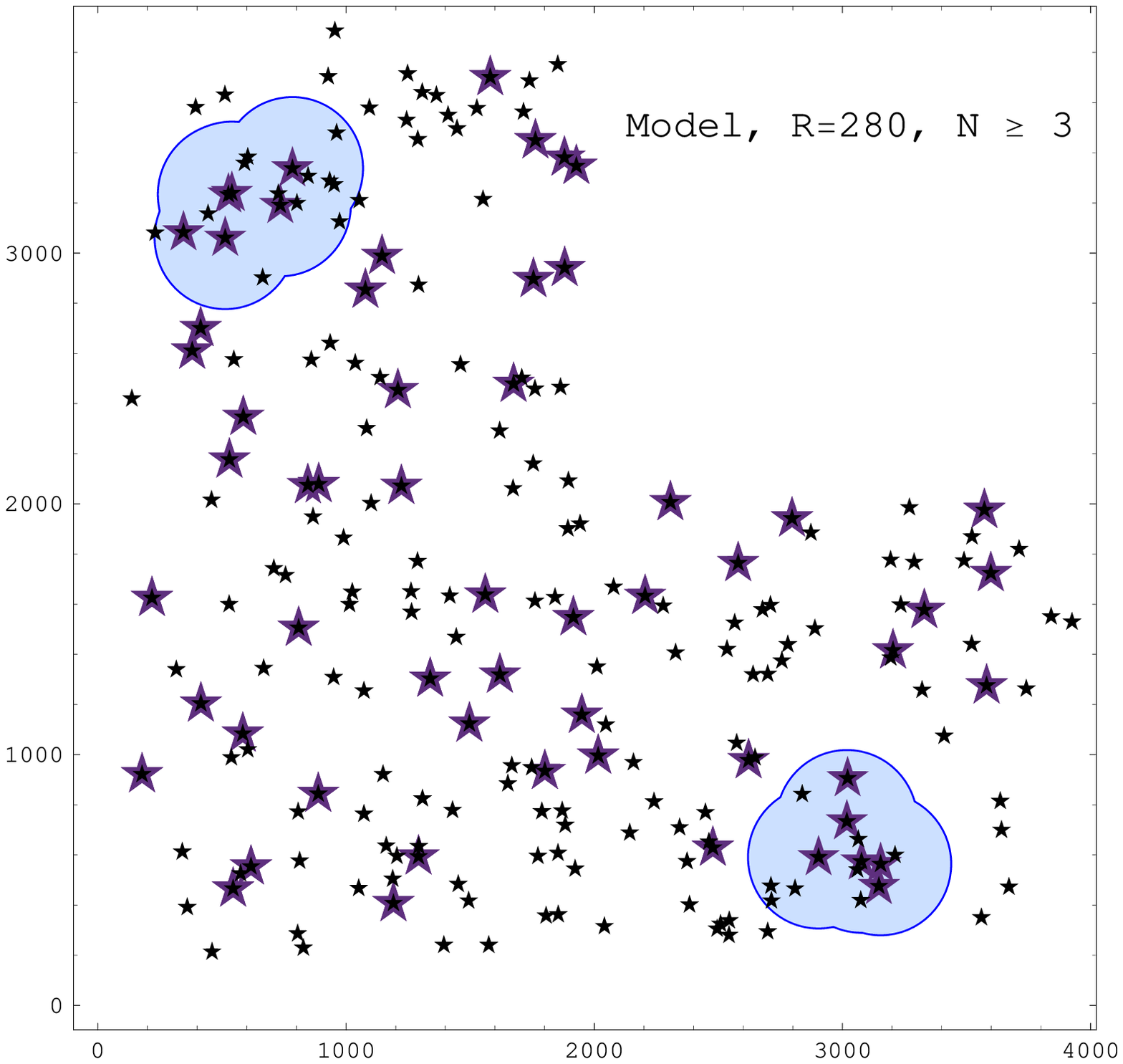}}
\caption[curvedCircles] {A field plot showing the spatial location
of ``curved" galaxies in the north HDF using the radial-fit (left
panel) and model (right panel) methods to determine the map
coefficients. Large ``stars" indicate ``curved" galaxies and small
``stars" indicate the remaining background galaxies that survived
the joint-variable signal-to-noise cut.  Circles are shown for 3
or more neighbors.} \label{curvedMomFitModBothFields}
\end{figure}
We now turn to the field plots.  Figure
\ref{curvedMomFitModBothFields} shows ``curved" galaxies in the
Hubble north field and for galaxies having three or more
neighbors, their neighborhood circles of radius R=280 pixels.
Panel (a) shows clumping of ``curved" galaxies as determined using
the radial-fit method. This can be compared with (b) the same plot
with ``curved" as determined using model method. In these plots,
large ``stars" indicate ``curved" galaxies, and small ``stars"
indicate remaining background galaxies.

As seen in fig. \ref{deltaBarAll} the change in $\delta$ is large
enough so that many galaxies would move across the boundary
defining ``curved" galaxies.  Still, the overall pattern remains
remarkably similar for the two methods.

\begin{figure}[h!]
\centerline{\includegraphics[width=2.0in,height=2.6in]{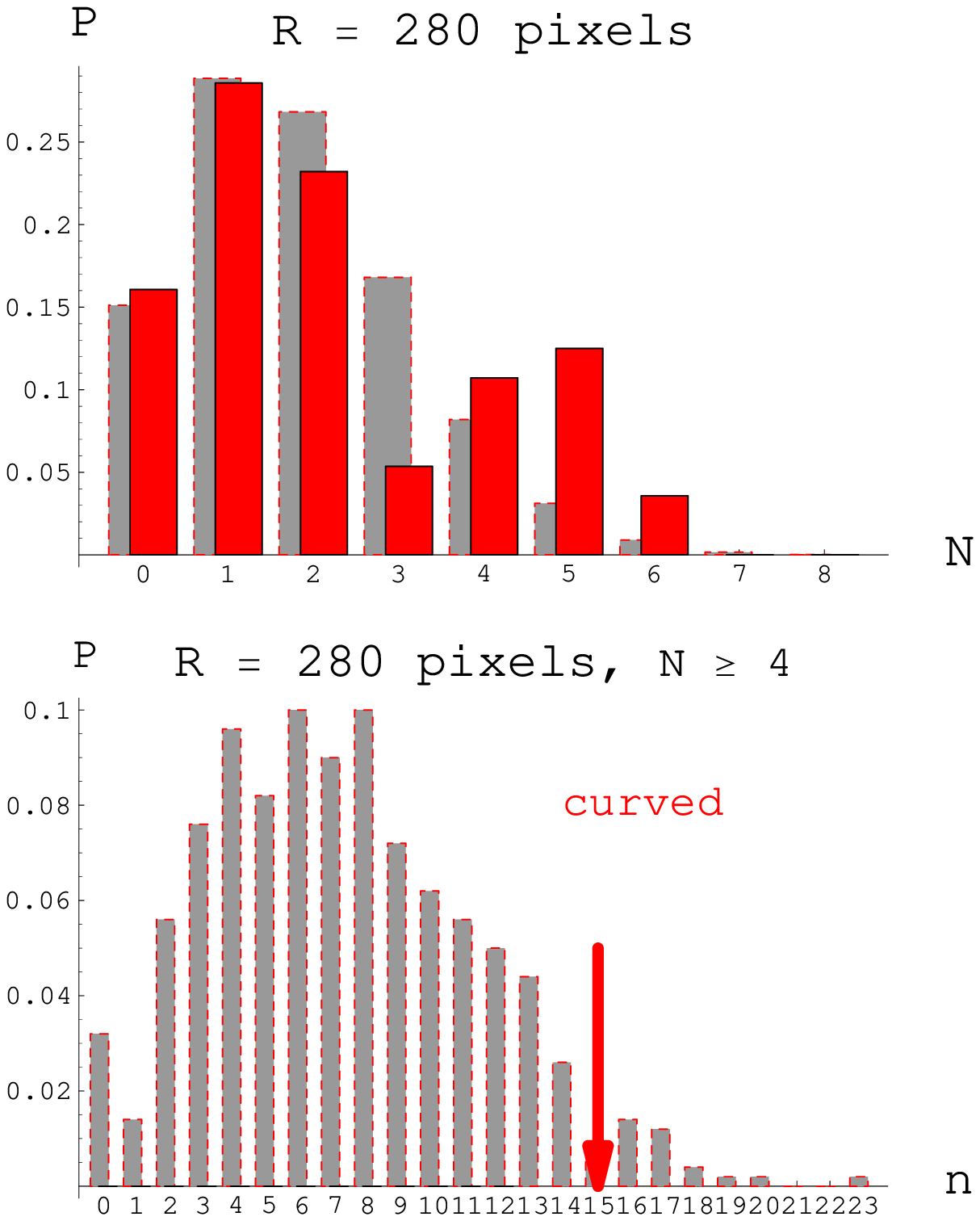}\,\,
\includegraphics[width=3in,height=3in]{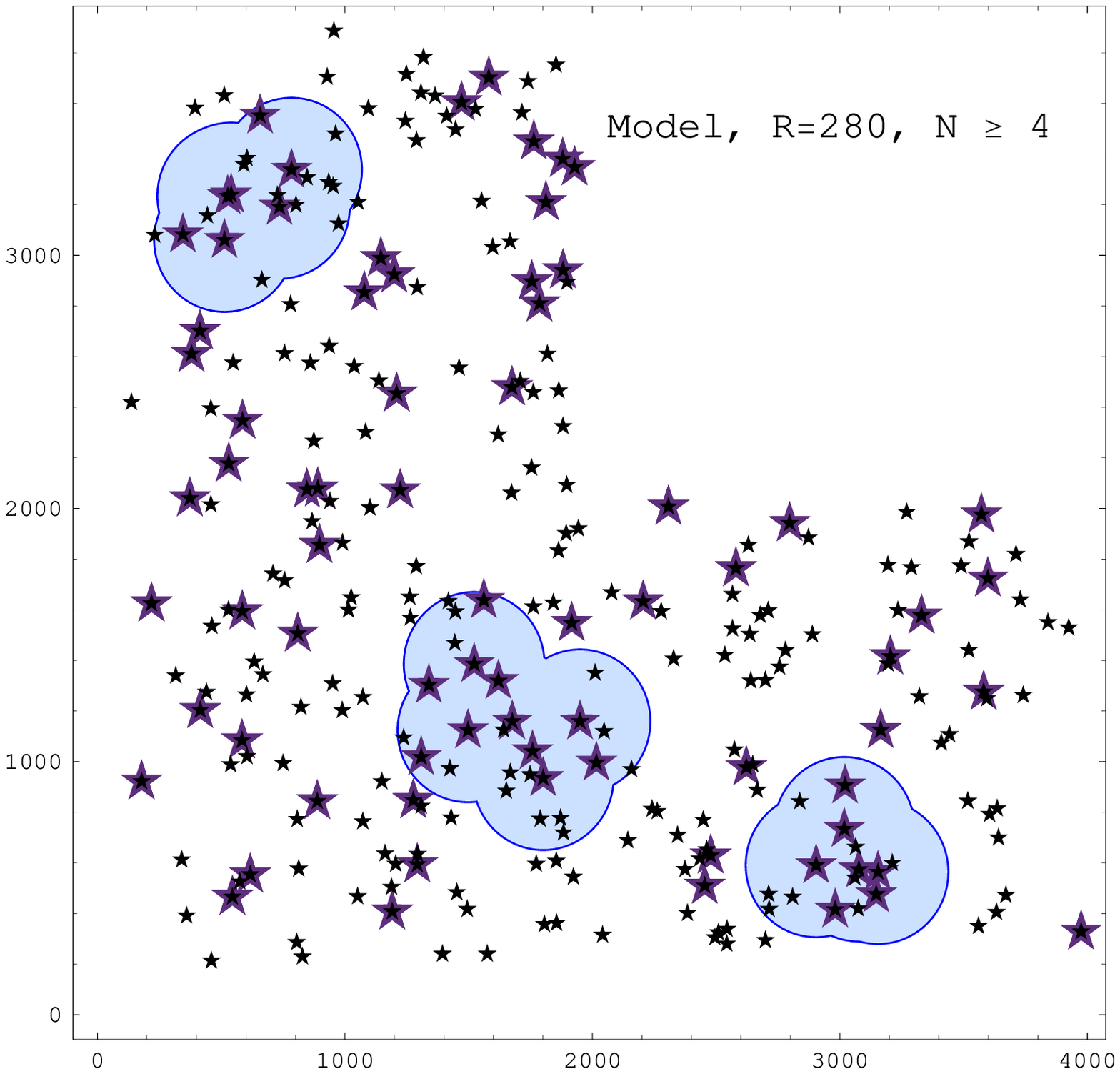}}
\caption[weakenedNoiseCut] {A ``neighbors'' histogram and a field
plot showing the spatial location of ``curved" galaxies in the
north HDF using the model method to determine the map
coefficients. Circles are shown for 4 or more neighbors.}
\label{weakenedNoiseCut}
\end{figure}

Next we carry through the same analysis for a less stringent noise
cut.  See figures \ref{weakenedNoiseCut} a-c.  Remarkably the
probability to randomly achieve the resultant clumping decreases
to $3\%$ for the model method as we increase the radius of the
noise cut condition of eq. \ref{cutCondition} from 0.17 to 0.25.
For an orientation to this cut change, see fig. \ref{combinedCut}.
If we were truly adding noise, one would expect the distribution
to become more random, not less. The new field plot is shown in
fig. \ref{weakenedNoiseCut} c. There is now a $3^{rd}$ major
clump.

Finally we note that the random conditions often include rather
improbable z-distributions.  If we require the randomly chosen
galaxy subsets to have z-distributions approximating the actual
z-distribution, the probability decreases from $3\%$ to $1\%$.

\subsection{ ``Aligned" galaxies}\label{alignedGal}

Dark matter clusters with haloes in the sextupole lensing mass
range will statistically move the observed shapes of background
galaxies to a ``more curved" condition. See sections
\ref{fieldMass} and \ref{fieldOverdensity}.  The regions of the
field occupied by haloes slightly enhance the number of curved
galaxies and deplete the number of aligned galaxies. For this
reason we now apply our analysis of ``curved" galaxies to
``aligned galaxies".

\begin{figure}[h!]
\centerline{\leavevmode
\includegraphics[width=2in,height=3.5cm]{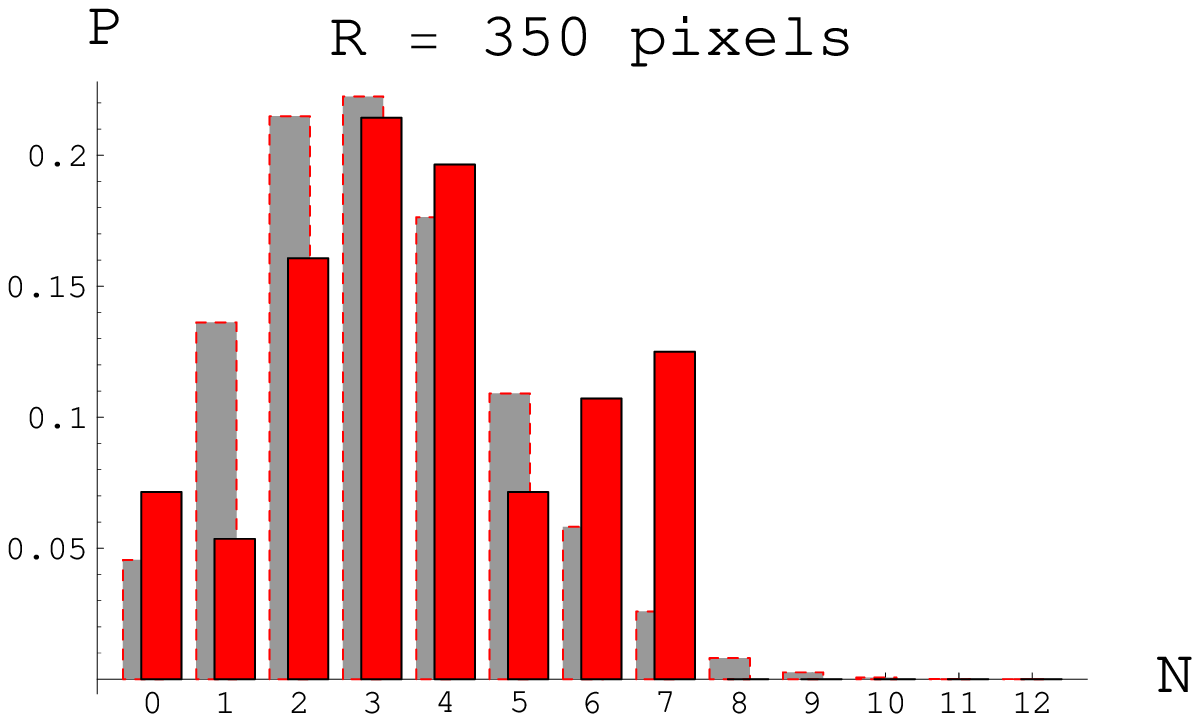}\,\,\,\,
\includegraphics[width=2in,height=3.5cm]{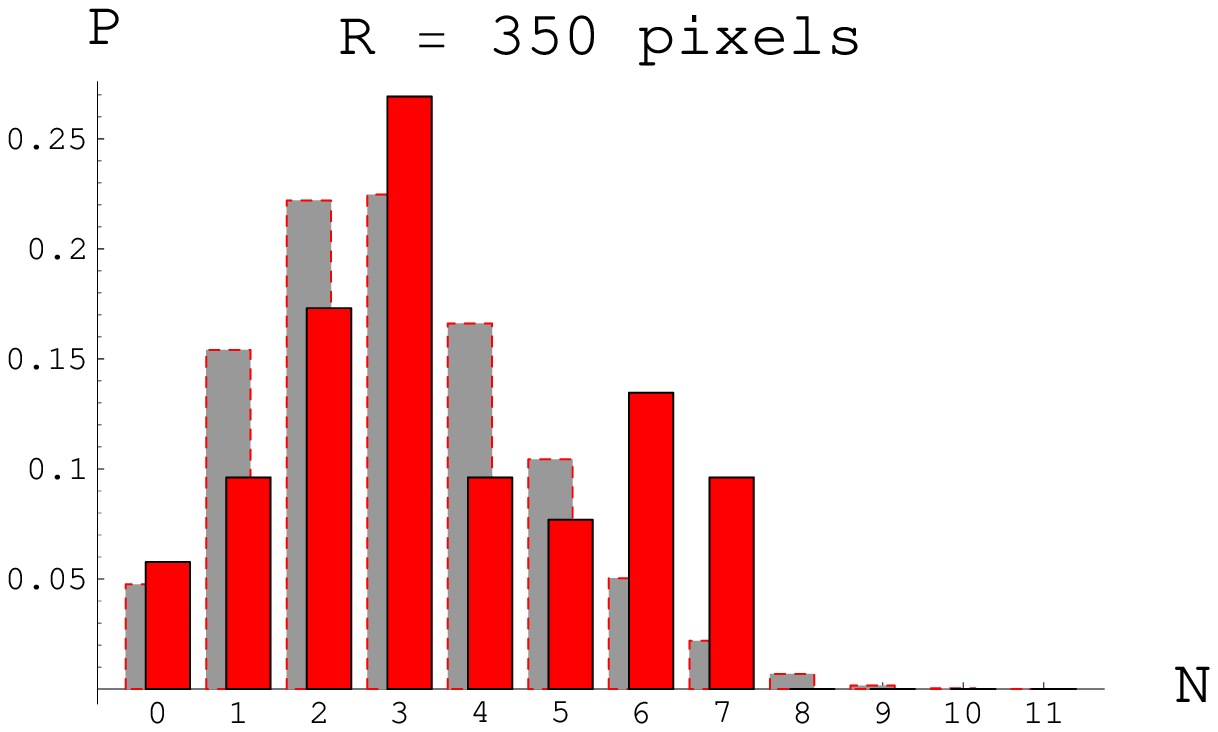}}
\centerline{\leavevmode
$$  \hskip1cm (a)\,\, {\rm ``Radial-fit"} \hskip3cm  (b)\,\, {\rm ``Model"} $$}
\caption[alignedNeighBarMomFitMod] { Histograms show the
distribution of the number of ``aligned" neighbors of ``aligned"
galaxies within a circle of radius 350 pixels, compared with the
average of such a distribution for 500 randomly chosen subsets of
galaxies in the north HDF having the same number of members.}
\label{alignedNeighBarMomFitMod}
\end{figure}

In  fig. \ref{alignedNeighBarMomFitMod}(a)we show an example of
type $1$. The distribution shows the number of ``aligned" galaxies
 in a circle of radius $R=350$ using the radial-fit method (which we take to
be $\delta > 20^\circ$.) In fig. \ref{alignedNeighBarMomFitMod}(b)
we show the same distribution using the model method.

\begin{figure}[h!]
\centerline{\leavevmode
\includegraphics[width=2in,height=3.5cm]{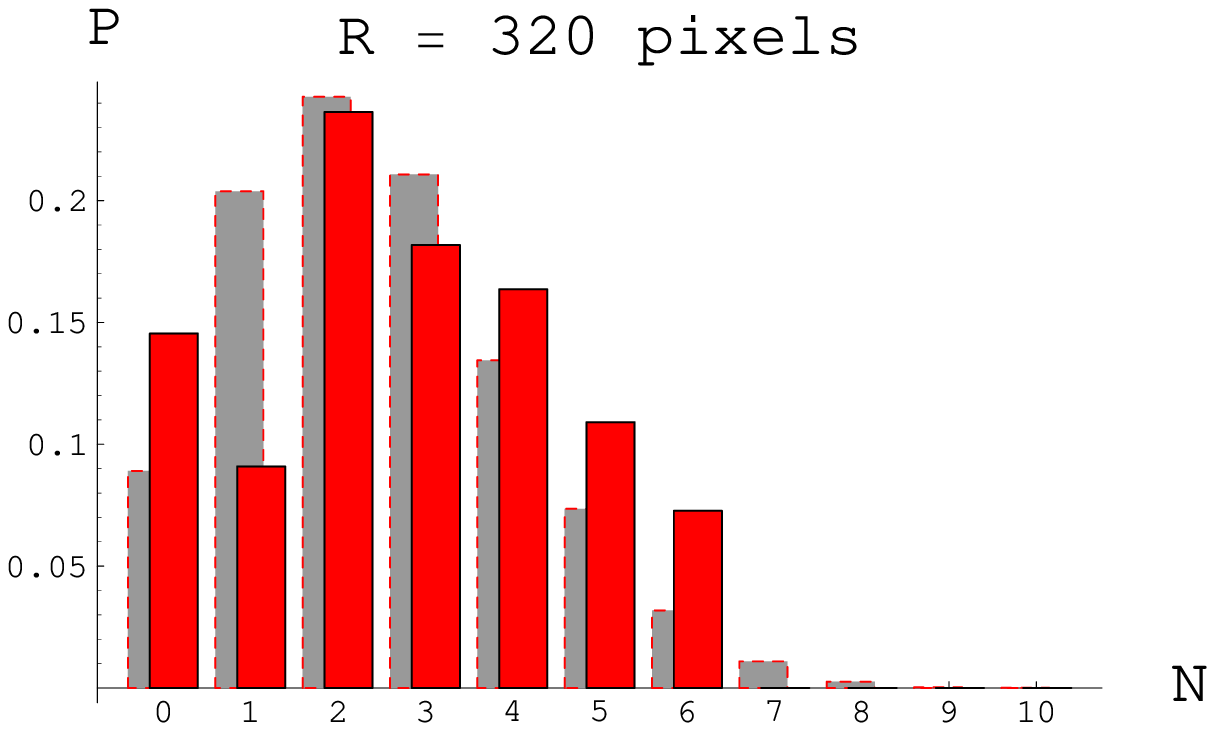}\,\,
\includegraphics[width=2in,height=3.5cm]{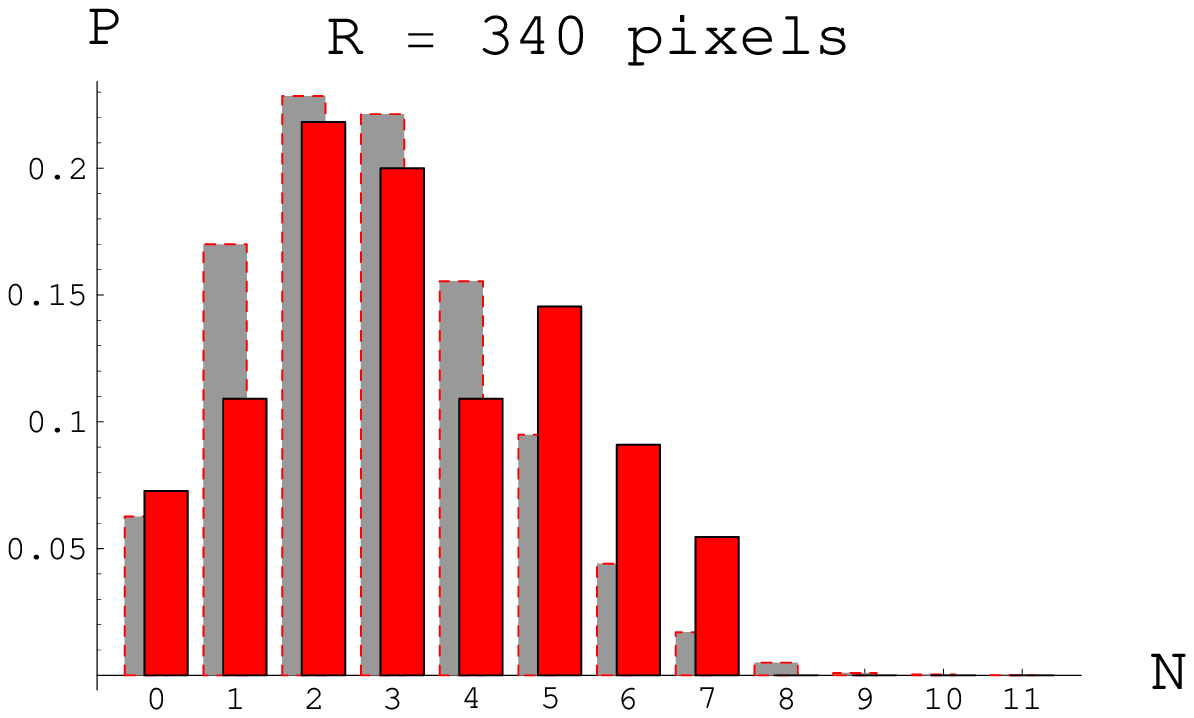}\,\,
\includegraphics[width=2in,height=3.5cm]{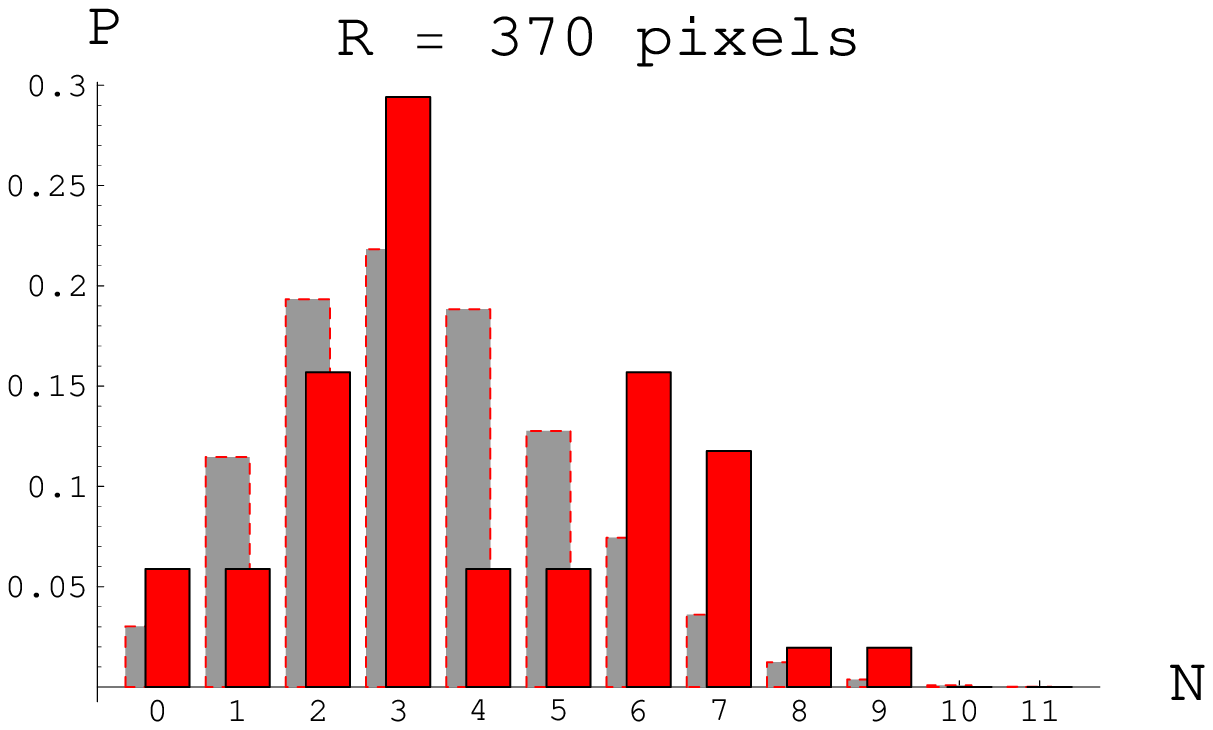}}
\caption[alignedFourRadii] {A ``neighbors'' histogram for
``aligned" galaxies as defined using the model method for the
north HDF North in circle of radius a) R = 320, b) 340 and c) 370
pixels, respectively.} \label{alignedFourRadii}
\end{figure}

\begin{figure}[h!]
\centerline{\leavevmode
\includegraphics[width=2in,height=3.5cm]{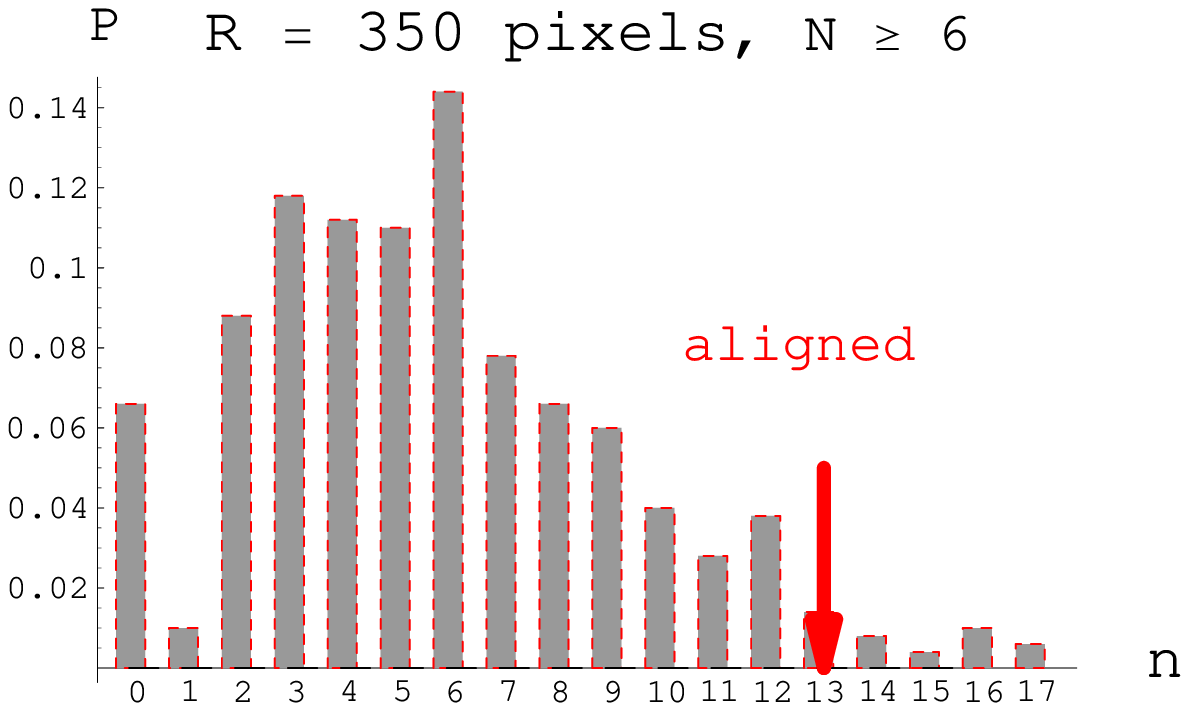}\,\,\,\,
\includegraphics[width=2in,height=3.5cm]{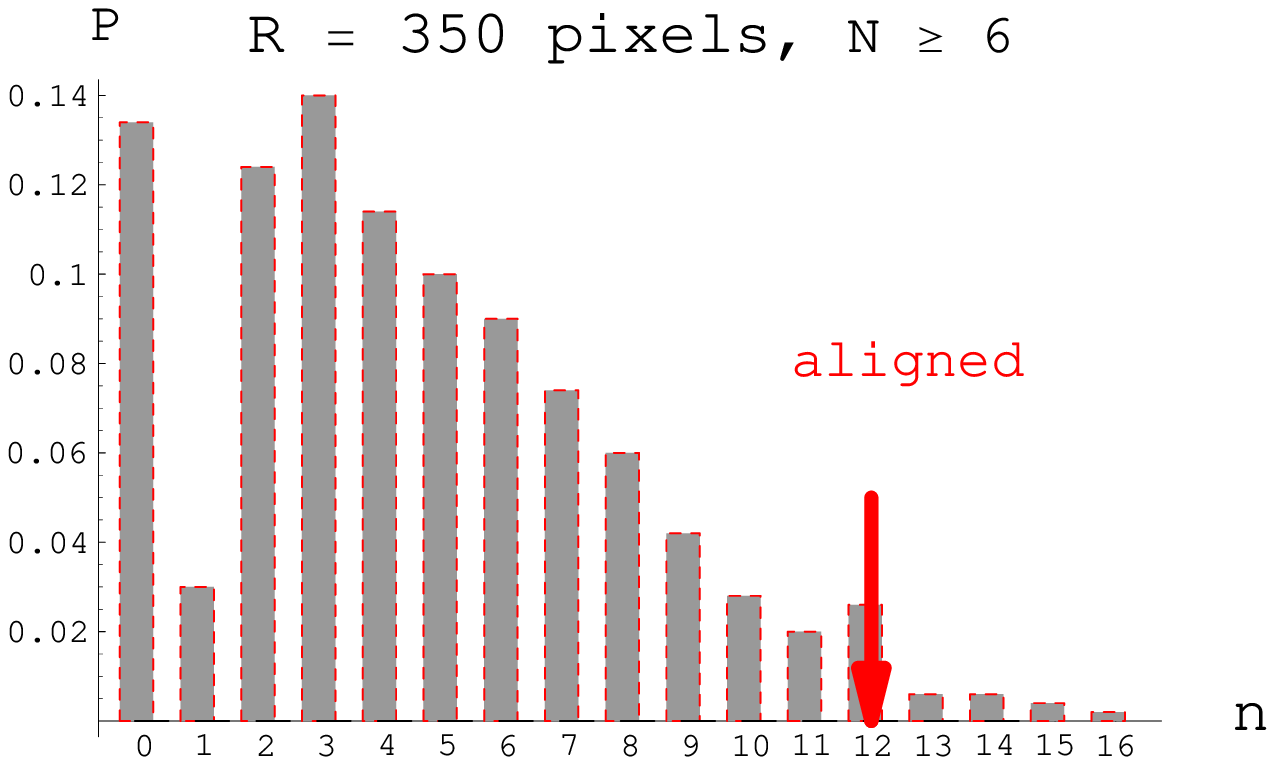}}
\centerline{\leavevmode
$$  \hskip1cm (a)\,\, {\rm ``Radial-fit"} \hskip3cm  (b)\,\, {\rm ``Model"} $$}
\caption[alignedThreeOrMoreMomFitMod] { Histogram of numbers of
galaxies in 500 randomly chosen sets having 6 or more neighbors in
a circle of 340 pixels. The red arrow indicates the number of
``aligned" galaxies with 6 or more neighbors for the ``aligned''
set.} \label{alignedThreeOrMoreMomFitMod}
\end{figure}

In fig. \ref{alignedFourRadii} we show what happens as the radius
of the circle is varied, using the model method (which is typical
of the other methods).
Fig. \ref{alignedThreeOrMoreMomFitMod}(a, b) displays an analysis
of the type $2$ for the  radial-fit and model methods. Typically,
for the optimum radius, which is usually  350-370 pixels, there
will be less than 25 out of 500 sets that have as many
galaxy-circles with counts equal to or greater than the original
curved set. In other words, the probability of achieving the
``aligned" set by chance is equal to or smaller than 4{\%}.

\begin{figure}[h!]
\centerline{\includegraphics[width=3in,height=3in]{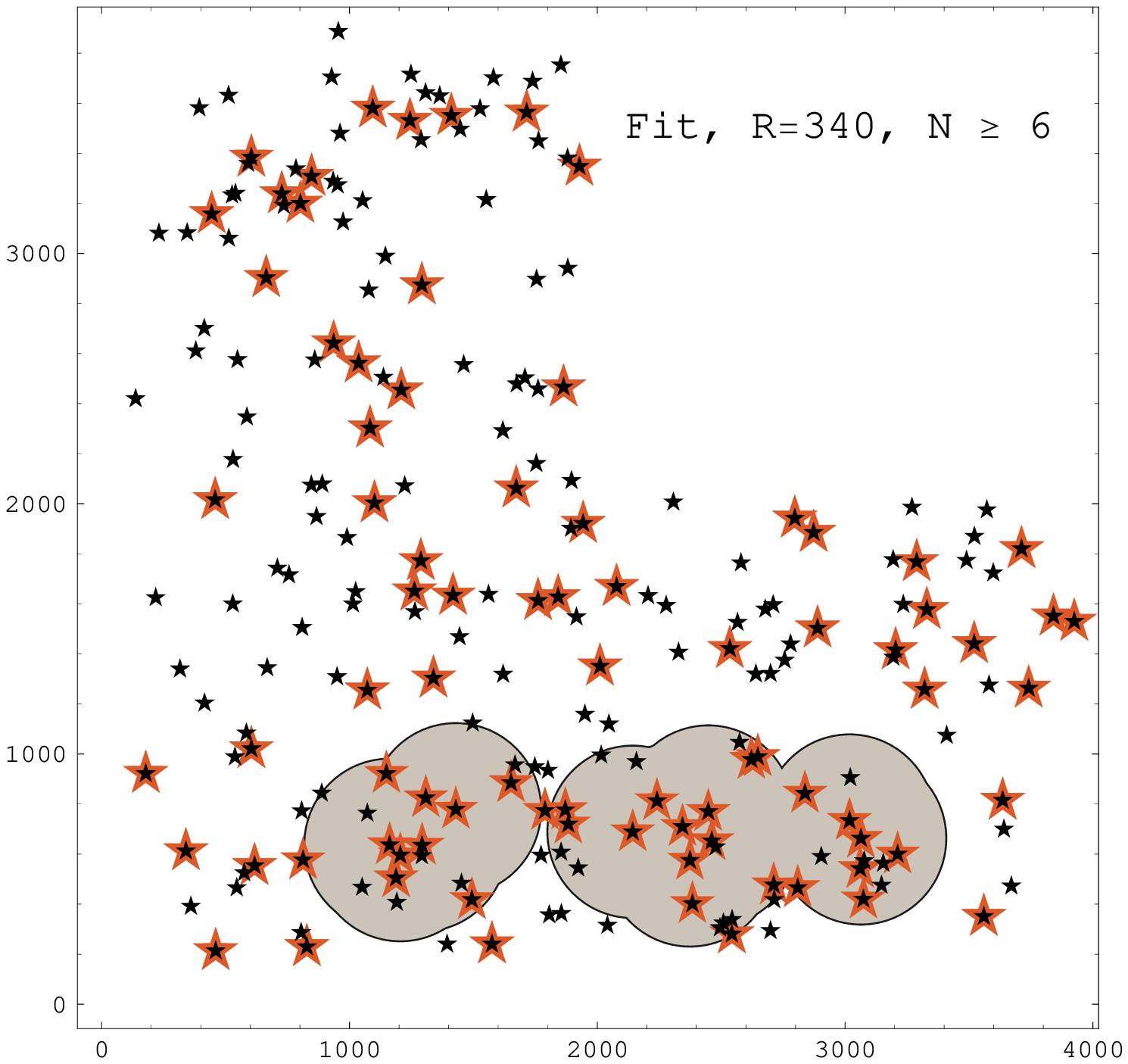}\,\,
\includegraphics[width=3in,height=3in]{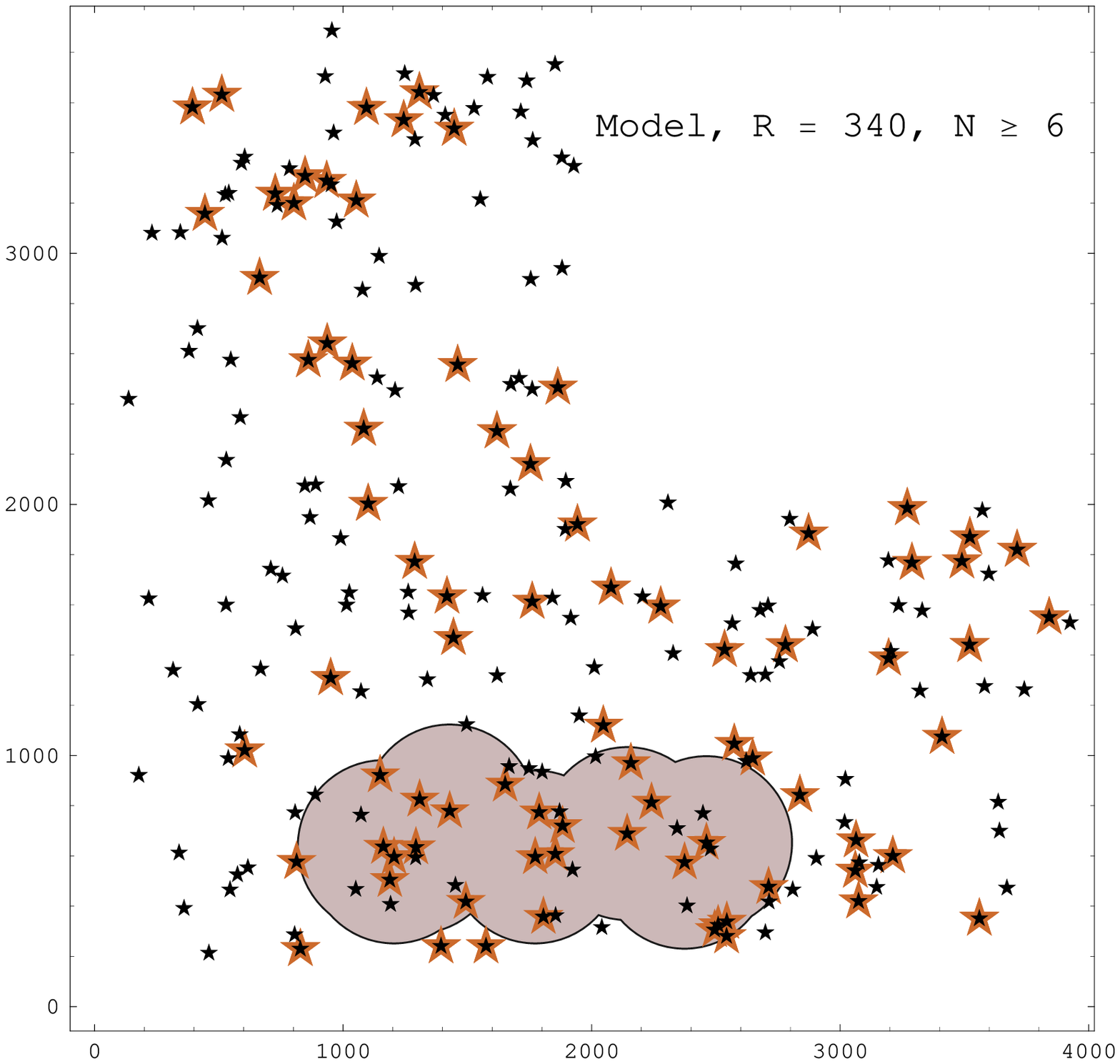}} \caption[curvedCircles]
{``Aligned" galaxies in the north HDF field comparing radial-fit
and model methods to determine the map coefficients.  Neighborhood
circles have a radius of 340 pixels, and must contain 4 or more
neighbors.  The left panel used the model method and right panel
used the radial-fit method. Green shaded areas show the combined
interiors of the ``aligned" neighborhood circles.}
\label{alignedMomFitModBothFields}
\end{figure}

Finally fig. \ref{alignedMomFitModBothFields}(a) shows ``aligned"
galaxies in the Hubble north field and their neighborhood circles
of radius R=350 pixels, using the radial-fit method. This can be
compared with (b) the same plot using model method. In these
plots, large ``stars" indicate ``aligned" galaxies and small
``stars" indicate background galaxies. Shaded areas show the
combined interiors of the ``aligned" neighborhood circles.

That aligned galaxies are also clumped is not a trivial
consequence of the fact that curved galaxies are clumped. Indeed
galaxies midway between curved and aligned are not clumped. And
though the aligned set is not completely independent of the curved
set (it represents $\raise.5ex\hbox{$\scriptstyle 1$}\kern-.1em/
\kern-.15em\lower.25ex\hbox{$\scriptstyle 2$} $ of the complement
of the curved galaxies), we maintain that it is independent enough
to assert that the probability of finding both of these sets to be
clumped by chance is the product of the probability of each.
\footnote{We are not claiming the clumping is due to lensing, only
that the observed clumping is unlikely to occur by chance.}

\subsection{``Mid-range" galaxies}\label{midRangeClump}
\begin{figure}[h!]
\centerline{\leavevmode
\includegraphics[width=2in,height=3.5cm]{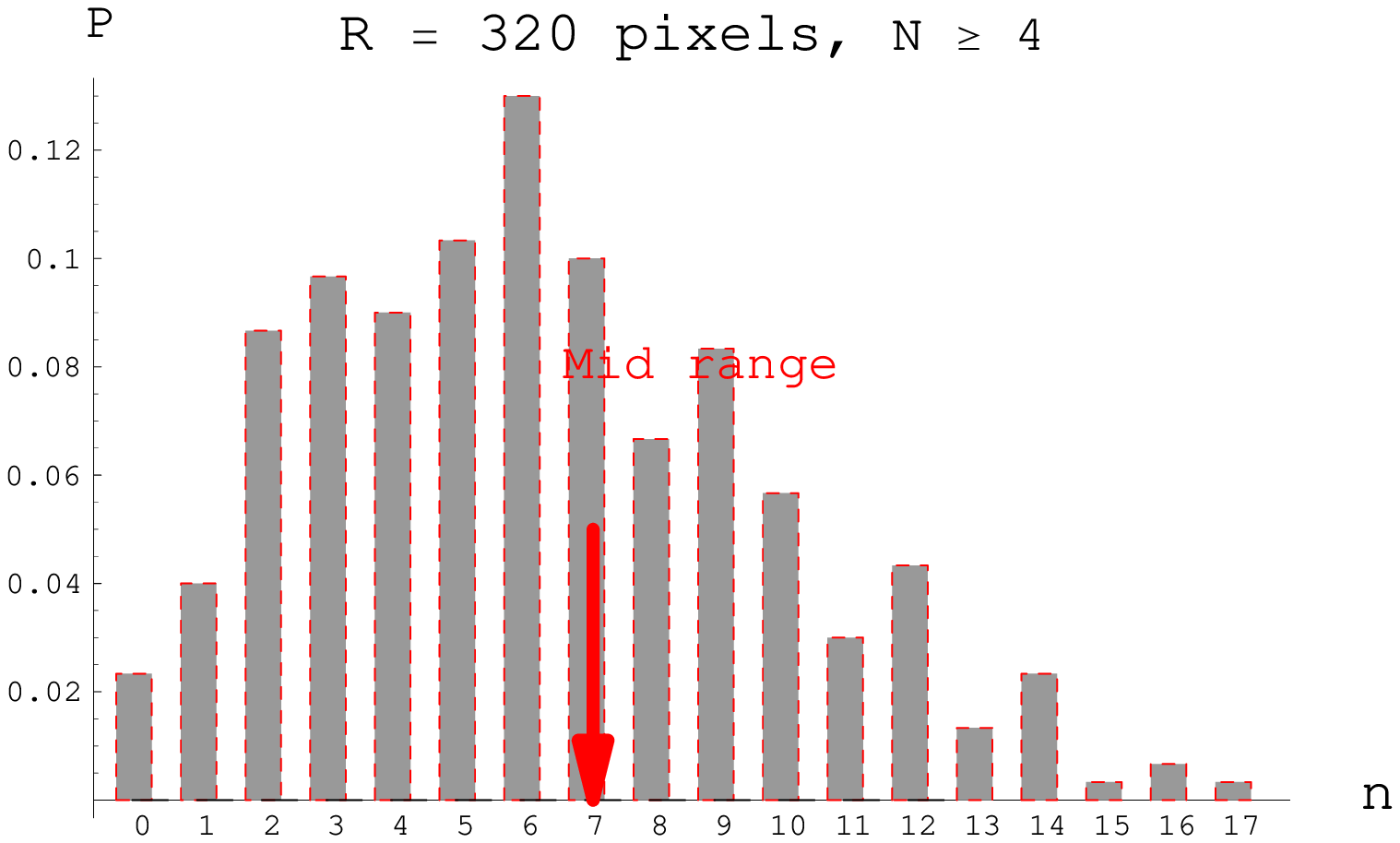}\,\,\,\,
\includegraphics[width=2in,height=3.5cm]{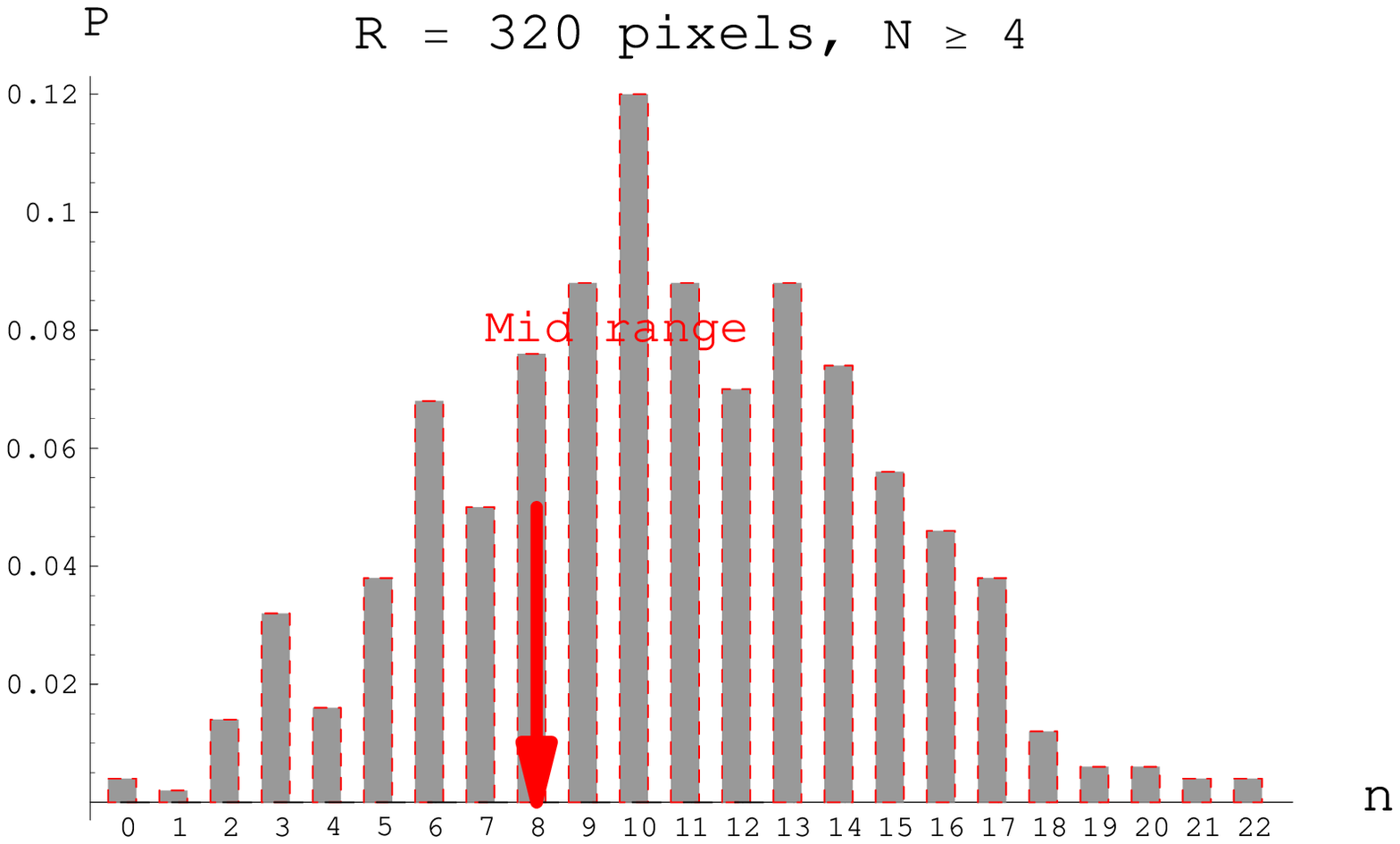}}
\centerline{\leavevmode
$$ \hskip1cm (a)\,\, {\rm ``Radial-fit"} \hskip3cm  (b)\,\, {\rm ``Model"} $$}
\caption[midRangeThreeOrMoreMomFitMod] { Histogram of numbers of
galaxies in 300 randomly chosen sets having 4 or more neighbors in
a circle of 320 pixels. The red arrow indicates the number of
``mid-range'' galaxies with 4 or more neighbors.}
\label{midRangeThreeOrMoreMomFitMod}
\end{figure}

A look at mid-range galaxies shows they are behaving quite
differently.  Fig. \ref{midRangeThreeOrMoreMomFitMod} displays an
analysis of the type $2$ for (a) the radial-fit method, and (b)
the model method. (The number corresponding to the ``mid-range''
set is indicated by an arrow in this bar graph.) There is no
indication of the presence of clumping.

\section{Dark matter sextupole lensing?}\label{darkMatterHaloes}

The previous sections have outlined a theory for sextupole lensing
and described map coefficient measurements and the search for
correlations among coefficient orientation present in the north
Hubble deep field. This section inquires whether those
observations can be explained by the concordance $\Lambda CDM$
model.

The first subsection discusses the mass and impact parameter
requirements of the constituent lens halo. The second subsection
makes an analytical estimate of the total lensing mass required,
and describes  simple simulations supporting this estimate.

The third subsection uses standard $\Lambda CDM$ to compute the
halo mass distribution in the Hubble foreground, showing for
example that $80\%$ of the range should be visible.

The fourth subsection finds that based on the observed over- and
underdensities  of foreground visible galaxies, the imputed dark
matter overdensities are  adequate to account for the clumping
signal observed.

Finally in the fourth subsection we compare the location of the
over- and underdensities with the location of the curved galaxy
clumps.

\subsection{Mass range of lensing haloes }\label{constituentMass}

\begin{figure}[h]
\centering\leavevmode\epsfysize= 7cm \epsfbox{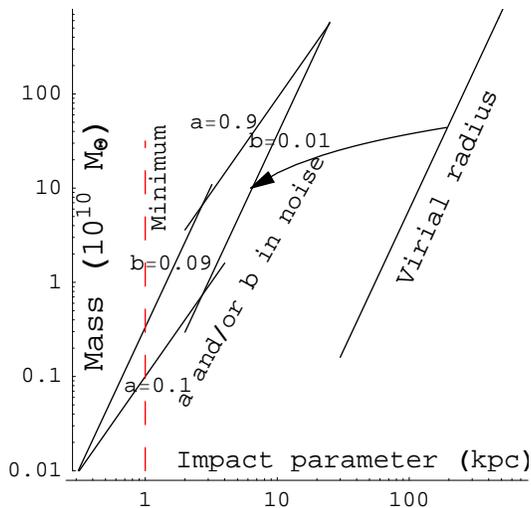}
\caption[massVSimpactParam] {A plot of the constituent lensing
mass versus the impact parameter for a lensing plane at $z=0.6$.
Lines of constant $a$ and $b$ delineate the lensing envelope
required to compete with the range of background values observed
with the model method. The angular size of background galaxies
establishes the minimum possible impact parameter. In the standard
$\Lambda CDM$ the virial radius goes as $M^{1/3}$. This
establishes a line on which dark matter haloes reside.  The
induced map parameters $a$ and $b$ for light streams that
penetrate this halo will fall along curved lines. An example of
such a curve is shown.} \label{massVSimpactParam}
\end{figure}

In order to establish correlations in the quadrupole and sextupole
map-coefficient alignment, lensing-induced coefficients must be
competitive with background values, hence  at least comparable
with or larger than the minimum observed background values. This
simple fact creates limits on the range of acceptable lensing mass
and impact parameters.

Fig. \ref{massVSimpactParam} shows a plot of mass $M$ and impact
parameter $r_0$ for lensing haloes located at $z=0.6$ for
$D_{LS}=D_{LT}$.  The situation on this lensing plane is typical.
Constant values of $a$ and $b$ from lensing are straight lines on
this log-log diagram with slopes of 2 and 3 respectively. For
example, quadrupole coefficients less than 0.1 or sextupole
coefficients smaller than 0.01 would not be observable because
they would be lost in the noise.

Impact parameters smaller than about 1 kpc are excluded because
the radius of the background-galaxy light-path footprints are at
least 2.5 pixels, which is about 0.7 kpc on the $z=0.6$ plane.
This minimum impact parameter is shown in
fig.\ref{massVSimpactParam}.

According to the concordance model the outer radius of dark matter
haloes, referred to as the virial radius, $r_V$, depends on the
total (virial) mass simply as $r_V \, \varpropto \, {M_V}^{1/3}$.
The constant of proportionality may be established by taking
$r_V=300 $ kpc for $M_V = 1.6 \; 10^{12} M_\odot$.  Thus the locus
of dark matter haloes is the straight line indicated in
fig.\ref{massVSimpactParam}. The $a$ and $b$ induced in a light
stream at the virial radius is far too small to be of interest.

However these halo radii are large and light streams will surely
traverse their interiors.  Using a standard NFW profile
\citep{Navarro:1996gj} one can integrate out the longitudinal
variable perpendicular to the lensing plane and calculate the
induced $a$ and $b$ within the interior of the integrated mass
distribution. The result is quite simple, namely $a$ and $b$ are
given to good accuracy by constants times $M(r)/r^2$ and
$M(r)/r^3$, respectively. This is true because of the radial
logarithmic behavior of the integrated interior mass. Thus we can
follow the behavior of $a$ and $b$ by following the interior mass
vs. radius. An example curve is shown in
fig.\ref{massVSimpactParam}.

One finds that the observable lensing region is entered just about
when the radius reaches the so-called ``scale" radius, the radius
at which the profile shifts from logarithmic. In other words, our
finding is that the cores of dark matter haloes are dense enough
to cause sextupole lensing, and this is true for all haloes with
mass greater than about $5 \, 10^{9} M_\odot$ up to a mass of $2
\, 10^{13} M_\odot$.


\subsection{Required lensing mass overdensities}\label{overdensityMass}

We continue to assume the lensing mass lies on the z=0.6 plane. We
emphasize that any mass in the foreground field could be projected
onto this plane if the lensing mass is modified to compensate for
the change in the lever arm coefficient.

Let us assume for simplicity that we are dealing with a single
constituent with a lensing mass of $10^{10} M_\odot$, and that its
spatial distribution is approximately a top-hat. These assumptions
can easily be lifted and generalized in simulations. According to
the considerations of the previous section the radius of such a
lens can be no larger than 7 kpc and should be more like 6 kpc.

Because of the paucity of background galaxies, the probability to
have a scattering event must be large where clumping is seen. For
example, an area of $\pi 280^2$ pixels\footnote{280 pixels
corresponds to 75 kpc at z=0.6.} would typically contain 9 to 10
background galaxies. Assuming these are equally divided between
curved, mid-range and aligned, then of the 6 non-curved background
galaxies, to achieve a total of 5 curved at least 2 must get
curved by lensing. So the probability for a background galaxy to
be lensed by a sufficiently small impact parameter must be the
order of $33\%$.

We can factor the probability to be transformed from not-curved to
curved into a probability that the background galaxy is suitable
(the background coefficient is small enough to easily alter and/or
its orientation is conducive to change) times the probability that
the lensing kick is large enough. If we assume that these
probabilities are about equal, then each of these probabilities
must be near $60\%$.  This probability will be the fractional area
occupied by lenses which produce lensing. In other words we have
to occupy a hefty $60\%$ of the field with haloes capable of
lensing ``suitable" background galaxies.

Taking the radius of the lens to be 5 kpc, then to achieve $60\%$
we need $0.6 (75/3)^2 = 375$ lenses, or in other words a lensing
mass totaling $4 \; 10^ {12} M_\odot$.  This translates to a total
virial mass of $1.8 \, 10^{13} M_\odot$.

To check this estimate we constructed two simulations. One
simulation placed 300 lensing masses of $1.5 \, 10^{10} M_\odot$
randomly in a circle of radius 280 pixels at $z=0.6$.  1000
light-paths were randomly chosen through this lens and the lensing
coefficients were calculated for each path.  Then random
background noise values of the map coefficients, based on the
measured values, were added to the lensing coefficients for each
path. Finally 1000 subsets of these 1000 light paths were randomly
selected. The number of paths in each subset was randomized based
on a Poisson distribution with a mean equal to the average density
of galaxies in the field.  For each subset the number of curved
and aligned galaxies were determined. Then subsets with 9 or more
members were selected.   Of these subsets $50\%$ had 5 or more
curved galaxies as compared to $20\%$ if the lensing mass was set
to zero.  In other words, a lens with a total mass of $4.5 \,
10^{12} \, M_\odot$ dramatically increases the probability of
finding 5 or more curved galaxies when there are 9 or more
background galaxies in the lens area.  This result was verified to
hold as well for the range of other constituent mass values.

A second simulation actually placed lenses of radius 280 pixels
within the field, and circles were used to identify clumping, as
in section \ref{clumping}. Analagous results on lens requirements
were obtained.

The bottom line of our simple simulations and analytical estimate
is that a lensing mass in the range of  $5 \,10^{12}\, M_\odot$ is
both necessary and sufficient to produce the observed clumping of
curved galaxies. Since aligned galaxies are depleted in these
simulations, they support clumping of aligned galaxies as well.

\subsection{Mass in the Hubble deep field foreground }\label{fieldMass}

\begin{figure}[h]
\centering\leavevmode\epsfysize= 6cm \epsfbox{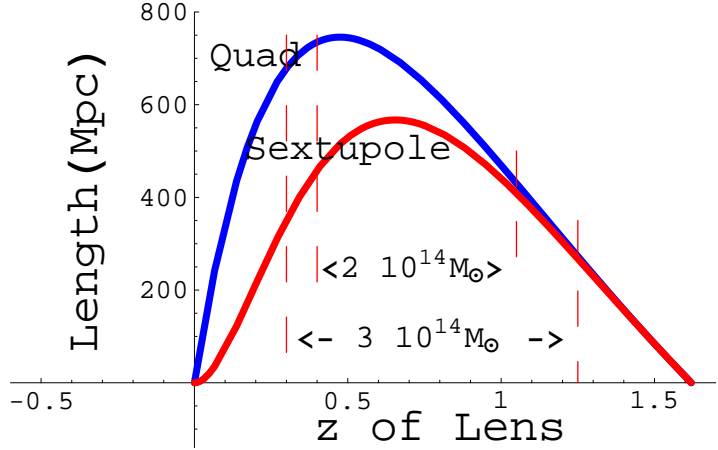}
\caption[leverArms] {A plot of the lever arm coefficients as a
function of the position of the lensing plane for the quadrupole
and sextupole map coefficients for the median $z=1.64$ background
galaxy. These coefficients are given by $D_{LS}
\frac{D_{TL}}{D_{TS}}$ and $ D_{LS} \left({\frac{D_{TL} }{D_{TS}
}} \right)^2$ respectively. The original mass residing in the
regions of large lever-arm and small lever-arm coefficient are
indicated.} \label{leverArms}
\end{figure}

In fig. \ref{leverArms} we have plotted the geometric lever-arm
coefficients (the product of the $D$'s in
eq.\ref{coeffFromPotential}) for the quadrupole and sextupole map
coefficients for a ``median z=1.64" background galaxy.

The geometric coefficients are large for the range $ 0.3 < z <
1.25$, and hence dark matter haloes in this z-region could give
rise to lensing. We will refer to this region as the foreground.
The co-moving volume is a needle-like truncated pyramid, having an
angular width of only 0.75 milliradian. The co-moving distance to
z=1.25 is 3.9 Gpc, and to z=0.3 is 1.2 Gpc. The co-moving volume
``behind" the 3 WFPC chips between z=0.3 and z=1.25 is about $8000
\, Mpc^3$. Using a matter density of 0.27 times the critical
density ($3.7 \, 10^{10}\frac{M_\odot}{Mpc^3}$) for the original
mass density, would imply that this volume originally contained a
dark plus baryonic mass of $3 \, 10^{14} M_\odot$.\footnote{The
total mass is insensitive to the lower z limit. By reducing the
upper limit to z=1.05 the mass decreases by $1/3^{rd}$.}

To get an estimate for the number of haloes within the acceptable
mass range as determined in the previous subsection, one can use
the Sheth-Tormen distribution \citep{Sheth:1999mn}, which is
supported by N-body simulations \citep{Reed:2003hp}. The result is
that about $1.6 \, 10^{14} M_\odot$ lies in the mass range between
$7 \; 10^{9}$ and  $2.3 \, 10^{13} M_\odot$, amounting to $40\%$
of the total halo mass.

Breaking down the $1.6 \, 10^{14} M_\odot$ by decade,  gives
roughly: $6 \; 10^{13} M_\odot$ (9 haloes) between $ 2.3 \,
10^{12}$ and $2.3 \, 10^{13} M_\odot$; $4.7 \, 10^{13} M_\odot$
(64 haloes) between $ 2.3 \, 10^{11}$  and  $2.3 \, 10^{12}
M_\odot$; $3.5 \, 10^{13} M_\odot$ (480 haloes) between $2.3 \,
10^{10}$ and $2.3 \, 10^{11} M_\odot$; and $1.5 \, 10^{13}
M_\odot$ (1140 haloes) between $7 \, 10^{9}$ and $2.3 \, 10^{10}
M_\odot$.

A circle of radius 280 pixels represents about $2.2\%$ of the
field, hence on the average such a circle has a foreground mass of
$3.6 \, 10^{12} M_\odot$.  Dividing that mass by 4.5 yields the
core mass of these halos, $8 \, 10^{11} M_\odot$.  With an
overdensity of 3 this would give a core lensing mass of $2.4 \,
10^{12} M_\odot$, pretty close to the required $5 \, 10^{12}
M_\odot$.\footnote{Our sense is that a full field simulation, in
which the regions adjoining the lens were also populated with
lensing haloes, albeit at lower densities, would increase the
probability of obtaining a lensing signal, and could be
responsible for a factor of 2. There are other uncertainties. We
are pleased with finding agreement within a factor of 2.}

\subsection{Visible foreground galaxy distribution }\label{fieldOverdensity}

Dark matter haloes within the top end of the mass range described
in section \ref{constituentMass} should contain visible galaxies.
There are 420 observable galaxies in the foreground z-range,
indicating haloes down to perhaps $3 \, 10^{10} M_\odot$ are
bright enough to be seen. According to the considerations of the
previous section this represents $80\%$ of the mass in the range
important for sextupole lensing.  Hence we can take the visible
galaxy overdensities as markers for the invisible dark matter halo
overdensities.

\begin{figure}[h!]
\centerline{\includegraphics[width=3.5in,height=3.5in]{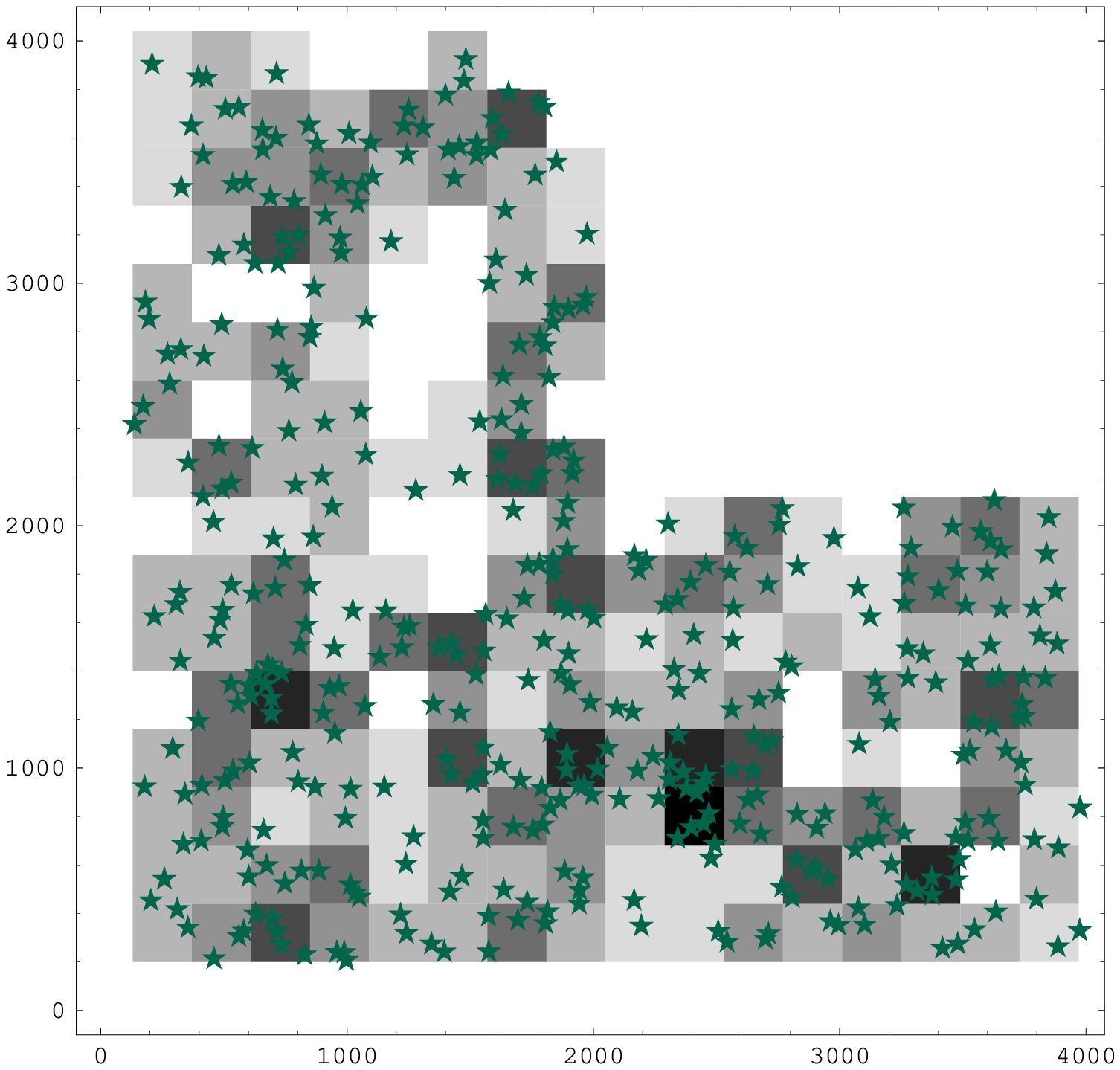}\,\,
\includegraphics[width=3.5in,height=3.5in]{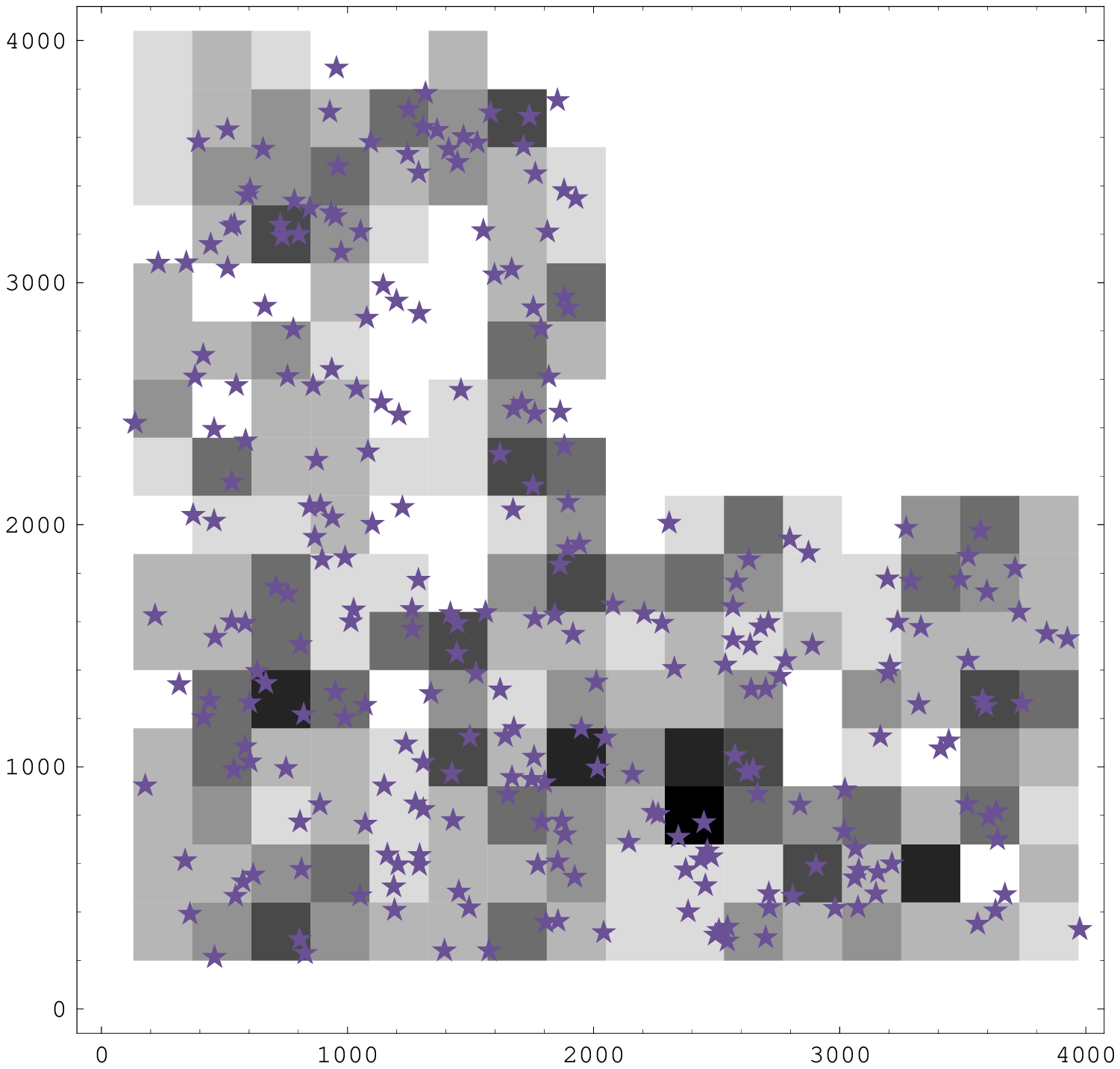}}
\centerline{\leavevmode
$$ (a)\,\, \hskip8cm  (b)$$}\caption[foregroundGalaxies]
{In the left panel foreground galaxies are shown as (dark green)
stars on a grid of 192 square boxes. The boxes are shaded
according to galaxy density, black being the most dense areas.  In
the right panel we show the same grid with the background galaxies
shown as (dark blue) stars.} \label{foregroundGalaxies}
\end{figure}

In fig. \ref{foregroundGalaxies}(a) we plot the positions of the
420 foreground galaxies on a grid of 192 boxes. The occupancy
count varies from 1 to 7, with a mean of 2.17 galaxies per box.
Darker boxes indicate larger numbers of foreground galaxies.
There are 5 boxes with an overdensity of 3 or more times the
average density.

In fig. \ref{foregroundGalaxies}(b) we show the background
galaxies on the same grid as fig. \ref{foregroundGalaxies}(a).
According to our conjecture, to obtain a clump of curved galaxies
one would need to have i) an overdensity of foreground haloes and
ii) at least 9 or more background galaxies probing the overdense
region.  In fig. \ref{foregroundGalaxies}(a) one can identify five
regions with large foreground overdensities (excluding border
regions).  Upon checking out these regions in fig.
\ref{foregroundGalaxies}(b) one sees that two of those five do not
have the prerequisite number of background galaxies probing the
region.  The three curved clumps of fig. \ref{weakenedNoiseCut}
intersect the remaining three regions.  This does not prove our
conjecture, but is certainly estalishes that a lensing scenario is
consistent with our observations and $\Lambda CDM$.

\section{ Systematic error sources}\label{systematics}

In this section we turn our attention to possible systematic
non-lensing sources of the observed clumping.

\subsection{Background galaxy clumping}\label{backgroundClumps}

Could the observed clumping of curved galaxies originate in the
background galaxies themselves?  Since the galaxies at any given
slice in $z$ are known to be spatially correlated, clumping would
naturally result if the galaxies in particular galaxy-groups
possessed the features we are measuring. This could arise in two
distinct ways: i) there might be some age-dependent process at
work.  For example, perhaps old galaxies are more curved.  Or ii)
perhaps some galaxy groups tend to be more curved than others,
because of some common history or composition.

\begin{figure}[h!]
\centering {\leavevmode
\includegraphics[width=8cm,height=5cm]{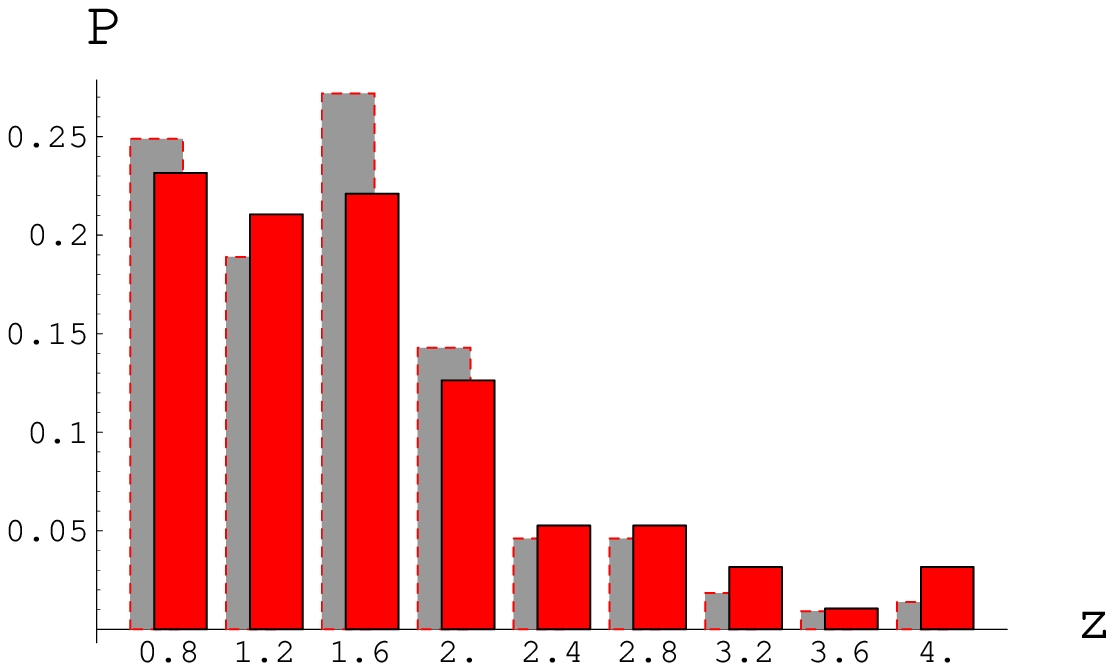}\,\,
\includegraphics[width=8cm,height=5cm]{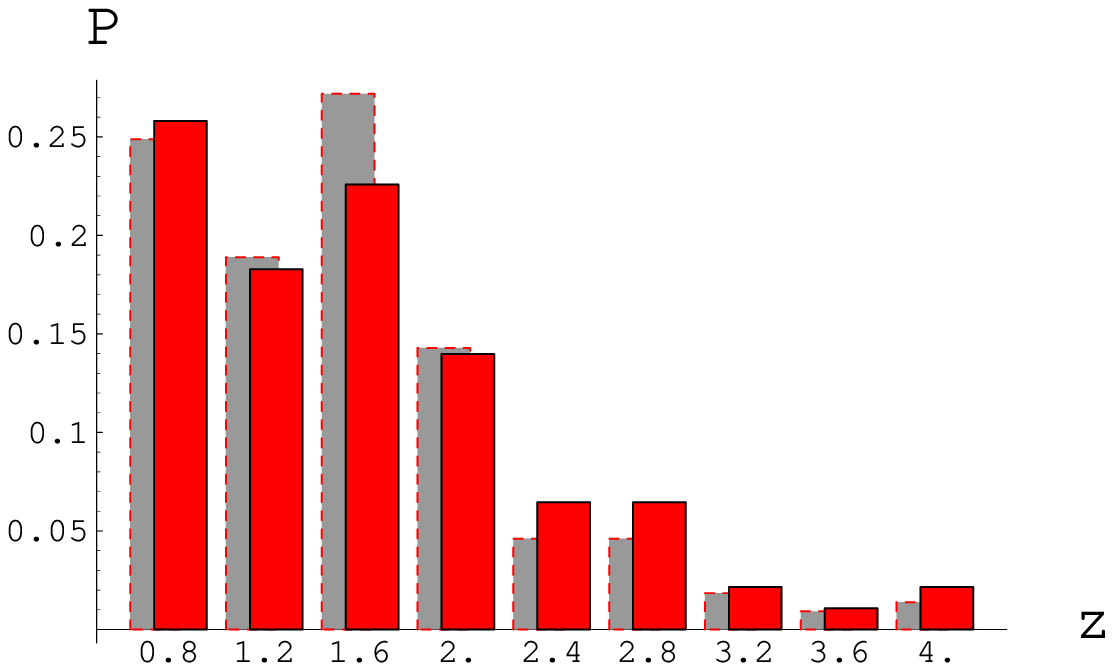}}
\caption[curvedzGroups] {\centering The left bar graph compares
the z-distribution of ``curved'' galaxies in the foreground bars
and all galaxies in the background bars.  The right bar graph
compares the z-distribution of ``aligned'' galaxies in the
foreground bars and all galaxies in the background bars.).}
\label{curvedzGroups}
\end{figure}

In response to item i) one can look at the z-distribution of
``curved" or ``aligned" galaxies to see if there is any evidence
of a bias in the population.  Figure \ref{curvedzGroups} compares
the z-distribution of ``curved" galaxies with all galaxies. and
the z-distribution of ``aligned" galaxies with all galaxies. There
is no particular evidence that older galaxies are more or less
curved or aligned, according to our criteria for curved and
aligned.

\begin{figure}[h!]
\centerline{\includegraphics[width=3.5in,height=3.5in]{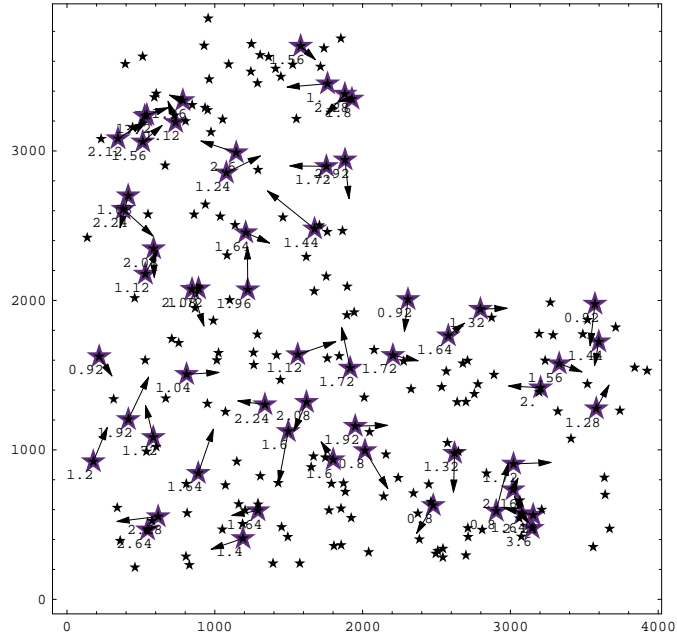}}
\caption[curvedCircles] {Curved galaxies in the north HDF using
the model methods to determine the map coefficients: arrows for
each curved galaxy point to the center of curvature and the
z-values of the ``curved" galaxies are indicated.}
\label{curvedArrowsNzValueFields}
\end{figure}

To address item ii) we first note that z-values are measured in
intervals of $\Delta z = 0.02$. At $z=1.0,\, 2.0, \mbox{ and }
3.0$, a separation of $\Delta z = 0.02$ corresponds to a co-moving
separation of $50, \, 30, \, \mbox{ and } \, 20 \, Mpc$,
respectively. Therefore, galaxies in different z-bins are
well-separated in 3-dimensional space, and an event in one group
should not be able to influence another group.  Hence item ii)
also appears unlikely, for 3 reasons:

\begin{enumerate}
\item None of the z-bin groups seems to be especially curved.
Plotting fig. \ref{curvedzGroups} for $\Delta z = 0.02$ bins,
shows no striking evidence that any group is particularly curved.

\item Same z-bin galaxy groups are spread across large regions of
the Hubble field. Their inter-galaxy separation is typically
larger than the scale of the correlation we are noticing.

\item For each actual ``curved" group we have observed, all
members have distinct z-values. Therefore their curvature could
not arise from a single causative event. This can be see in the
left panel of fig. \ref{curvedArrowsNzValueFields}, where the
z-values are printed alongside the curved galaxies. One can check
the groups to verify that in each group there is no z-value
occurring more than one time.  The z-values for the upper left
``curved" clump are $z=1.12, \, 1.16, \, 1.56, \, 1.72, \, 2.12,
\mbox{ and } 3.00.$  The z-values for the lower right ``curved"
clump are  $z=0.08, \, 1.64, \, 1.72, \, 2.16, \, 2.24, \mbox{ and
} 3.10.$
\end{enumerate}

In summary, we find it to be highly unlikely that the excess
clumping we are seeing for both ``curved" and ``aligned" galaxies
could result from correlated shapes present in the background
galaxies themselves.

\subsection{PSF residuals}\label{PSFclumps}

We know that the PSF correction plays a role in the determination
of which galaxies are ``curved" and we have made an effort to
remove the PSF effects, but are there significant residuals?
Unfortunately the PSF can vary with time, as the temperature of
the Hubble telescope changes. However there are reasons to doubt
the observed clumping is coming from the PSF.

Figures \ref{quadPSFcoeff} and \ref{sextPSFcoeff} show the
orientation and magnitude of the quadrupole and sextupole
coefficient for the Tiny Tim PSF, respectively. We note that the
orientation of the sextupole coefficient is almost constant in
each chip and the magnitude varies only slightly. This is in
strong contrast to the quadrupole moment which gets stronger at
the edge of the chips and whose orientation follows large circles,
centered roughly on the chip. Combining these two would result in
a pattern that would repeat six times around each circle (curved
$\rightarrow$ mid-range $\rightarrow$ aligned $\rightarrow$
mid-range $\rightarrow$ curved $\rightarrow$ ...). For 3
orientations of the quadrupole a minimum is aligned with a minimum
of the sextupole, and each such orientation occurs twice, on
opposite sides of the circle. The radius of this circle is $\sim
800$ pixels, hence the circumference is $\sim 4800$ pixels. Each
repeat has a length of $\sim 800$ pixels, hence the curved region
width would be $\sim 270$ pixels.  In other words these regions
would have half the radius of the clumped regions we have found,
and one-quarter the area.

No such pattern is apparent, and the regions predicted to be
curved do not correspond to the location of our clumped regions.
Since the PSF ``curvature" would vary continuously across the
field, galaxies within one of the curved regions would all curve
the same way, curving galaxies the same way within each clump.
Furthermore there is no evidence of the uni-directional curving
that would be predicted. See fig. \ref{curvedArrowsNzValueFields}.

The time-variation is thought to change the quadrupole strength,
but not flip its direction.  Therefore, even in the case of time
variation, the pattern could be expected to follow the description
of the preceding paragraph.

\subsection{ Image composition}\label{compCLUMPS}

Since we suspect we are processing images in a way not originally
anticipated by the creators of the image composition process, we
were concerned that our observations could be due to a feature of
that process.  We raised this concern with the Space Telescope
Science Institute, who assured us that: ``To (our) knowledge,
there are no instrument artifacts or any portion of the image
processing and stacking which would mimic the curvature you are
noting in the background galaxies." \footnote{We sent the STSCI a
draft of our paper (astro-ph 0803007) which used only moment
methods. This reply from the STSCI Help Desk, was received on
Sept. 9, 2003 with call number CNSHD330235.}

The final image is the co-addition and ``drizzling" of images that
were taken with nine distinctly ``dithered" pointings, enabling
the resolution of the composite image to be higher than any single
image. The dither offsets have to be known accurately at each
point on the focal plane, or else the co-addition will indeed
yield image moments not present in the original images. But this
is a question of magnitude.  As we have repeatedly pointed out,
the magnitude of any distortion must be the order of the observed
moments themselves. In the case of the quadrupole, suppose we have
a false separation of two symmetrical images each offset by
$\Delta$ but in opposite directions. Then a false quadrupole map
coefficient of magnitude $ 1/2 (\Delta/r_0)^2$ will be generated.
For this to have a magnitude of a typical $a \approx 0.2$ would
require $\Delta/r_0 \sim 0.6$. Since our typical $r_0 \geq 3$ ,
this would require $\Delta \geq 1.8$ HDF pixels.  In our opinion,
this is an improbably large offset. We would estimate false
offsets to be at most 1/4 this size (0.2 original camera pixels).
With the effect going as the square that would be a factor of 16
too small.  Similarly the false sextupole generated by three
images each offset in a triangular configuration by $\Delta$ would
have a sextupole coefficient of magnitude $1/6(\Delta^3/r_0^4)$.
For this to have a magnitude of 0.02 even for $r_0=3$ would
require $\Delta \geq 2 $ HDF pixels, again at least 4x larger than
expected false offsets.

If there were errors in the dither amplitudes, one would expect
these errors to be coherent across the field, that is all of the
galaxies in a local region would have the same errors in the
co-addition process.  This would result in a spatial clumping of
curvature, but all the galaxies in any given local region would
appear to be curved in the same direction.  This is not what we
see.

We also note that the regions where curved galaxies clump do not
appear to have any particular identifiable pattern, i.e. they do
not appear to coincide with chip boundaries.

\subsection{ Pixel-derived effects}\label{pixelCLUMPS}

Early on we had a concern that spurious sextupole moments could be
generated by the simple process of pixelating an image.  To probe
this, we took known typically sized bi-Gaussian distributions and,
varying the centroid, dropped them onto a pixel grid. To our
surprise, the falsely induced sextupole map coefficients had
strengths less than $10^{-5}$. Square pixel grids do give rise to
spurious octupole moments, and sensitivity limits are set in that
case.

Of course pixel defects will give rise to spurious changes in
galaxy shapes. By taking a hole out of one side of an image, a
spurious curvature would be induced.  Defects in the quantity and
arrangement required would seem like a serious camera defect.

\subsection{ Galaxy selection effects}\label{selectCLUMPS}

Galaxy selection software would appear to be blind to position in
space and incapable of leading to the spatial clumping of curved
or aligned galaxies

\section{Summary}\label{summary}

We visually examined galaxy images selected by the SExtractor
software from the Hubble deep fields. After filtering images with
two or more major maxima, we imputed sextupole and quadrupole map
coefficient strengths using a moment method, a radial-fit method
and a model method.  The model method includes convolution with
the PSF of the Hubble, charge diffusion and an emulation of the
dither and drizzle process. The ``curved'' galaxies we sought were
identified as those whose sextupole coefficient was oriented with
one of its minima within $10^\circ$ of a quadrupole minima. The
``aligned" galaxies have maxima aligned within $10^\circ$. We then
looked for and found a spatial clumping in the Hubble deep field
of ``curved" and ``aligned" galaxies. The probability of ``curved"
clumping to occur by chance was found to be about 1{\%} for a weak
noise cut and constrained z-distributions on randomly selected
galaxy subsets.  The probability of ``aligned" clumping to occur
by chance was found to be about 4{\%}.

Our lensing hypothesis proposes that within the observed clumps of
curved background galaxies a couple images are curved by close
collisions with dark matter haloes residing in the field
foreground. Simulations and analytical estimates indicate the
required total lensing mass $\sim 5 \,10^{12} M_\odot$. Such a
mass is similar to the total mass expected in overdensities of the
foreground dark matter halo cores. Furthermore there appears to be
a correlation between foreground galaxy overdensities and the
observed curved clumps.

Unfortunately the quadrupole moment of the PSF is known to vary
with time and temperature.  We are hesitant to claim an
observation of small scale structure without a better
characterization of the PSF and without better statistics.
Fortunately the Hubble ACS camera is providing larger fields, and
efforts are underway to characterize the time dependence of it's
PSF. \footnote{Private communication. Richard Massey and Jason
Rhodes}.

The Hubble deep fields are less than 2 min by 2 min, $\sim
10^{-3}$ sq. deg.  A space-based project in the planning stages
(the SuperNova /Acceleration Probe (SNAP) \citep{unknown:2002dp,
SNAP1, SNAP2, SNAP3}) has a weak lensing program of $\sim$ 1000
sq. degrees ($10^6$ times the Hubble deep field) with comparable
resolution, better characterization of the PSF, and deep enough to
provide $\sim 100$ background galaxy images per sq. min.

\vskip 1.5cm

 {\bf\large\centerline{Acknowledgements}}

We would like to acknowledge the encouragement and support of Tony
Tyson and David Wittman of Bell Labs. Visits by both of us to
Bell-Labs prepared us for this undertaking, especially introducing
us to existing software and weak-lensing techniques.  We would
also like to thank Pisin Chen and Ron Ruth at SLAC for trusting in
our judgment and encouraging us to proceed. This work was
supported by DOE grant  DE-AC03-76SF00515.

After presenting our ideas to Tony Tyson in January 2002, he
suggested we look at a posting by D.M. Goldberg and P. Natarajan
titled, ``The Galaxy Octupole Moment as a Probe of Weak Lensing
Shear Fields" \citep{Goldberg:2001nf}. The mathematical formulae
presented there were actually what we have been calling the
sextupole. Our nomenclature derives from beam physics, where the
sextupole field is created by a magnet with six poles, 3 of
positive polarity and 3 of negative polarity. To our knowledge,
there was no mention by Goldberg and Natarajan of the induced
"sextupole/octupole" shape or using its correlation with the
quadrupole shape to distinguish galaxies.

We thank Ken Shen for his assistance in providing a program to
identify double-maximum galaxies, and Konstantin Shmakov for help
with software.

The first version of this present work was posted in August 2003.
\citep{Irwin:2003qw}.

\vskip 1.5cm

 {\bf\large\centerline{Appendix}}

Imagine that we are considering a background galaxy (S) that has
not yet been lensed, and suppose the radial-fit method is used to
find $a=a_S$. One can imagine a radially symmetric root-galaxy (R)
that had a zero $a$ parameter mapped to match the radial behavior
and the quadrupole coefficient of background galaxy (S). The
corresponding map would be $w_R=w_S + a_S \bar{w}_S$.

Now suppose that there is a lensing event described by $w_S=w_T +
a_L \bar{w}_T.$  The composition of these 2 maps gives
\begin{equation}
\label{compositon} w_R=(1+\bar{a}_L a_S) w_T + (a_L+a_S)\bar{w}_T.
\end{equation}
The map from the telescope observation (T) to the root-galaxy (R)
has $a=a_L+a_S$.  In addition there is a slight rotation and scale
change represented by the coefficient of $w_T$. Since our measured
estimates are based on a unit coefficient for the $w_T$
coefficient, we must remove this implicit (undetectable) scaling
and rotation.  Define a Q (quelle) galaxy that is a scaled root
galaxy by $w_Q=(1+\bar{a}_L a_S)^{-1} w_R.$  Then  we have
\begin{equation}
\label{compositon} w_Q= w_T + (1+\bar{a}_L a_S)^{-1}
(a_L+a_S)\bar{w}_T.
\end{equation}

In the case of weak lensing $a_L$ is small, so $(1+\bar{a}_L a_S)$
is close to unity and we obtain the simple vector addition $a =
a_L + a_S$.  In the case that $a_L$ is comparable with $a_S$ (say
they are both of magnitude 0.4), then the coefficient is still a
small adjustment since $\vert a_L a_S \vert \sim 0.16$.  The
correction is smaller if $a_L$ and $a_S$ are not aligned.  With
these parameters the average deviation of $(1+\bar{a}_L a_S)$ from
unity is a tiny $1/2 \vert a_L a_S \vert^2 \sim  .01$. The bottom
line is that the lensing may be added vectorially with the
background.  The bias error with this approximation can be the
order of $1\%$.

To be complete one can carry this through for the sextupole as
well.  Define \BA \label{composition} w_R &=& w_S + a_S \bar{w}_S
+ b_S {\bar{w}_S}^2 \cr w_S &=& w_T + a_L \bar{w}_T + b_L
{\bar{w}_T}^2 \EA yielding \BA \label{composition} w_R =
(1+\bar{a}_L a_S) w_T &+& (a_L + a_S) \bar{w}_T + (b_L + b_S)
{\bar{w}_T}^2 \cr &+& 2 b_S \bar{a}_L w_T \bar{w}_T + a_S
\bar{b}_L {w_T}^2 \cr &+& b_S \bar{b}_L \bar{w}_T {w_T}^2 + 2
\bar{a}_L b_S \bar{b}_L {w_T}^3 + b_S \bar{b}_L {w_T}^4. \EA

The last line can be dropped as all have products of 2 or more b's
which are the order of $\sim .001$. On the second line we have a
new pair of quadratic order terms that we have referred to as
d-terms or gradient terms.  They will not contribute to the
sextupole or quadrupole term.  They will be noise in the d-term if
that is pursued later. If we had included such terms in the
background galaxies these terms could have generated sextupole
terms but all would be small since the measured d-terms are
typically a factor of 3 smaller than the sextupole terms.

Scaling as before, and dropping the very small terms we get
\begin{equation}
\label{compositon} w_Q = w_T + (1+\bar{a}_L a_S)^{-1} (a_L + a_S)
\bar{w}_T + (1+\bar{a}_L a_S)^{-1} (b_L + b_S) {\bar{w}_T}^2 +
\cdots
\end{equation}
The final result is ``the measured map coefficient is the simple
sum of the lensing coefficient and the background coefficient",
where the background coefficients are those that would have been
measured for the galaxy image in the absence of lensing.

\end{document}